\documentclass[aip,graphicx,reprint]{revtex4-1}
\usepackage{xcolor}
\usepackage{graphicx}
\usepackage{physics}
\usepackage{hyperref}
\usepackage{upgreek}
\usepackage[T1]{fontenc}

\hypersetup{
    colorlinks=true,
    linkcolor=blue,
    filecolor=black,      
    urlcolor=blue,
    allcolors=blue,
}

\draft

\begin{document}

\title{Routing single photons with quantum emitters coupled to nanostructures}

\author{Mateusz Duda}
\email{mduda1@sheffield.ac.uk}
\affiliation{School of Mathematical and Physical Sciences, University of Sheffield, Sheffield S3 7RH, UK}

\author{Nicholas J. Martin}
\affiliation{School of Mathematical and Physical Sciences, University of Sheffield, Sheffield S3 7RH, UK}

\author{Eve O. Mills}
\affiliation{School of Mathematical and Physical Sciences, University of Sheffield, Sheffield S3 7RH, UK}

\author{Luke R. Wilson}
\affiliation{School of Mathematical and Physical Sciences, University of Sheffield, Sheffield S3 7RH, UK}

\author{Pieter Kok}
\affiliation{School of Mathematical and Physical Sciences, University of Sheffield, Sheffield S3 7RH, UK}

\date{\today}

\hyphenation{wave-guide}


\begin{abstract}
Quantum emitters coupled to nanophotonic structures are an excellent platform for controllable single-photon scattering. The tunable light-matter interaction enables the construction of a single-photon switch---a device that can route a single photon from an input port to a selected output port. Such single-photon switching devices can be integrated into reconfigurable photonic circuits to actively control the photon propagation direction in a quantum network. Ideally, a single-photon switch should operate with high speed, efficiency, and fidelity, preserving the state of the input photon in the routing process. This review brings together key input-output methods from quantum optics, theoretical proposals of emitter-based single-photon routing mechanisms, and experimental demonstrations of single-photon switching devices across different physical platforms, including semiconductor quantum dots, neutral atoms, superconducting qubits, and color centers. We highlight the need for reporting the key figures of merit (speed/efficiency/fidelity) in future single-photon switch demonstrations to support further developments in the field.
\end{abstract}

\maketitle 

\newpage
\tableofcontents

\newpage
\section{Introduction}\label{sec:intro}

Photons are considered the most promising candidates for transferring quantum information, due to their fast propagation and weak interactions with their environment~\cite{Kimble2008, Northup2014}. This makes photonics the leading platform for quantum communications, where a secret key is encoded in quantum states of photons and transferred between distant parties~\cite{Gisin2002, Pirandola2020}. More generally, any quantum network consisting of local information-processing nodes and communication channels is likely to use photons as the information carrier.

The development of photonic quantum technologies has seen rapid progress in recent years~\cite{OBrien2009}. This is due to the potential that quantum technologies have to outperform their classical counterparts, for example in terms of speed (computing), security (communication), and resolution (imaging), which can be increased by exploiting superposition, measurement uncertainty, and entanglement, respectively. In the context of {\it{photonic}} quantum technologies, a key research focus has been to reduce the footprint of optical setups, leading to a gradual transition from bulk optics to chip-based photonic devices (integrated quantum photonics~\cite{Wang2020}). This enables the key steps of quantum information processing---state generation, processing, and measurement---to be performed in chip-scale nanostructures, which is a necessary requirement for developing more complex quantum networks that can move us from proof-of-principle experiments with few components to real-world applications potentially requiring millions of optical components.

\begin{figure}
    \centering
    \includegraphics[width=\linewidth]{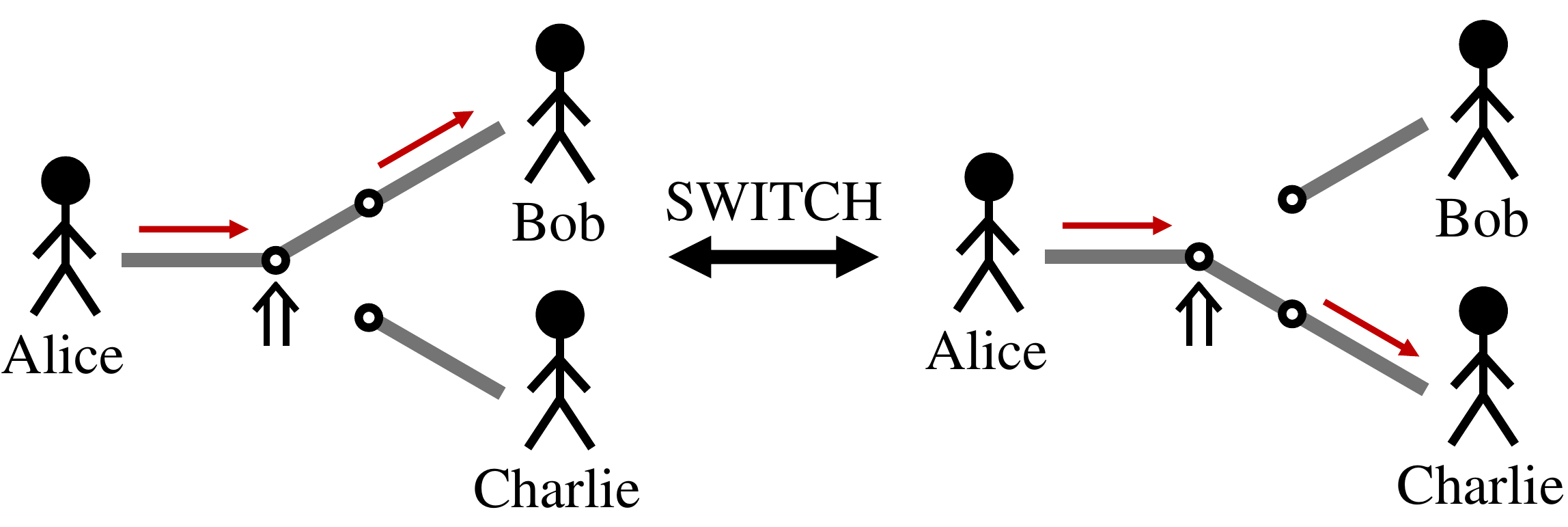}
    \caption{Diagram of a simple three-node network. Alice sends photons into the network, and uses a switch to route photons to one of the two receivers (Bob or Charlie). The red arrows indicate the direction of photon flow, and the black arrows indicate a control mechanism used by Alice to flip the switch.}
    \label{fig:switch_diagram}
\end{figure}

In quantum networks with multiple receiver nodes, controlling photon propagation is essential. For example, consider the simple three-node network shown in Fig.~\ref{fig:switch_diagram}, where Alice represents a sender node, and Bob and Charlie represent receiver nodes. If Alice wants to send information to Bob and then Charlie via a string of photons, she needs a photon switch that allows her to route photons to the desired target location, such that Bob and Charlie receive the correct information. A photon switch is therefore a basic component that is needed to actively control photon flow in a network. A photon router is a related device that sends photons between different nodes without an active control mechanism.

The ability to switch single photons is a central requirement in various quantum technology proposals utilizing photonic qubits. For example, controlling the photon propagation direction enables device sharing in quantum communication protocols~\cite{Karavias2024}, fusion-based quantum computing~\cite{Bartolucci2021, Chan2025, PsiQuantum2025}, and ultimately constructing the quantum internet~\cite{Kimble2008, DiAdamo2022}. Switching also allows photons to be directed into a storage loop, which could be used in a photonic quantum memory~\cite{Pittman2002}, and it enables multiplexing, where photons generated by different single-photon sources are directed onto the same path~\cite{Meyer-Scott2020}. As the complexity of quantum networks grows and the number of possible photon paths increases, the need for components like single-photon switches becomes more apparent. Such components form the basis of reconfigurable photonic integrated circuits, where the circuitry can be rearranged using switches that are triggered by, e.g., previous photon detection events (feedforward).

Light-matter interactions are a natural way to control photon propagation in nanophotonic devices, by means of single-photon scattering from localized systems. In quantum optics, the fundamental interaction between light and matter is that of a single photon with a single atom. A single photon can scatter from an atom in a certain direction with some probability that depends on the physical parameters of the atom, such as its energy level structure. By controlling these parameters (e.g., via feedforward from photon detection), it becomes possible to route photons in a chosen direction, as the interference between different scattering pathways can be manipulated. Further control over photon propagation is achieved by confining photons to waveguides~\cite{Roy2017}, which provide channels for transporting photons and can be easily interfaced with artificial atoms such as semiconductor quantum dots (QDs)~\cite{Lodahl2015} or superconducting qubits~\cite{Blais2021}. We will collectively refer to atoms and atomic-like systems like these as quantum emitters.

Emitter-based single-photon switches are attractive because they offer a routing primitive that is not available in purely linear optics. The optical response can depend on a single microscopic quantum degree of freedom, such as a single emitter excitation or a long-lived internal state like a spin qubit, rather than only on a macroscopic classical control signal. In waveguide and cavity quantum electrodynamics (QED), a single quantum emitter that is strongly coupled to a guided optical mode can coherently scatter incident photons, and the single-excitation saturability of the emitter gives rise to a few-photon nonlinear response~\cite{Ralph2015, Pettersson2026}, including photon-number-dependent transmission, reflection, and phase shifts~\cite{Chang2014, RevModPhys.95.015002}. In solid-state nanophotonic platforms, experiments have demonstrated measurable single-photon-level nonlinear modification of guided light, while near-unity coupling efficiencies indicate that high-contrast operation is plausible in principle. However, the performance of such devices remains strongly contingent on suppressing noise sources such as spectral wandering, blinking, pure dephasing, and other loss channels~\cite{Javadi2015, Arcari2014, Lodahl2015, Hallacy2025}.

In contrast, classical integrated switching platforms, including Mach--Zehnder interferometer meshes and programmable photonic circuits~\cite{Bogaerts2020ProgrammablePhotonicCircuits,LeeDupuis2019SwitchFabrics}, thermo-optic phase shifters~\cite{Parra2024ThermoOpticReview}, electro-optic carrier-depletion phase shifters~\cite{Reed2014CarrierDepletionReview}, resonator-based switches~\cite{Chen2023ReviewIntegratedOpticalSwitches}, and MEMS-actuated switches~\cite{Quack2023IntegratedSiliconPhotonicMEMS,LeeDupuis2019SwitchFabrics}, implement linear-optical transformations where the configuration is set by classical control signals. For a fixed device setting, photon transport remains linear, while heater currents, applied voltages, or mechanical actuation determine the selected routing state~\cite{LeeDupuis2019SwitchFabrics}. Such devices can route single photons with high fidelity because single-photon states evolve linearly in passive and actively-tuned linear circuits. However, they do not by themselves generate deterministic photon--photon interactions, and effective nonlinear functionality in linear-optics schemes instead relies on additional resources due to, e.g., postselecting on successful measurement outcomes~\cite{Knill2001}.

As engineering platforms, classical switching architectures are compelling because they support scalable, high-port-count, and potentially broadband routing fabrics, especially in interferometric implementations where the performance can be analyzed and optimized at the level of elementary switching cells. At the same time, these advantages should not be conflated with negligible system overhead, because large-scale switch systems can still incur substantial insertion loss and calibration complexity~\cite{LeeDupuis2019SwitchFabrics, Qiao2017}.

Switches based on quantum emitters are most compelling in settings where the relevant control variable must itself remain microscopic and, ideally, quantum in nature, for example when photon transport is conditioned on an internal spin state rather than a classical actuator. Spin-state-dependent photon switching has been demonstrated in nanophotonic waveguides, but the usefulness of this approach as a genuinely quantum routing primitive depends on whether coherence of the control degree of freedom can be preserved throughout the scattering process, in addition to achieving strong and spectrally-stable light-matter coupling~\cite{Javadi2018, Aharonovich2016}.

These complementary strengths motivate hybrid architectures in which classical integrated switching provides the reconfigurable routing backbone, while emitter-based devices are used at selected nodes where state-dependent scattering or genuine single-emitter nonlinearity is required. A hybrid strategy would therefore not aim to replace mature linear photonics, but to reserve emitter-based switches for situations where a microscopic, state-dependent, nonlinear few-photon response provides functionality that would otherwise be probabilistic, measurement-conditioned, or substantially more resource-intensive.

In this review, we summarize theoretical proposals and experimental demonstrations of single-photon switches based on photon scattering from quantum emitters coupled to various nanostructures. We also discuss the key figures of merit that quantify the performance of a single-photon switch: speed, efficiency, and fidelity. A high speed implies a fast response to a control mechanism, which is required for a high device operation rate. A high efficiency corresponds to minimal leakage to undesired waveguide channels and minimal scattering loss out of the system, ensuring that photons are always routed in the desired direction (i.e., deterministically). Finally, a high fidelity is essential to ensure that the routed photon wave packets are not perturbed by the switching process, which would correspond to a loss of information stored in the wave packets. The minimum required speed, efficiency, and fidelity will depend on the use case; for example, the required speed of a single-photon switch will be determined by the maximum photon flux in the system. An efficiency below unity will limit the number of switches that can be used in a photonic circuit, as the probability of photon loss will grow after each switching event. The number of switches that are needed will determine the minimum efficiency/fidelity that each switch should operate with. There are also other practical considerations that need to be taken into account for single-photon switches to be implemented in useful technologies, including  how long the switch can remain in one configuration before it resets (operation time), how easily the switch can be integrated into a larger network with other photonic devices (scalability/compatibility), operating conditions, and power consumption per switching event.

The remainder of this review is organized as follows. In Section~\ref{sec:methods}, we outline the most common theoretical methods that have been used to model single-photon scattering from quantum emitters in waveguide-based structures. As an example, we apply these methods to one of the simplest scenarios that has been considered in the literature, namely a two-level quantum emitter coupled to a one-dimensional waveguide. At the end of the section, we define the efficiency and fidelity of a single-photon switch in terms of single-photon scattering amplitudes, and discuss the trade-off between speed and efficiency/fidelity. We then review theoretical single-photon routing proposals in Section~\ref{sec:theory}, covering different types of routing mechanisms. In Section~\ref{sec:FOM_experiment} we change focus from theory to experiment, and start by expanding on the figures of merit from Section~\ref{sec:methods} by discussing other switch parameters that are relevant in practical implementations and may depend on the material platform. Then, in Section~\ref{sec:structures}, we provide an overview of solid-state nanostructures that have been considered for emitter-based photon switching across different platforms, including photonic crystal waveguides and cavities, and other resonator structures like ring resonators and superconducting microwave resonators. In Section~\ref{sec:experiment}, we review experimental demonstrations of single-photon switching with quantum emitters, covering different types of emitters including semiconductor QDs, neutral atoms, superconducting qubits, and solid-state defects (e.g., color centers). We conclude the review with a summary and outlook in Section~\ref{sec:conclusion}.


\section{Theoretical methods}\label{sec:methods}

Before summarizing the proposals of single-photon switches, we outline the theoretical methods that have been used to model these systems. The goal of single-photon scattering calculations is typically to derive the transmission and reflection amplitudes of a system, from which the efficiency and fidelity of routing a single photon can be calculated. In the following, we focus on three of the most widely used techniques for calculating single-photon transmission and reflection amplitudes in the systems of interest---the real-space approach (Section~\ref{subsec:methods_1}), the input-output formalism combined with the scattering matrix (Section~\ref{subsec:methods_2}), and the discrete coordinate scattering approach (Section~\ref{subsec:methods_3}). We also mention other theoretical methods that have been used in the literature (Section~\ref{subsec:methods_4}). We then show how the transmission and reflection amplitudes can be used to calculate the efficiency and fidelity of a single-photon switch, giving an example of how the efficiency and fidelity scale with the width of the input photon wave packet and the emitter loss rate (Section~\ref{subsec:methods_5}). At the end of this section, we discuss other noise sources in realistic quantum emitters, including pure dephasing and spectral wandering (Section~\ref{subsec:methods_6}).

In order to illustrate how each of the three main methods works, we apply them to the simple case of a two-level quantum emitter coupled to a one-dimensional waveguide, as shown in Fig.~\ref{fig:two_level_emitter_waveguide}. The waveguide is treated as continuous in Sections~\ref{subsec:methods_1} and \ref{subsec:methods_2} [Fig.~\ref{fig:two_level_emitter_waveguide}(a)], and it is treated as a discrete set of coupled resonators in Section~\ref{subsec:methods_3} [Fig.~\ref{fig:two_level_emitter_waveguide}(b)]. Both types of waveguides have been widely used in theoretical investigations of single-photon routing. While the real-space approach and input-output formalism are often used to analyze photon transport properties in continuous waveguides, the discrete coordinate scattering approach is a discrete version of the real-space approach that allows the same transport properties to be derived in coupled-resonator waveguides.


\subsection{Real-space approach}\label{subsec:methods_1}

\begin{figure}
    \centering
    \includegraphics[width=\linewidth]{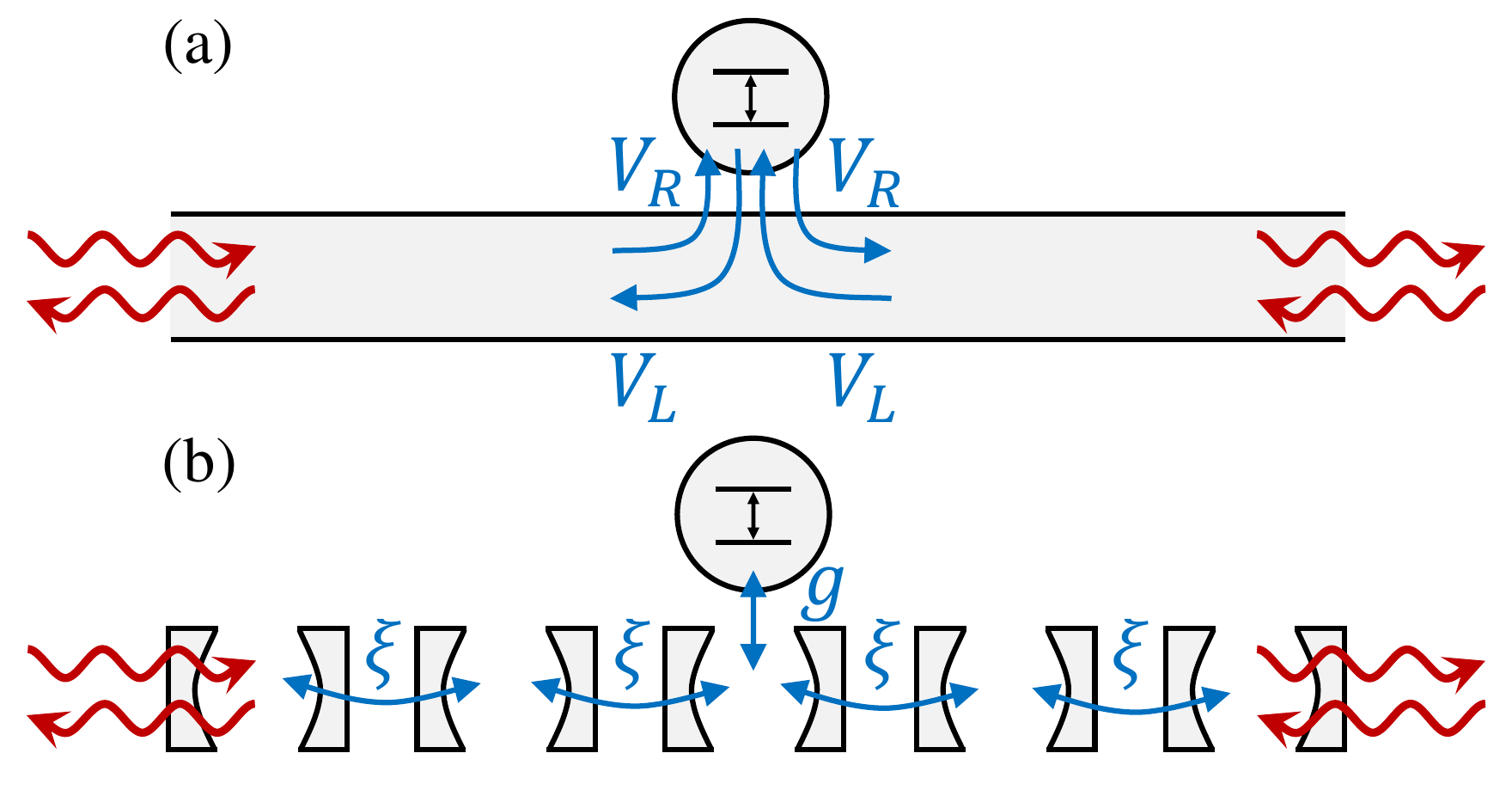}
    \caption{A two-level quantum emitter coupled to (a) a continuous waveguide and (b) a discrete coupled-resonator waveguide. In (a), the emitter couples to the right- and left-moving modes of the waveguide with coupling rates $V_R$ and $V_L$, respectively. In (b), the emitter couples to one resonator with coupling rate $g$, and the nearest-neighbor hopping rate between the resonators is $\xi$.}
    \label{fig:two_level_emitter_waveguide}
\end{figure}

The real-space approach is the most widely used analytical method for studying single-photon scattering in quantum optics, and was first used in this context by Shen and Fan in 2005~\cite{Shen2005, Shen2005_2}. This method involves solving the time-independent Schr\"{o}dinger equation with the Hamiltonian of the system expressed in the position (i.e., real-space) basis. The single-photon transmission and reflection amplitudes $t$ and $r$ are obtained by assuming plane-wave solutions for the incoming and outgoing photon amplitudes, which are expressed in terms of $t$ and $r$. In the following, we outline the derivation of the transmission amplitude $t_{\text{wg}}$ and the reflection amplitude $r_{\text{wg}}$ for the system in Fig.~\ref{fig:two_level_emitter_waveguide}(a), where the subscript indicates that we are considering a two-level emitter coupled to a continuous waveguide.

\subsubsection{Hamiltonian}

We start from the $k$-space Hamiltonian (${\hbar = 1}$)
\begin{align}
\begin{split}
    H =&\; \tilde{\omega}_e \sigma^+ \sigma^- + \int_0^{\infty} \omega(k) a_R^{\dagger}(k) a_R(k) dk\\
    &+ \int_{-\infty}^0 \omega(k) a_L^{\dagger}(k) a_L(k) dk\\
    &+ \int_0^{\infty} \left[ \tilde{V}_R(k) a_R^{\dagger}(k) \sigma^- + \tilde{V}^*_R(k) a_R(k) \sigma^+\right]dk\\
    &+ \int_{-\infty}^0 \left[ \tilde{V}_L(k) a_L^{\dagger}(k) \sigma^- + \tilde{V}^*_L(k) a_L(k) \sigma^+\right]dk,
\end{split}
\end{align}
where $\tilde{\omega}_e$ is the transition frequency of the two-level emitter, and ${\sigma^+ = \ketbra{e}{g}}$ and ${\sigma^- = \ketbra{g}{e}}$ are the raising and lowering operators, respectively, in terms of the ground state $\ket{g}$ and the excited state $\ket{e}$ of the emitter. In addition, $\omega(k)$ is the waveguide dispersion relation, $a_R^{\dagger}(k)$ and $a_R(k)$ [$a_L^{\dagger}(k)$ and $a_L(k)$] are the bosonic creation and annihilation operators for right-moving (left-moving) photons with wave number $k$, and $\tilde{V}_R(k)$ and $\tilde{V}_L(k)$ are the ($k$-dependent) coupling rates between the emitter and the right- and left-moving waveguide modes, respectively.

To simplify the problem, we assume that the emitter-waveguide coupling rates are independent of $k$, such that ${\tilde{V}_{\mu}(k) = V_{\mu}/\sqrt{2\pi}}$ for ${\mu \in \{L,R\}}$, where $V_\mu$ does not depend on $k$. This is routinely done in the literature, and is called the Markov approximation~\cite{Roy2017}. A uniform (nonlocal) coupling in $k$-space corresponds to a local interaction in position space. The Markov approximation leads to equations of motion that depend only on the current time~\cite{Gardiner1985} (i.e., the waveguide acts as a memoryless reservoir). With this approximation, the $k$-space Hamiltonian becomes
\begin{align}\label{eq:H_k}
\begin{split}
    H =&\; \tilde{\omega}_e \sigma^+ \sigma^- + \int_0^{\infty} \omega(k) a_R^{\dagger}(k) a_R(k) dk\\
    &+ \int_{-\infty}^0 \omega(k) a_L^{\dagger}(k) a_L(k) dk\\
    &+ \int_0^{\infty} \left[ \frac{V_R}{\sqrt{2\pi}} a_R^{\dagger}(k) \sigma^- + \frac{V^*_R}{\sqrt{2\pi}} a_R(k) \sigma^+\right]dk\\
    &+ \int_{-\infty}^0 \left[ \frac{V_L}{\sqrt{2\pi}} a_L^{\dagger}(k) \sigma^- + \frac{V^*_L}{\sqrt{2\pi}} a_L(k) \sigma^+\right]dk.
\end{split}
\end{align}

\begin{figure}
    \centering
    \includegraphics[width=0.8\linewidth]{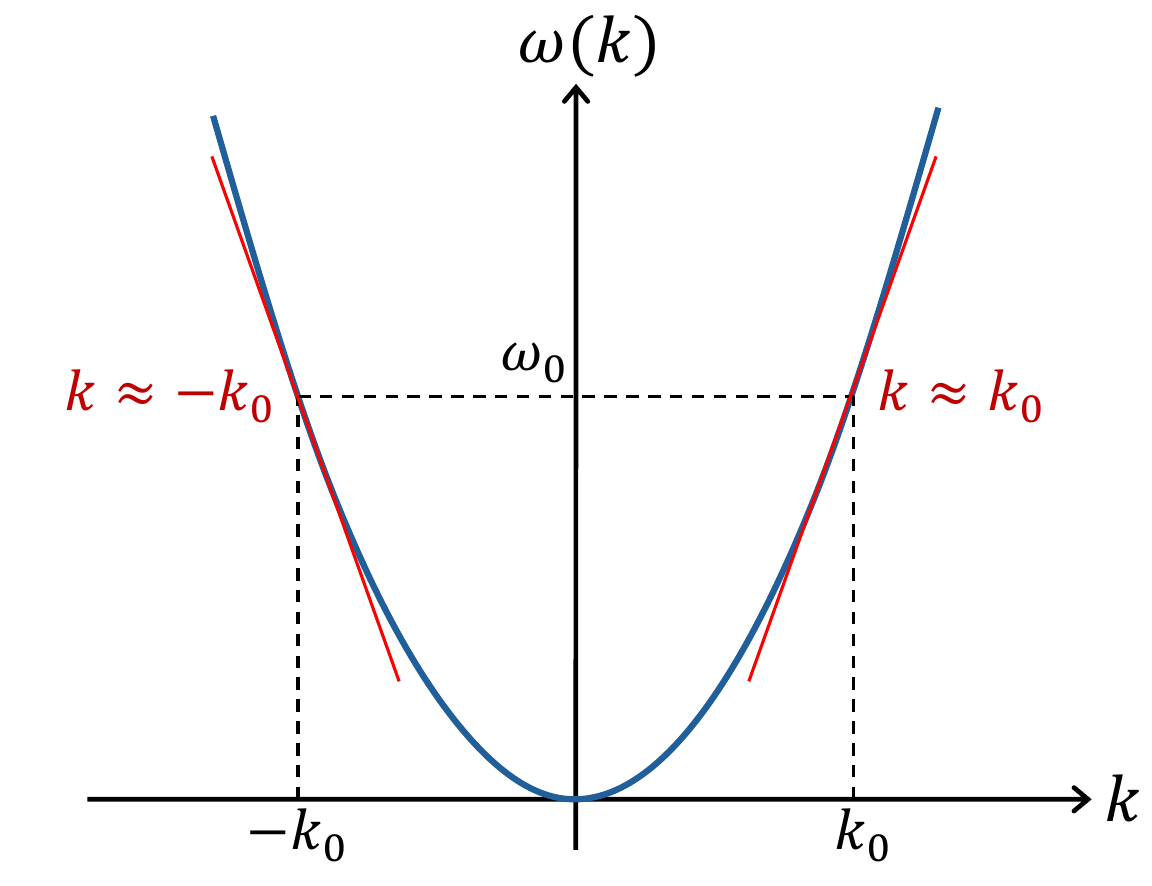}
    \caption{Linearization of the waveguide dispersion $\omega(k)$ for the system in Fig.~\ref{fig:two_level_emitter_waveguide}(a). The waveguide dispersion (solid blue curve) is approximated as being linear (solid red lines) near the wave numbers $\pm k_0$ corresponding to the frequency $\omega_0$.}
    \label{fig:dispersion_linear}
\end{figure}

To use this Hamiltonian in the real-space approach, we need to transform from $k$-space to position space. In order to do this, we use the linear dispersion approximation to linearize $\omega(k)$ near the frequency ${\omega_0 = \omega(k_0) = \omega(-k_0)}$ (the central frequency of interest), which is a good approximation for frequencies near $\omega_0$. This linearization is shown in Fig.~\ref{fig:dispersion_linear}. When we use the Taylor series expansion of $\omega(k)$ at ${k = \pm k_0}$, we obtain:
\begin{equation}
\omega(k) \approx \left\{ \begin{array}{lr}
\omega_0 - v_g(k + k_0),\hspace{0.05in} k \approx -k_0, \\[0.05in]
\omega_0 + v_g(k - k_0),\hspace{0.05in}  k \approx k_0, \end{array}\right.
\end{equation}
where $\smash{v_g = \frac{d\omega}{dk}\bigr|_{k_0}}$ is the group velocity. When we use this in Eq.~(\ref{eq:H_k}), the approximation for ${k \approx -k_0}$ applies in the integrals over negative $k$, and the approximation for ${k \approx k_0}$ applies in the integrals over positive $k$, leading to
\begin{align}
\begin{split}
    H =&\; \tilde{\omega}_e \sigma^+ \sigma^- + \int_0^{\infty} \left[ \omega_0 + v_g(k - k_0) \right] a_R^{\dagger}(k) a_R(k) dk\\
    &+ \int_{-\infty}^0 \left[ \omega_0 - v_g(k + k_0) \right] a_L^{\dagger}(k) a_L(k) dk\\
    &+ \int_0^{\infty} \left[ \frac{V_R}{\sqrt{2\pi}} a_R^{\dagger}(k) \sigma^- + \frac{V^*_R}{\sqrt{2\pi}} a_R(k) \sigma^+\right]dk\\
    &+ \int_{-\infty}^0 \left[ \frac{V_L}{\sqrt{2\pi}} a_L^{\dagger}(k) \sigma^- + \frac{V^*_L}{\sqrt{2\pi}} a_L(k) \sigma^+\right]dk.
\end{split}
\end{align}
We eliminate the frequency $\omega_0$ using the transformation ${H \rightarrow H - \omega_0N_E}$, where
\begin{equation}
    N_E = \sigma^+ \sigma^- + \int_0^{\infty} a_R^{\dagger}(k) a_R(k) dk + \int_{-\infty}^0 a_L^{\dagger}(k) a_L(k) dk
\end{equation}
is the total excitation number operator. This transformation is simply a constant energy shift that does not affect the dynamics of the system because $N_E$ is a conserved quantity (i.e., ${[H, N_E] = 0}$). We absorb the remaining factor of $\omega_0$ into the transition frequency of the emitter by defining ${\omega_e = \tilde{\omega}_e - \omega_0}$.

After the linearization, we extend the integration limits in $H$ to $\pm \infty$, which is valid because, within the linear dispersion approximation, we only consider photons with frequencies close to $\omega_0$ (i.e., only wave numbers close to $\pm k_0$ are of interest). We then use the substitution ${k' = k-k_0}$ in integrals containing $a_R(k)$, and ${k' = k+k_0}$ in integrals containing $a_L(k)$. Finally, we relabel $k'$ with $k$, which gives
\begin{align}\label{eq:H_k_linearised}
\begin{split}
    H =&\; \omega_e \sigma^+ \sigma^- + \int_{-\infty}^{\infty} v_g k \; a_R^{\dagger}(k+k_0) a_R(k+k_0) dk\\
    &- \int_{-\infty}^{\infty} v_g k \; a_L^{\dagger}(k-k_0) a_L(k-k_0) dk\\
    &+ \int_{-\infty}^{\infty} \left[ \frac{V_R}{\sqrt{2\pi}} a_R^{\dagger}(k+k_0) \sigma^- + \frac{V^*_R}{\sqrt{2\pi}} a_R(k+k_0) \sigma^+\right]dk\\
    &+ \int_{-\infty}^{\infty} \left[ \frac{V_L}{\sqrt{2\pi}} a_L^{\dagger}(k-k_0) \sigma^- + \frac{V^*_L}{\sqrt{2\pi}} a_L(k-k_0) \sigma^+\right]dk.
\end{split}
\end{align}

We now define the Fourier transforms~\cite{Shen2009}
\begin{subequations}
\begin{equation}\label{eq:Fourier_transform}
    a_{\mu}(k) = \frac{1}{\sqrt{2\pi}} \int_{-\infty}^{\infty} a_{\mu}(x) e^{-ikx} dx,
\end{equation}
\begin{equation}\label{eq:Fourier_transform_2}
    a_{\mu}(x) = \frac{1}{\sqrt{2\pi}} \int_{-\infty}^{\infty} a_{\mu}(k) e^{ikx} dk,
\end{equation}
\end{subequations}
where $a_{\mu}(x)$ is the annihilation operator for photons at position $x$ in the waveguide, moving in the ${\mu \in \{L,R\}}$ direction. We substitute Eq.~(\ref{eq:Fourier_transform}) and its Hermitian conjugate into Eq.~(\ref{eq:H_k_linearised}), leading to the real-space Hamiltonian~\cite{Shen2005, Shen2005_2}
\begin{align}\label{eq:H_x}
\begin{split}
    H =&\; \int_{-\infty}^{\infty} dx \left[ -iv_g a_R^{\dagger}(x) \frac{\partial}{\partial x}a_R(x) + iv_g a_L^{\dagger}(x) \frac{\partial}{\partial x}a_L(x) \right]\\
    &+ \omega_e \sigma^+ \sigma^-+ V_R a_R^{\dagger}(0) \sigma^- + V_R^* a_R(0) \sigma^+\\
    &+ V_L a_L^{\dagger}(0) \sigma^- + V_L^* a_L(0) \sigma^+,
\end{split}
\end{align}
as shown in Appendix~\ref{app:real-space_model}. The term on the first line corresponds to free photon propagation in the waveguide, the second term corresponds to free evolution of the two-level emitter, and the rest of the terms describe a local interaction at the position of the emitter (${x=0}$).

\subsubsection{Single-photon transmission and reflection}\label{subsubsec:methods}

We now use Eq.~(\ref{eq:H_x}) to solve for the single-photon transmission and reflection amplitudes $t_{\text{wg}}$ and $r_{\text{wg}}$, following Refs.~\cite{Shen2005,Shen2005_2} where the same system was considered. The general state of a single photon in the system can be written as
\begin{align}\label{eq:state}
\begin{split}
    \ket{\psi_k} =&\; \int_{-\infty}^{\infty} dx\; u_{R,k}(x) a_R^{\dagger}(x)\ket{0,g}\\
    &+ \int_{-\infty}^{\infty} dx\; u_{L,k}(x) a_L^{\dagger}(x)\ket{0,g} + u_e\ket{0,e},
\end{split}
\end{align}
where $u_{\mu,k}(x)$ is the probability amplitude for finding a photon moving in the $\mu$ direction at position $x$, $u_e$ is the excitation amplitude of the two-level emitter, and $\ket{0}$ is the vacuum state of the waveguide modes. We are interested in the scattering of a photon with wave number $k$, with corresponding frequency ${\omega_k = v_gk}$ within the linear dispersion approximation. The single-photon state $\ket{\psi_k}$ therefore has the energy eigenvalue ${E_k = \omega_k = v_gk}$ (since ${\hbar = 1}$), and hence satisfies the time-independent Schr\"{o}dinger equation
\begin{equation}\label{eq:Schrodinger}
    H\ket{\psi_k} = v_gk\ket{\psi_k},
\end{equation}
which we want to solve for the amplitudes $u_{R,k}(x)$ and $u_{L,k}(x)$. We find $H\ket{\psi_k}$ by applying the real-space Hamiltonian in Eq.~(\ref{eq:H_x}) to the state in Eq.~(\ref{eq:state}) (see Appendix~\ref{app:real-space_model}). When we substitute this into Eq.~(\ref{eq:Schrodinger}) and compare coefficients of orthogonal states, we obtain a set of three simultaneous equations for the probability amplitudes:
\begin{subequations}
\begin{equation}\label{eq:u_e}
    -u_e \Delta + V_R^*u_{R,k}(0) + V_L^*u_{L,k}(0) = 0,
\end{equation}
\begin{equation}\label{eq:u_R}
    -iv_g \frac{\partial}{\partial x}u_{R,k}(x) - v_g k u_{R,k}(x) + V_Ru_e\delta(x) = 0,
\end{equation}
\begin{equation}\label{eq:u_L}
    iv_g \frac{\partial}{\partial x}u_{L,k}(x) - v_g k u_{L,k}(x) + V_Lu_e\delta(x) = 0,
\end{equation}
\end{subequations}
where ${\Delta = v_gk - \omega_e}$ is the photon-emitter detuning. From Eq.~(\ref{eq:u_e}), we have
\begin{equation}
    u_e = \frac{V_R^*}{\Delta}u_{R,k}(0) + \frac{V_L^*}{\Delta}u_{L,k}(0),
\end{equation}
which can be substituted into Eqs.~(\ref{eq:u_R}) and (\ref{eq:u_L}) to obtain a pair of equations involving the photon amplitudes $u_{R,k}(x)$ and $u_{L,k}(x)$:
\begin{subequations}
\begin{align}\label{eq:u_R_2}
\begin{split}
    &-iv_g \frac{\partial}{\partial x}u_{R,k}(x) - v_g k u_{R,k}(x) + \frac{|V_R|^2}{\Delta}u_{R,k}(0)\delta(x)\\[0.05in]
    &+ \frac{V_RV_L^*}{\Delta}u_{L,k}(0)\delta(x) = 0,
\end{split}
\end{align}
\begin{align}\label{eq:u_L_2}
\begin{split}
    &iv_g \frac{\partial}{\partial x}u_{L,k}(x) - v_g k u_{L,k}(x) + \frac{V_LV_R^*}{\Delta}u_{R,k}(0)\delta(x)\\[0.05in]
    &+ \frac{|V_L|^2}{\Delta}u_{L,k}(0)\delta(x) = 0.
\end{split}
\end{align}
\end{subequations}

To make further progress, we use the following ansatz for the photon amplitudes:
\begin{subequations}
\begin{equation}\label{eq:u_R_solution}
u_{R,k}(x) = e^{ikx}\Theta(-x) + t_{\text{wg}}e^{ikx}\Theta(x) =  \left\{ \begin{array}{lr}
e^{ikx},\hspace{0.05in} x < 0, \\[0.05in]
t_{\text{wg}}e^{ikx},\hspace{0.05in} x > 0, \end{array}\right.
\end{equation}
\begin{equation}\label{eq:u_L_solution}
u_{L,k}(x) = r_{\text{wg}}e^{-ikx}\Theta(-x) =  \left\{ \begin{array}{lr}
r_{\text{wg}}e^{-ikx},\hspace{0.05in} x < 0, \\[0.05in]
0,\hspace{0.05in} x > 0, \end{array}\right.
\end{equation}
\end{subequations}
where $t_{\text{wg}}$ and $r_{\text{wg}}$ are the transmission and reflection amplitudes, and $\Theta(x)$ is the Heaviside step function. Here we have assumed that the solutions for ${x\neq0}$ are plane waves, and the incoming plane wave is moving to the right. When the photon is scattered by the emitter at ${x=0}$, it acquires a left-moving (reflected) amplitude and a right-moving (transmitted) amplitude. The continuity relations at ${x=0}$ are ${u_{R,k}(0) = [u_{R,k}(0^+) + u_{R,k}(0^-)]/2 = (t_{\text{wg}}+1)/2}$ and ${u_{L,k}(0) = [u_{L,k}(0^+) + u_{L,k}(0^-)]/2 = r_{\text{wg}}/2}$.

When we use the ansatz from Eqs.~(\ref{eq:u_R_solution}) and (\ref{eq:u_L_solution}) in Eqs.~(\ref{eq:u_R_2}) and (\ref{eq:u_L_2}), we obtain two equations involving $t_{\text{wg}}$ and $r_{\text{wg}}$. In particular, after applying the derivatives, using ${\partial_x \Theta(\pm x) = \pm \delta(x)}$, and integrating over $x$, we find
\begin{subequations}
\begin{equation}
    \left( \frac{|V_R|^2}{2\Delta} - iv_g \right)t_{\text{wg}} + \frac{V_RV_L^*}{2\Delta}r_{\text{wg}} = - \left( \frac{|V_R|^2}{2\Delta} + iv_g \right),
\end{equation}
\begin{equation}
    \frac{V_LV_R^*}{2\Delta}t_{\text{wg}} + \left( \frac{|V_L|^2}{2\Delta} - iv_g \right)r_{\text{wg}} = - \frac{V_LV_R^*}{2\Delta},
\end{equation}
\end{subequations}
which can straightforwardly be solved for the transmission amplitude
\begin{equation}\label{eq:t_real_space}
    t_{\text{wg}} = \frac{\Delta - \frac{i}{2}(|\Gamma_R| - |\Gamma_L|)}{\Delta + \frac{i}{2}(|\Gamma_R| + |\Gamma_L|)}
\end{equation}
and the reflection amplitude
\begin{equation}\label{eq:r_real_space}
    r_{\text{wg}} = \frac{-i\sqrt{\Gamma_L \Gamma_R^*}}{\Delta + \frac{i}{2}(|\Gamma_R| + |\Gamma_L|)},
\end{equation}
where ${\Gamma_{\mu} = V_{\mu}^2/v_g}$ is the decay rate of the emitter in the $\mu$ direction in the waveguide. As expected, when ${\Gamma_R = 0}$ or ${\Gamma_L= 0}$ the incoming photon is transmitted perfectly (${r_{\text{wg}}=0}$), because in the former case the right-moving input photon does not interact with the emitter and propagates freely through the waveguide, and in the latter case the photon can only be scattered to the right.

\begin{figure}
    \centering
    \includegraphics[width=\linewidth]{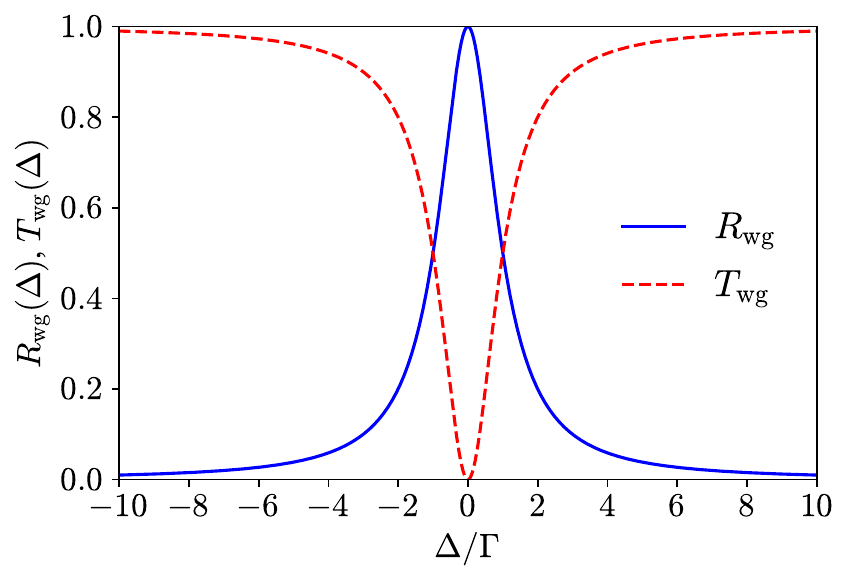}
    \caption{Single-photon transmission [dashed red curve, Eq.~(\ref{eq:T})] and reflection [solid blue curve, Eq.~(\ref{eq:R})] as a function of the photon-emitter detuning $\Delta$ for the system in Fig.~\ref{fig:two_level_emitter_waveguide}(a). Photon loss is neglected in these results, and we have ${T_{\text{wg}}(\Delta = 0) = 0}$ and ${R_{\text{wg}}(\Delta = 0) = 1}$ on resonance with the emitter.}
    \label{fig:R_T}
\end{figure}

We now focus on the case of real, symmetric decay rates, where ${\Gamma_R = \Gamma_R^* = \Gamma_L = \Gamma_L^* = \Gamma}$, and the single-photon transmission and reflection amplitudes become
\begin{equation}\label{eq:t_and_r}
    t_{\text{wg}} = \frac{\Delta}{\Delta + i\Gamma} \quad \text{and} \quad r_{\text{wg}} = \frac{-i\Gamma}{\Delta + i\Gamma}.
\end{equation}
This leads to the transmission probability
\begin{equation}\label{eq:T}
    T_{\text{wg}} = |t_{\text{wg}}|^2 = \frac{\Delta^2}{\Delta^2 + \Gamma^2},
\end{equation}
and the reflection probability
\begin{equation}\label{eq:R}
    R_{\text{wg}} = |r_{\text{wg}}|^2 = \frac{\Gamma^2}{\Delta^2 + \Gamma^2}.
\end{equation}
The reflection probability $R_{\text{wg}}$ is a Lorentzian function of the photon-emitter detuning $\Delta$, with full width at half-maximum (FWHM) $2\Gamma$, as shown in Fig.~\ref{fig:R_T} (${T_{\text{wg}} = 1 - R_{\text{wg}}}$). A photon that is on resonance with the two-level emitter (${\Delta = 0}$) is perfectly reflected~\cite{Shen2005}, with a $\pi$ phase shift (${r_{\text{wg}} = -1}$). Far from resonance, where ${|\Delta| \gg \Gamma}$, the photon is transmitted with probability ${T_{\text{wg}} \approx 1}$, since essentially no scattering occurs when the photon is detuned far from the linewidth of the emitter. 

\begin{figure}
    \centering
    \includegraphics[width=\linewidth]{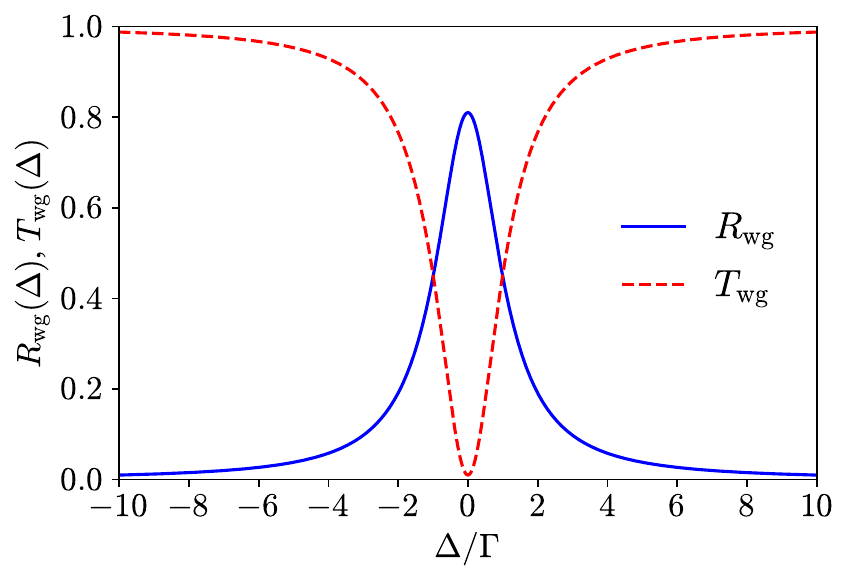}
    \caption{Single-photon transmission [dashed red curve, Eq.~(\ref{eq:T_loss})] and reflection [solid blue curve, Eq.~(\ref{eq:R_loss})] as a function of the photon-emitter detuning $\Delta$ for the system in Fig.~\ref{fig:two_level_emitter_waveguide}(a). The emitter loss rate $\gamma$ is included in these results, and was chosen such that ${\gamma = (2/9)\Gamma}$. This leads to ${T_{\text{wg}}(\Delta = 0) = 0.01}$ and ${R_{\text{wg}}(\Delta = 0) = 0.81}$ on resonance.}
    \label{fig:R_T_loss}
\end{figure}

In practical implementations, photon loss is inevitable as the two-level emitter couples to non-guided modes in its environment. This imperfect emitter-waveguide coupling can be accounted for in scattering calculations by introducing a dissipation rate $\gamma$, corresponding to the photon loss rate out of the system due to processes like spontaneous emission. In 2013, Rephaeli and Fan showed that including emitter loss amounts to a simple substitution where an imaginary part is added to the transition frequency~\cite{Rephaeli2013}: ${\omega_e \rightarrow \omega_e - i\gamma/2}$. This is a shorthand way of adding a reservoir (waveguide) to which the emitter couples with rate $\gamma$. Since ${\Delta = v_gk - \omega_e}$, this substitution is equivalent to ${\Delta \rightarrow \Delta + i\gamma/2}$. When we use this in Eq.~(\ref{eq:t_and_r}) and then take the modulus squared, the transmission and reflection probabilities become
\begin{subequations}
\begin{equation}\label{eq:T_loss}
    T_{\text{wg}} = \frac{\Delta^2 + (\gamma/2)^2}{\Delta^2 + (\Gamma + \gamma/2)^2},
\end{equation}
\begin{equation}\label{eq:R_loss}
    R_{\text{wg}} = \frac{\Gamma^2}{\Delta^2 + (\Gamma + \gamma/2)^2}.
\end{equation}
\end{subequations}
Now, when ${\Delta = 0}$, we no longer have perfect reflection---losses from the emitter reduce the maximum reflection, and increase the minimum transmission. An example of this is shown in Fig.~\ref{fig:R_T_loss} with ${\gamma = (2/9)\Gamma}$, where ${T_{\text{wg}}(\Delta = 0) = 0.01}$ and ${R_{\text{wg}}(\Delta = 0) = 0.81}$. This corresponds to a beta factor (coupling efficiency) of ${\beta = 2\Gamma/(2\Gamma + \gamma) = 0.9}$~\cite{Lund-Hansen2008, Scarpelli2019}. It is clear that ${T_{\text{wg}} + R_{\text{wg}} < 1}$ due to the nonzero probability of photon loss.

\subsection{Input-output formalism and scattering matrix}\label{subsec:methods_2}

The input-output formalism is another method that is commonly used to model photon scattering. Developed by Gardiner and Collett in 1985~\cite{Gardiner1985}, the input-output formalism relates fields leaving a system to fields entering the system. In contrast to the real-space approach, which is based on the Schr\"{o}dinger picture, the input-output formalism is based on the Heisenberg picture, where the time evolution of field operators is used to study the scattering dynamics. In 2010, Fan, Kocaba\c{s}, and Shen made a connection between the input-output formalism and the scattering matrix $S$, enabling the calculation of few-photon scattering properties using the toolbox of the input-output formalism~\cite{Fan2010}.

In this section, we use the procedure from Ref.~\cite{Fan2010} to derive the single-photon transmission and reflection amplitudes for a two-level emitter coupled to a waveguide, as in Fig.~\ref{fig:two_level_emitter_waveguide}(a). In particular, we will obtain the input-output relations from the Heisenberg equations of the waveguide operators, and use these relations to derive the single-photon scattering matrix, the elements of which correspond to the transmission and reflection amplitudes. We will show that this calculation leads to the same results as the real-space approach from the previous section.

\subsubsection{Hamiltonian}

To derive the input-output relations, we transform the Hamiltonian in Eq.~(\ref{eq:H_k}) from $k$-space to frequency space. This transformation initially follows the transformation from $k$-space to position space from the previous section, in that we linearize the waveguide dispersion to obtain the Hamiltonian in Eq.~(\ref{eq:H_k_linearised}). Next, instead of using a Fourier transform, we use the substitution ${\omega = v_gk}$ in integrals containing $a_R(k+k_0)$, and ${\omega = -v_gk}$ in integrals containing $a_L(k-k_0)$. We also define the frequency-space waveguide operators
\begin{equation}
    a_R(\omega) = \frac{a_R(k+k_0)}{\sqrt{v_g}} \quad \text{and} \quad a_L(\omega) = \frac{a_L(k-k_0)}{\sqrt{v_g}},
\end{equation}
with ${[a_{\mu}(\omega), a_{\nu}^{\dagger}(\omega')] = \delta_{\mu\nu}\delta(\omega-\omega')}$, which follows from the commutation relations ${[a_{\mu}(k), a_{\nu}^{\dagger}(k')] = \delta_{\mu\nu}\delta(k-k')}$ in $k$-space. Using the substitutions and the above definitions in Eq.~(\ref{eq:H_k_linearised}) leads to the frequency-space Hamiltonian
\begin{align}\label{eq:H_omega}
\begin{split}
    H =&\; \omega_e \sigma^+ \sigma^- + \sum_{\mu = L,R} \int_{-\infty}^{\infty} \omega a_{\mu}^{\dagger}(\omega) a_{\mu}(\omega) d\omega\\
    &+ \sum_{\mu = L,R} \int_{-\infty}^{\infty} \left[ \sqrt{\frac{\Gamma_{\mu}}{2\pi}} a_{\mu}^{\dagger}(\omega) \sigma^- + \sqrt{\frac{\Gamma^*_{\mu}}{2\pi}} a_{\mu}(\omega) \sigma^+ \right]d\omega,
\end{split}
\end{align}
where ${\Gamma_{\mu} = V_{\mu}^2/v_g}$ is the decay rate of the emitter in the $\mu$ direction in the waveguide, defined as in the previous section.

\subsubsection{Single-photon transmission and reflection}

The input-output relations for our system are obtained from the Heisenberg equations of the waveguide operators $a_{\mu}(\omega,t)$ (in the Heisenberg picture, operators become dependent on time $t$). Using the frequency-space Hamiltonian in Eq.~(\ref{eq:H_omega}), we find that the Heisenberg equation for $a_{\mu}(\omega,t)$ is
\begin{align}
\begin{split}
    \frac{d}{dt} a_{\mu}(\omega,t) =&\; i\left[ H, a_{\mu}(\omega,t) \right]\\
    =& -i \omega a_{\mu}(\omega,t) - i\sqrt{\frac{\Gamma_{\mu}}{2\pi}} \sigma^-(t).
\end{split}
\end{align}
Multiplying both sides of this equation by $e^{i\omega t}$ and rearranging leads to
\begin{equation}\label{eq:wg_Heisenberg}
    \frac{d}{dt} \left[ a_{\mu}(\omega,t) e^{i\omega t} \right] = - i\sqrt{\frac{\Gamma_{\mu}}{2\pi}} \sigma^-(t) e^{i\omega t}.
\end{equation}
Relabeling $t$ with $t'$, and integrating from an “input time” $t_0$ to some time $t$ gives
\begin{equation}
    a_{\mu}(\omega,t)e^{i\omega t} - a_{\mu}(\omega,t_0)e^{i\omega t_0} = - i\sqrt{\frac{\Gamma_{\mu}}{2\pi}} \int_{t_0}^t dt' \sigma^-(t') e^{i\omega t'}.
\end{equation}
We now multiply both sides by $e^{-i\omega t}$, divide by $\sqrt{2\pi}$, and then integrate over all $\omega$:
\begin{align}\label{eq:wg_Heisenberg_int}
\begin{split}
    \frac{1}{\sqrt{2\pi}}& \int_{-\infty}^{\infty} a_{\mu}(\omega,t)d\omega - a_{\mu,\text{in}}(t)\\
    &= -i\sqrt{\Gamma_{\mu}}\int_{t_0}^t dt' \sigma^-(t') \left( \frac{1}{2\pi} \int_{-\infty}^{\infty}d\omega\; e^{-i\omega(t-t')} \right)\\
    &= -i\sqrt{\Gamma_{\mu}}\int_{t_0}^t dt' \sigma^-(t') \delta(t-t')\\
    &= -\frac{i}{2} \sqrt{\Gamma_{\mu}} \sigma^-(t),
\end{split}
\end{align}
where in the last line we acquired a factor of $1/2$ because the delta function is centered at one of the integration limits, and we used the definition
\begin{equation}
    a_{\mu,\text{in}}(t) = \frac{1}{\sqrt{2\pi}} \int_{-\infty}^{\infty} a_{\mu}(\omega,t_0) e^{-i\omega(t-t_0)} d\omega
\end{equation}
of an input operator in the input-output formalism~\cite{Gardiner1985}. We now return to Eq.~(\ref{eq:wg_Heisenberg}), relabel $t$ with $t'$, integrate from some time $t$ to an “output time” $t_1$, and repeat the remaining steps outlined above to obtain
\begin{equation}\label{eq:wg_Heisenberg_int_2}
    a_{\mu,\text{out}}(t) - \frac{1}{\sqrt{2\pi}} \int_{-\infty}^{\infty} a_{\mu}(\omega,t)d\omega = -\frac{i}{2} \sqrt{\Gamma_{\mu}} \sigma^-(t),
\end{equation}
where we used the definition
\begin{equation}
    a_{\mu,\text{out}}(t) = \frac{1}{\sqrt{2\pi}} \int_{-\infty}^{\infty} a_{\mu}(\omega,t_1) e^{-i\omega(t-t_1)} d\omega
\end{equation}
of an output operator in the input-output formalism~\cite{Gardiner1985}. From Eqs.~(\ref{eq:wg_Heisenberg_int}) and (\ref{eq:wg_Heisenberg_int_2}), it follows that
\begin{subequations}
\begin{equation}\label{eq:input}
    a_{\mu,\text{in}}(t) = \frac{1}{\sqrt{2\pi}} \int_{-\infty}^{\infty} a_{\mu}(\omega,t)d\omega + \frac{i}{2} \sqrt{\Gamma_{\mu}} \sigma^-(t),
\end{equation}
\begin{equation}
    a_{\mu,\text{out}}(t) = \frac{1}{\sqrt{2\pi}} \int_{-\infty}^{\infty} a_{\mu}(\omega,t)d\omega - \frac{i}{2} \sqrt{\Gamma_{\mu}} \sigma^-(t),
\end{equation}
\end{subequations}
so the input-output relations are
\begin{equation}\label{eq:input_output}
    a_{\mu,\text{out}}(t) = a_{\mu,\text{in}}(t) - i \sqrt{\Gamma_{\mu}} \sigma^-(t)
\end{equation}
for ${\mu \in \{L,R\}}$.

Using these input-output relations, we can derive the single-photon scattering matrix elements
\begin{equation}\label{eq:S_matrix}
    S_{pk}^{\mu\nu} = \matrixel{p_{\mu}}{S}{k_\nu} = \braket{p_{\mu}^-}{k_{\nu}^+} = \expval{a_{\mu,\text{out}}(p)a_{\nu,\text{in}}^{\dagger}(k)}{0},
\end{equation}
where $\nu$ and $\mu$ label the input and output photon directions, and $k$ and $p$ label the input and output photon frequencies, respectively. In addition, $\smash{\ket{k_{\nu}^+} = a_{\nu,\text{in}}^{\dagger}(k)\ket{0}}$ is a scattering eigenstate that evolves from the free state $\ket{k_\nu}$ in the distant past, and $\smash{\ket{p_{\mu}^-} = a_{\mu,\text{out}}^{\dagger}(p)\ket{0}}$ is a scattering eigenstate that evolves into the free state $\ket{p_\mu}$ in the distant future~\cite{Fan2010}. Note that, for the remainder of Section~\ref{subsec:methods_2}, we use $k$ to denote frequency as opposed to wave number, in line with convention.

As was shown in Ref.~\cite{Fan2010}, the time-domain input and output operators $a_{\mu,\text{in/out}}(t)$ in the input-output formalism are related to the frequency-domain input and output operators $a_{\mu,\text{in/out}}(p)$ in the scattering matrix simply via Fourier transforms, provided that we take the limit ${t_0 \rightarrow -\infty}$ for the input time and ${t_1 \rightarrow \infty}$ for the output time. Using
\begin{subequations}
\begin{equation}\label{eq:input_FT}
    a_{\mu,\text{in}}(t) = \frac{1}{\sqrt{2\pi}} \int_{-\infty}^{\infty} a_{\mu,\text{in}}(p) e^{-ipt} dp,
\end{equation}
\begin{equation}
    a_{\mu,\text{out}}(t) = \frac{1}{\sqrt{2\pi}} \int_{-\infty}^{\infty} a_{\mu,\text{out}}(p) e^{-ipt} dp,
\end{equation}
\begin{equation}\label{eq:emitter_FT}
    \sigma^-(t) = \frac{1}{\sqrt{2\pi}} \int_{-\infty}^{\infty} \sigma^-(p) e^{-ipt} dp,
\end{equation}
\end{subequations}
we take the Fourier transform of Eq.~(\ref{eq:input_output}), and substitute the result into the scattering matrix in Eq.~(\ref{eq:S_matrix}):
\begin{align}\label{eq:S_matrix_2}
\begin{split}
    S_{pk}^{\mu\nu} =& \bra{0} \left[ a_{\mu,\text{in}}(p) - i\sqrt{\Gamma_{\mu}}\sigma^-(p) \right] a_{\nu,\text{in}}^{\dagger}(k) \ket{0}\\[0.05in]
    =& \expval{a_{\mu,\text{in}}(p) a_{\nu,\text{in}}^{\dagger}(k)}{0} - i\sqrt{\Gamma_{\mu}} \expval{\sigma^-(p) a_{\nu,\text{in}}^{\dagger}(k)}{0}\\[0.05in]
    =&\; \delta_{\mu\nu} \delta(p-k) - i\sqrt{\Gamma_{\mu}} \matrixel{0}{\sigma^-(p)}{k_{\nu}^+},
\end{split}
\end{align}
where we used ${[a_{\mu,\text{in}}(p), a_{\nu,\text{in}}^{\dagger}(k)] = \delta_{\mu\nu} \delta(p-k)}$. Hence, in order to find $S_{pk}^{\mu\nu}$, we need to calculate the matrix element $\matrixel{0}{\sigma^-(p)}{k_{\nu}^+}$. We start from the Heisenberg equation for the emitter operator $\sigma^-(t)$:
\begin{align}\label{eq:emitter_Heisenberg}
\begin{split}
    \frac{d}{dt} \sigma^-(t) =&\; i\left[ H, \sigma^-(t) \right]\\
    =& -i\omega_e\sigma^-(t)\\
    &+ i\sigma_z(t)\sum_{\mu=L,R} \sqrt{\frac{\Gamma^*_{\mu}}{2\pi}} \int_{-\infty}^{\infty} a_{\mu}(\omega,t)d\omega,
\end{split}
\end{align}
where we used ${\sigma^{\pm}(t) = e^{iHt}\sigma^{\pm}(0)e^{-iHt}}$ to evaluate the commutators, with ${\sigma^+(0) = \ketbra{e}{g}}$ and ${\sigma^-(0) = \ketbra{g}{e}}$. In addition, ${\sigma_z(t) = e^{iHt}\sigma_z(0)e^{-iHt}}$, where ${\sigma_z(0) = \ketbra{e}{e} - \ketbra{g}{g}}$. We can eliminate the integral in Eq.~(\ref{eq:emitter_Heisenberg}) using the result for the input operators in Eq.~(\ref{eq:input}), giving
\begin{align}
\begin{split}
    \frac{d}{dt} \sigma^-(t) =& - \left[ i\omega_e + \frac{1}{2}\left( |\Gamma_R| + |\Gamma_L| \right) \right] \sigma^-(t)\\
    &+ i \sum_{\mu=L,R} \sqrt{\Gamma^*_\mu} \sigma_z(t)a_{\mu,\text{in}}(t),
\end{split}
\end{align}
where we used ${\sigma_z(t) \sigma^-(t) = - \sigma^-(t)}$. To obtain the matrix element we require, we now pre-multiply each term by $\bra{0}$, and post-multiply by $\ket{k_{\nu}^+}$:
\begin{align}
\begin{split}
    \frac{d}{dt} \matrixel{0}{\sigma^-(t)}{k_{\nu}^+} =& - \left[ i\omega_e + \frac{1}{2}\left( |\Gamma_R| + |\Gamma_L| \right) \right] \matrixel{0}{\sigma^-(t)}{k_{\nu}^+}\\
    &- i \sum_{\mu=L,R} \sqrt{\Gamma^*_\mu} \matrixel{0}{a_{\mu,\text{in}}(t)}{k_{\nu}^+},
\end{split}
\end{align}
since ${\bra{0}\sigma_z(t) = -\bra{0}}$. Substituting in the Fourier transforms from Eqs.~(\ref{eq:input_FT}) and (\ref{eq:emitter_FT}) leads to
\begin{align}
\begin{split}
    -ip\matrixel{0}{\sigma^-(p)}{k_{\nu}^+} =& - \left[ i\omega_e + \frac{1}{2}\left( |\Gamma_R| + |\Gamma_L| \right) \right] \matrixel{0}{\sigma^-(p)}{k_{\nu}^+}\\
    &- i \sum_{\mu=L,R} \sqrt{\Gamma^*_\mu} \matrixel{0}{a_{\mu,\text{in}}(p)}{k_{\nu}^+}.
\end{split}
\end{align}
From the definition ${\ket{k_{\nu}^+} = a_{\nu,\text{in}}^{\dagger}(k)\ket{0}}$, it follows that ${\matrixel{0}{a_{\mu,\text{in}}(p)}{k_{\nu}^+} = \delta_{\mu\nu}\delta(p-k)}$. We can then rearrange for $\matrixel{0}{\sigma^-(p)}{k_{\nu}^+}$ to obtain
\begin{equation}\label{eq:emitter_matrix_element}
    \matrixel{0}{\sigma^-(p)}{k_{\nu}^+} = \frac{\sqrt{\Gamma_{\nu}^*}}{\Delta + \frac{i}{2}\left( |\Gamma_R| + |\Gamma_L| \right)} \delta(p-k),
\end{equation}
where ${\Delta = p-\omega_e}$ is the photon-emitter detuning. Finally, substituting the result in Eq.~(\ref{eq:emitter_matrix_element}) into Eq.~(\ref{eq:S_matrix_2}) gives the single-photon scattering matrix elements
\begin{equation}
    S_{pk}^{\mu\nu} = \left[ \delta_{\mu\nu} - \frac{i\sqrt{\Gamma_{\mu}\Gamma_{\nu}^*}}{\Delta + \frac{i}{2}\left( |\Gamma_R| + |\Gamma_L| \right)} \right] \delta(p-k),
\end{equation}
where $\delta(p-k)$ ensures that energy is conserved in the scattering process (scattering can only occur when ${p=k}$, i.e., the input and output photon frequencies are the same).

Consider a right-moving input photon, as in the previous section, such that ${\nu = R}$. For transmission, we set ${\mu = R}$ as well because the output photon would also be moving to the right. The corresponding scattering matrix element is
\begin{equation}
    S_{pk}^{RR} = t_{\text{wg}}\delta(p-k),
\end{equation}
where
\begin{equation}
    t_{\text{wg}} = \frac{\Delta - \frac{i}{2}(|\Gamma_R| - |\Gamma_L|)}{\Delta + \frac{i}{2}(|\Gamma_R| + |\Gamma_L|)}
\end{equation}
is the transmission amplitude for the system in Fig.~\ref{fig:two_level_emitter_waveguide}(a), which is identical to Eq.~(\ref{eq:t_real_space}). For reflection, we set ${\mu = L}$ as the output photon would be moving to the left, so the corresponding scattering matrix element is
\begin{equation}
    S_{pk}^{LR} = r_{\text{wg}}\delta(p-k),
\end{equation}
where
\begin{equation}
    r_{\text{wg}} = \frac{-i\sqrt{\Gamma_L \Gamma_R^*}}{\Delta + \frac{i}{2}(|\Gamma_R| + |\Gamma_L|)}
\end{equation}
is the reflection amplitude, which is identical to Eq.~(\ref{eq:r_real_space}). Hence, the results of the approach based on the input-output formalism and scattering matrix are consistent with the results of the real-space approach from the previous section. Note that the emitter loss rate $\gamma$ can be included as before, using the substitution ${\Delta \rightarrow \Delta + i\gamma/2}$. This can be seen by adding Lindblad terms to the Heisenberg equation for the emitter operator $\sigma^-$ [Eq.~(\ref{eq:emitter_Heisenberg})], corresponding to the Lindblad operator ${L = \sqrt{\gamma} \sigma^-}$ (adding these loss terms is equivalent to using ${\omega_e \rightarrow \omega_e - i\gamma/2}$ in the final results, which is the same as the substitution mentioned above)~\cite{Duda2024}.


\subsection{Discrete coordinate scattering approach}\label{subsec:methods_3}

The final theoretical method that we will present in detail is the discrete coordinate scattering approach. This method is widely used to model photon scattering in coupled-resonator waveguides, and was first used for this purpose by Zhou et al. in 2008~\cite{Zhou2008, Gong2008, Zhou2008_2}. The discrete coordinate scattering approach is essentially a discrete version of the real-space approach from Section~\ref{subsec:methods_1}, where the waveguide consists of a discrete set of coupled cavities rather than a continuous medium. One key difference between the two types of waveguide is the dispersion relation $\omega_k$. Whereas the linear dispersion approximation is usually sufficient in the treatment of single-photon scattering in continuous waveguides, this is not the case with coupled-resonator waveguides, where the photon frequency $\omega_k$ has a nonlinear, periodic dependence on the wave number $k$.

In this section, we will follow the derivation in Ref.~\cite{Zhou2008} and derive the transmission and reflection amplitudes for a two-level emitter in a coupled-resonator waveguide, as shown in Fig.~\ref{fig:two_level_emitter_waveguide}(b). Similarly to the real-space approach, this involves using a Hamiltonian in position space to solve the time-independent Schr\"{o}dinger equation by assuming plane-wave solutions for the incoming and outgoing photon amplitudes. We will use $t_{\text{crw}}$ and $r_{\text{crw}}$ to denote the transmission and reflection amplitudes, respectively, where the subscripts indicate that we are now considering the coupled-resonator waveguide in Fig.~\ref{fig:two_level_emitter_waveguide}(b).

\subsubsection{Hamiltonian}

The Hamiltonian for a one-dimensional chain of single-mode cavities with a coupled two-level quantum emitter can be expressed in position space as~\cite{Zhou2008}
\begin{align}\label{eq:H_discrete}
\begin{split}
    H_{\text{crw}} =&\; \omega_e \sigma^+ \sigma^- + \omega_c \sum_j a_j^{\dagger} a_j - \xi\sum_j \left( a_{j+1}^{\dagger} a_j + a_j^{\dagger} a_{j+1} \right)\\
    &+ g a_0^{\dagger} \sigma^- + g^* a_0 \sigma^+,
\end{split}
\end{align}
where $a_j^{\dagger}$ and $a_j$ are the creation and annihilation operators for cavity $j$. All the cavities are assumed to be identical, with the resonance frequency $\omega_c$ and the (real) nearest-neighbor hopping rate $\xi$. We assume that the two-level emitter couples to the central cavity (${j=0}$) with coupling rate $g$ [see Fig.~\ref{fig:two_level_emitter_waveguide}(b)]. As in the previous sections, $\omega_e$ is the transition frequency of the emitter, and ${\sigma^+ = \ketbra{e}{g}}$ and ${\sigma^- = \ketbra{g}{e}}$ are the raising and lowering operators, respectively.

\begin{figure}
    \centering
    \includegraphics[width=0.9\linewidth]{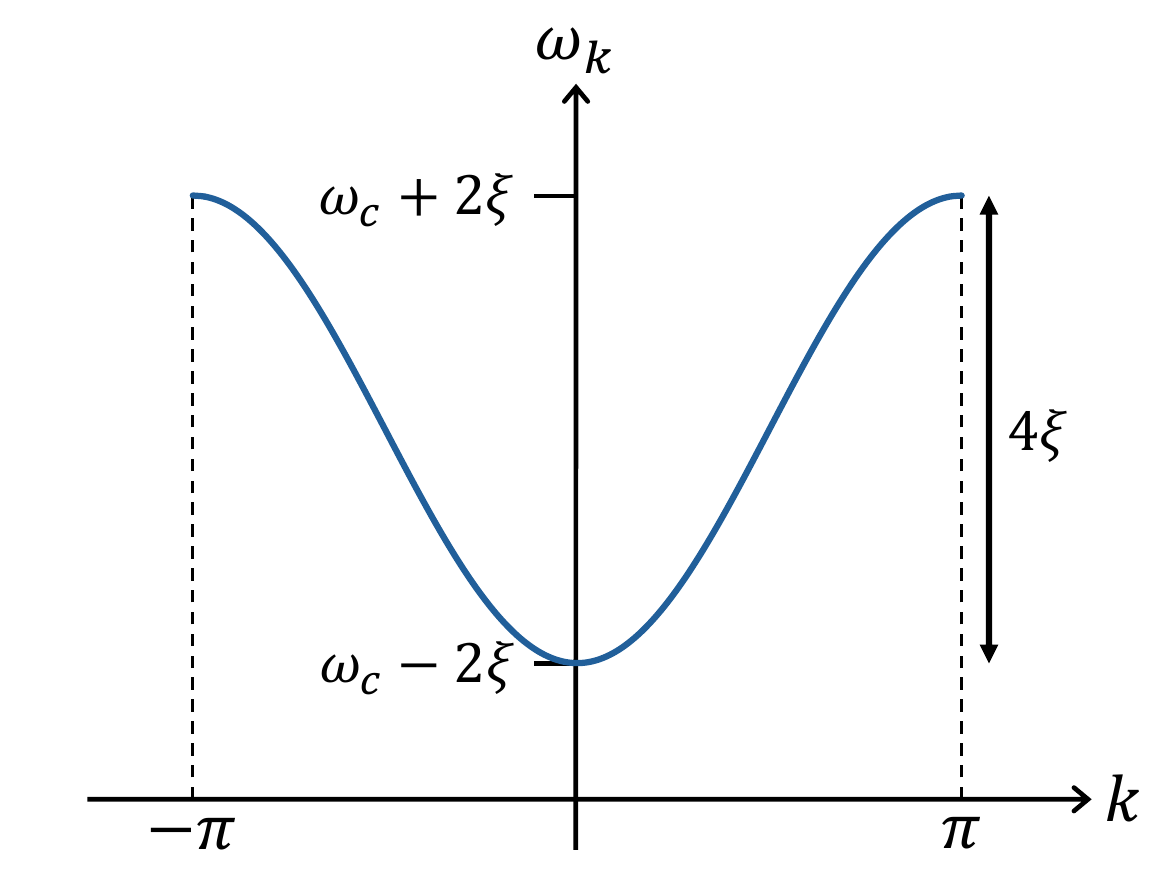}
    \caption{Dispersion relation ${\omega_k = \omega_c - 2\xi \text{cos}(k)}$ in the range ${k \in [-\pi,\pi]}$ for the coupled-resonator waveguide from Fig.~\ref{fig:two_level_emitter_waveguide}(b). There is a frequency band of width $4\xi$, corresponding to photon frequencies that are allowed to propagate through the waveguide.}
    \label{fig:dispersion_nonlinear}
\end{figure}

We can calculate the nonlinear dispersion relation in the coupled-resonator waveguide by taking the waveguide part of the Hamiltonian,
\begin{equation}\label{eq:H_cavities}
    \omega_c \sum_j a_j^{\dagger} a_j - \xi\sum_j \left( a_{j+1}^{\dagger} a_j + a_j^{\dagger} a_{j+1} \right),
\end{equation}
and using the Fourier transform
\begin{equation}\label{eq:discrete_FT}
    a_j = \frac{1}{\sqrt{N}} \sum_k b_k e^{ikj},
\end{equation}
which is the discrete version of Eq.~(\ref{eq:Fourier_transform_2}). Here $N$ is the number of cavities in the chain and $b_k$ is the annihilation operator for photons with wave number $k$ in the waveguide. In Appendix~\ref{app:discrete_model}, we show that the waveguide Hamiltonian in Eq.~(\ref{eq:H_cavities}) is diagonalized as $\sum_k \omega_k b_k^{\dagger}b_k$, where
\begin{equation}\label{eq:dispersion}
    \omega_k = \omega_c - 2\xi \text{cos}(k)
\end{equation}
is the dispersion relation. Hence, $\omega_k$ is periodic in $k$ with a period of $2\pi$ (note that we have assumed a lattice constant of unity). The coupled-resonator waveguide has a frequency band of $4\xi$ centered at $\omega_c$, as shown in Fig.~\ref{fig:dispersion_nonlinear}. Photons with frequencies inside this band can propagate through the chain of cavities. Outside of this band, photons cannot propagate as they are not resonant with the cavities and hence cannot be transmitted. The waveguide Hamiltonian in Eq.~(\ref{eq:H_cavities}) corresponds to a tight-binding boson model with nearest-neighbor hopping between lattice sites~\cite{Zhou2008_2}.

\subsubsection{Single-photon transmission and reflection}

Using the discrete position-space Hamiltonian in Eq.~(\ref{eq:H_discrete}), we calculate the transmission and reflection amplitudes for the system in Fig.~\ref{fig:two_level_emitter_waveguide}(b). The calculation closely follows the steps in Section~\ref{subsubsec:methods}, where the real-space approach was used for the continuous waveguide. In the case of the discrete coupled-resonator waveguide, the general state of a single photon with wave number $k$ in the system is given by
\begin{equation}\label{eq:state_discrete}
    \ket{\psi_k} = \sum_j u_{j,k} a_j^{\dagger} \ket{0,g} + u_e \ket{0,e},
\end{equation}
where $u_{j,k}$ is the probability amplitude for finding the photon in cavity $j$, $u_e$ is the excitation amplitude of the emitter, and $\ket{0}$ is the vacuum state of the waveguide (i.e., all the cavities). The frequency of a photon with wave number $k$ is given by the dispersion relation in Eq.~(\ref{eq:dispersion}). Hence, the state $\ket{\psi_k}$ has the energy eigenvalue ${E_k = \omega_k = \omega_c - 2\xi\text{cos}(k)}$, and satisfies the time-independent Schr\"{o}dinger equation
\begin{equation}\label{eq:Schrodinger_2}
    H_{\text{crw}}\ket{\psi_k} = \left[ \omega_c - 2\xi\text{cos}(k) \right] \ket{\psi_k}.
\end{equation}
After calculating $H_{\text{crw}}\ket{\psi_k}$ and comparing coefficients of orthogonal states on both sides of Eq.~(\ref{eq:Schrodinger_2}) (see Appendix~\ref{app:discrete_model}), we obtain two simultaneous equations involving the amplitudes $u_{j,k}$ and $u_e$:
\begin{subequations}
\begin{equation}\label{eq:u_e_discrete}
   -u_e \Delta_e + g^* u_{0,k} = 0,
\end{equation}
\begin{equation}\label{eq:u_j_discrete}
    2\xi\text{cos}(k)u_{j,k} - \xi\left( u_{j-1,k} + u_{j+1,k} \right) + g u_e \delta_{j0} = 0,
\end{equation}
\end{subequations}
where ${\Delta_e = \omega_k - \omega_e = \omega_c - 2\xi\text{cos}(k) - \omega_e}$ is the photon-emitter detuning. Rearranging Eq.~(\ref{eq:u_e_discrete}) for $u_e$ gives
\begin{equation}
    u_e = \frac{g^*}{\Delta_e} u_{0,k},
\end{equation}
which can be substituted into Eq.~(\ref{eq:u_j_discrete}) to obtain
\begin{equation}
    2\xi\text{cos}(k)u_{j,k} - \xi\left( u_{j-1,k} + u_{j+1,k} \right) + \frac{|g|^2}{\Delta_e} u_{0,k} \delta_{j0} = 0.
\end{equation}
This equation is valid for all $j$, but for the purpose of calculating the transmission and reflection amplitudes we focus on the case where ${j=0}$:
\begin{equation}\label{eq:u_0}
    \left[ 2\xi\text{cos}(k) + \frac{|g|^2}{\Delta_e} \right] u_{0,k} - \xi\left( u_{-1,k} + u_{1,k} \right) = 0.
\end{equation}

We solve Eq.~(\ref{eq:u_0}) using the ansatz
\begin{equation}\label{eq:u_j_solution}
u_{j,k} =  \left\{ \begin{array}{lr}
e^{ikj} + r_{\text{crw}} e^{-ikj},\hspace{0.05in} j < 0, \\[0.05in]
t_{\text{crw}}e^{ikj},\hspace{0.05in} j > 0, \end{array}\right.
\end{equation}
where $t_{\text{crw}}$ and $r_{\text{crw}}$ are the transmission and reflection amplitudes to be determined. In analogy with the real-space approach, we have assumed that the solutions for ${j \neq 0}$ are plane waves. For ${j < 0}$, the photon amplitude $u_{j,k}$ includes the input plane wave coming from the left and the reflected plane wave moving to the left. For ${j>0}$, the photon amplitude is given by the plane wave transmitted to the right. The continuity condition at ${j=0}$, ${u_{0^-,k} = u_{0^+,k}}$, implies that ${1+r_{\text{crw}} = t_{\text{crw}}}$. From Eq.~(\ref{eq:u_j_solution}) and the continuity condition, it follows that
\begin{subequations}
\begin{equation}
    u_{0,k} = 1 + r_{\text{crw}},
\end{equation}
\begin{equation}
    u_{-1,k} = e^{-ik} + r_{\text{crw}}e^{ik},
\end{equation}
\begin{equation}
    u_{1,k} = t_{\text{crw}}e^{ik} = \left(1 + r_{\text{crw}}\right)e^{ik}.
\end{equation}
\end{subequations}
These can be substituted into Eq.~(\ref{eq:u_0}) to obtain
\begin{align}
\begin{split}
    \left[ 2\xi\text{cos}(k) + \frac{|g|^2}{\Delta_e} \right] \left( 1 + r_{\text{crw}} \right) = \xi &\Bigl[ e^{-ik} + r_{\text{crw}}e^{ik} \\
    &+ \left( 1 + r_{\text{crw}} \right)e^{ik} \Bigr],
\end{split}
\end{align}
which can be easily solved for $r_{\text{crw}}$. The transmission amplitude is then simply ${t_{\text{crw}} = 1+r_{\text{crw}}}$. In terms of the photon wave number $k$, we therefore have the transmission amplitude
\begin{equation}
    t_{\text{crw}} = \frac{2i\xi\text{sin}(k)[\omega_c - 2\xi\text{cos}(k) - \omega_e]}{2i\xi\text{sin}(k)[\omega_c - 2\xi\text{cos}(k) - \omega_e] - |g|^2}
\end{equation}
and the reflection amplitude
\begin{equation}
    r_{\text{crw}} = \frac{|g|^2}{2i\xi\text{sin}(k)[\omega_c - 2\xi\text{cos}(k) - \omega_e] - |g|^2},
\end{equation}
in agreement with Ref.~\cite{Zhou2008}. When ${g=0}$, the photon is transmitted perfectly (assuming its frequency lies within the allowed band), as the emitter is decoupled from the waveguide and the photon propagates freely through the cavities. The transmission and reflection probabilities can be expressed in terms of frequencies as
\begin{subequations}
\begin{equation}
    T_{\text{crw}} = |t_{\text{crw}}|^2 = \frac{\Delta_e^2 (4\xi^2 - \Delta_c^2)}{\Delta_e^2 (4\xi^2 - \Delta_c^2) + |g|^4},
\end{equation}
\begin{equation}
    R_{\text{crw}} = |r_{\text{crw}}|^2 = \frac{|g|^4}{\Delta_e^2 (4\xi^2 - \Delta_c^2) + |g|^4},
\end{equation}
\end{subequations}
where $\Delta_e$ is the photon-emitter detuning, defined previously in the derivation, and ${\Delta_c = \omega_k - \omega_c = - 2\xi\text{cos}(k)}$ is the photon-cavity detuning.

\begin{figure}
    \centering
    \includegraphics[width=\linewidth]{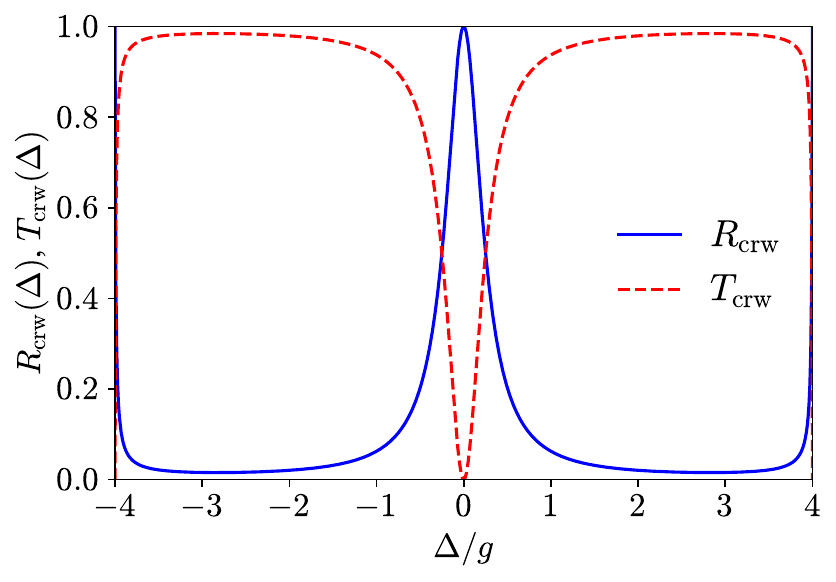}
    \caption{Single-photon transmission [dashed red curve, Eq.~(\ref{eq:T_discrete})] and reflection [solid blue curve, Eq.~(\ref{eq:R_discrete})] as a function of the photon-emitter detuning $\Delta$ for the system in Fig.~\ref{fig:two_level_emitter_waveguide}(b). Photon loss is neglected in these results.}
    \label{fig:R_T_discrete}
\end{figure}

For simplicity, we now focus on the case of real emitter-cavity coupling (${g^* = g}$), and where the emitter is on resonance with the cavities (${\omega_e = \omega_c}$). This means that ${\Delta_e = \Delta_c = \Delta}$, so
\begin{equation}\label{eq:T_discrete}
    T_{\text{crw}} = \frac{\Delta^2 (4\xi^2 - \Delta^2)}{\Delta^2 (4\xi^2 - \Delta^2) + g^4}
\end{equation}
and
\begin{equation}\label{eq:R_discrete}
    R_{\text{crw}} = \frac{g^4}{\Delta^2 (4\xi^2 - \Delta^2) + g^4}.
\end{equation}
In Fig.~\ref{fig:R_T_discrete}, we show $T_{\text{crw}}$ and $R_{\text{crw}}$ as a function of the detuning $\Delta$. These spectra look very similar to the case of a continuous waveguide in Fig.~\ref{fig:R_T}---when ${\Delta = 0}$, ${R_{\text{crw}} = 1}$ and ${T_{\text{crw}} = 0}$, meaning that the input photon is perfectly reflected on resonance with the emitter. As the detuning from the emitter increases, the transmission increases because the photon is less likely to be scattered. The main difference from the continuous waveguide is that we have a frequency band outside of which photons cannot propagate through the coupled-resonator waveguide. This can be seen in Fig.~\ref{fig:R_T_discrete} where, at the band edge (${\Delta/g = \pm 2\xi/g = \pm 4}$), the transmission falls back to zero and the reflection increases to one.

Both continuous waveguides and coupled-resonator waveguides have been widely studied in theoretical proposals of single-photon switches and routers. However, in experimental implementations of single-photon switches, continuous waveguides are used as opposed to waveguides consisting of coupled cavities. For example, photonic crystal waveguides are used with QDs, optical fibers are used with neutral atoms, and transmission lines are used with superconducting qubits (see Section~\ref{sec:experiment}). The lack of experimental results for single-photon routing in coupled-resonator waveguides is likely due to the difficulty associated with fabricating an array of identical cavities. Fabrication imperfections at the nanoscale make it challenging to produce cavities with identical resonance frequencies. Although post-fabrication tuning of cavities is possible (e.g., optical~\cite{Fushman2007}, electromechanical~\cite{Brunswick2025}, or strain tuning~\cite{Luxmoore2012}), correcting frequencies of individual cavities is difficult when they are coupled together and closely spaced. On the other hand, fabrication of continuous waveguides is generally more straightforward (especially simple slab/nanobeam waveguides and optical fibers), making them more suitable for efficiently guiding photons within current fabrication capabilities.


\subsection{Other theoretical methods}\label{subsec:methods_4}

We now briefly mention other theoretical approaches that have been used to study single-photon scattering in the literature. The methods outlined above are concerned with the asymptotic input and output states, which is sufficient for calculating transmission and reflection amplitudes but it does not provide information about the temporal dynamics of the interactions. In order to look at the full time evolution rather than simply the initial and final states, a time-dependent theory must be used. For example, the time-dependent Schr\"{o}dinger equation can be solved to obtain the temporal dynamics of the photon and the emitter~\cite{Liao2015}. The time-dependent Schr\"{o}dinger equation for scattering of a single photon has been solved via direct integration~\cite{Liao2015, Manzoni2014}, a Laplace transform~\cite{Yan2014, Li2016}, and a perturbative approach~\cite{Qin2016}. Photon loss can be included either with a non-Hermitian Hamiltonian in the Schr\"{o}dinger equation (via the substitution used in Section~\ref{subsubsec:methods})~\cite{Liao2015, Manzoni2014}, or using a full master equation with Lindblad terms that account for loss~\cite{Manzacca2007, Rinaldi2025}. The temporal dynamics can also be studied using Heisenberg or quantum Langevin equations in the Heisenberg picture~\cite{Zhu2019_2, Stolyarov2020}. Taking a master equation or Langevin equation approach allows for the inclusion of other noise sources in quantum emitters, for example pure dephasing.

Some authors have adopted a Green tensor approach to deriving master equations that give the time evolution of a density matrix describing a system~\cite{Dung2002, Lang2020}, or quantum Langevin equations describing the evolution of field operators~\cite{Gruner1996, Caneva2015}. Starting from a classical electromagnetic Green's function, an effective master equation can be derived, where the Hamiltonian and Lindblad operators are expressed in terms of the Green's function~\cite{Chang2018}. The Green tensor method has also been used to derive the single-photon scattering matrix for a two-level emitter in a cavity (via LSZ reduction)~\cite{Shi2011} and for an array of two-level emitters coupled to a waveguide (via a generalized input-output formalism)~\cite{Caneva2015}.

In addition, the methods outlined in Sections~\ref{subsec:methods_1}-\ref{subsec:methods_3} can be extended to scattering from multiple localized systems (e.g., quantum emitters) that are spatially separated. This is commonly done using the transfer matrix approach, which can be combined with the real-space approach~\cite{Tsoi2009, Berndsen2024}, the input-output formalism~\cite{Duda2024}, and the discrete coordinate scattering approach~\cite{Huang2019_2}. In particular, once the scattering solutions are found for each individual localized subsystem, the transfer matrix for each subsystem can be constructed, and the total transfer matrix for the full system is then the product of the transfer matrices for the localized subsystems. Note that the transfer matrix approach is only applicable in the single-photon regime. When multi-photon scattering is considered, the SLH formalism can be used~\cite{Rinaldi2025, Combes2017}, which enables the construction of the $N$-photon scattering matrix for systems with multiple scattering elements.


\subsection{Efficiency and fidelity of a single-photon switch}\label{subsec:methods_5}

Single-photon scattering calculations like those outlined above give us the transmission amplitude $t$ and the reflection amplitude $r$ for the scattering process. Using these quantities, we can define figures of merit that are relevant to the analysis of single-photon routing devices. The transmission and reflection amplitudes that we calculated above are functions of the input photon frequency (since they are functions of the photon-emitter detuning $\Delta$), which we will denote here by $\omega$. If we consider the case of an input photon with a single frequency $\omega$, we can define the efficiency of the routing process simply as the probability that a photon of this frequency is sent in the desired direction. This means that the efficiency $E_{t,\omega}$ of routing an input photon with frequency $\omega$ down the transmission path is given by the transmission probability:
\begin{equation}\label{eq:E_t_one_frequency}
    E_{t,\omega} = |t(\omega)|^2,
\end{equation}
and the efficiency $E_{r,\omega}$ of routing an input photon down the reflection path is given by the reflection probability:
\begin{equation}\label{eq:E_r_one_frequency}
    E_{r,\omega} = |r(\omega)|^2.
\end{equation}

These definitions need to be modified when we take into account the finite width of the input photon wave packet. Consider a wave packet that has a spectral envelope given by $f(\omega)$, which is normalized such that
\begin{equation}
    \int_{-\infty}^{\infty} |f(\omega)|^2 d\omega = 1.
\end{equation}
The amplitudes $t(\omega)$ and $r(\omega)$ apply to a single frequency component of the photon wave packet. To calculate the efficiency for transmitting or reflecting the full wave packet, we need to integrate over all the frequency components~\cite{Duda2024}. This leads to the transmission efficiency
\begin{equation}\label{eq:E_t}
    E_t = \left| \int_{-\infty}^{\infty} |t(\omega)|^2 |f(\omega)|^2 d\omega \right|^2
\end{equation}
and the reflection efficiency 
\begin{equation}\label{eq:E_r}
    E_r = \left| \int_{-\infty}^{\infty} |r(\omega)|^2 |f(\omega)|^2 d\omega \right|^2.
\end{equation}
When the transmission (reflection) probability is one over the full extent of the wave packet, we obtain ${E_t = 1}$ (${E_r = 1}$). In other words, we must have ${|t(\omega)| = 1}$ [${|r(\omega)| = 1}$] over the frequency range where the wave packet $f(\omega)$ is nonzero to achieve a perfect transmission (reflection) efficiency.

Taking into account the frequency distribution of the photon wave packet also allows us to define the fidelity of the routing process, which quantifies the similarity between the input and output wave packets. In general, the wave packet is modified during the scattering process, either because the transmission/reflection is not perfect and the wave packet is split into a superposition of both propagating directions, or due to a frequency-dependent phase that is acquired in the interaction and manifests in the amplitudes $t(\omega)$ and $r(\omega)$. The fidelity of the output wave packet can be defined as
\begin{equation}\label{eq:F_t}
    F_t = \left| \int_{-\infty}^{\infty} t(\omega) |f(\omega)|^2 d\omega \right|^2
\end{equation}
for transmission, and
\begin{equation}\label{eq:F_r}
    F_r = \left| \int_{-\infty}^{\infty} r(\omega) |f(\omega)|^2 d\omega \right|^2
\end{equation}
for reflection~\cite{Duda2024}. The fidelity $F_t$ ($F_r$) will only be equal to one if ${t(\omega) = 1}$ [${r(\omega) = 1}$] over the frequency range where $f(\omega)$ is nonzero. Unlike efficiency, the fidelity is sensitive to the phases of the amplitudes $t(\omega)$ and $r(\omega)$. A fidelity of zero implies that the output wave packet is orthogonal to the input wave packet.

For an ideal single-photon switch, we should be able to switch between ${E_t = 1}$ and ${E_r = 1}$ deterministically. If we require the input wave packet to be preserved during the switching process, we must also have ${F_t = 1}$ when the switch is transmitting and ${F_r = 1}$ when the switch is reflecting. This would be desirable if we only want the switch to route photons without altering the pulse shape, which is a necessary requirement to maintain indistinguishability between photons in a quantum network (e.g., to realize photon-photon interference~\cite{Mandel1987} in quantum computing schemes~\cite{Kok2007}). The efficiencies and fidelities defined here apply to a simple two-port setup with a transmission path and a reflection path, but they can be generalized to systems with more output ports by calculating the scattering amplitudes for all the relevant ports.

There inevitably exists a trade-off between the efficiency/fidelity of a switch and the switch speed. To maximize the efficiency and fidelity, we require the input photon wave packet $f(\omega)$ to be narrower than the switching bandwidth over which $t(\omega)$ and $r(\omega)$ reach their optimal values. However, a wave packet that is spectrally narrow will have a long time duration, which limits the device operation rate. Conversely, pulses that have a short duration will have a broader frequency distribution, which will limit the efficiency/fidelity if the wave packet extends beyond the switching bandwidth.

\begin{figure}
    \centering
    \includegraphics[width=\linewidth]{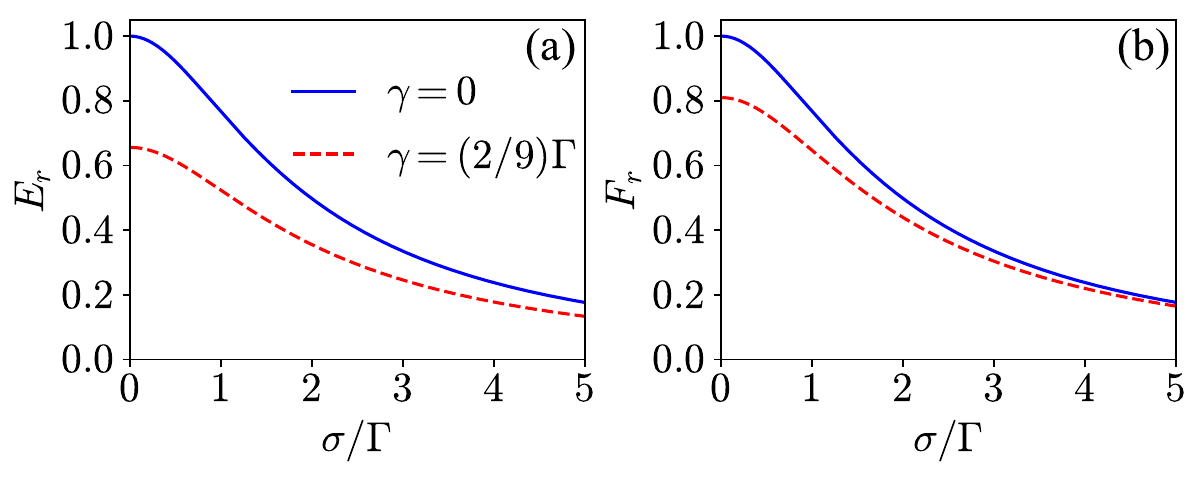}
    \caption{(a) Reflection efficiency $E_r$ and (b) reflection fidelity $F_r$ for a two-level emitter symmetrically coupled to a waveguide, calculated as a function of the bandwidth $\sigma$ of a Gaussian wave packet [Eq.~(\ref{eq:Gaussian})]. The wave packet is centered at the transition frequency of the emitter, where perfect reflection occurs in the absence of loss. The solid blue curves correspond to the lossless case from Fig.~\ref{fig:R_T} (${\gamma = 0}$), and the dashed red curves correspond to the losses from Fig.~\ref{fig:R_T_loss} [${\gamma = (2/9)\Gamma}$].}
    \label{fig:E_F_vs_sigma}
\end{figure}

In Fig.~\ref{fig:E_F_vs_sigma}, we show the reflection efficiency $E_r$ and the reflection fidelity $F_r$ as a function of the FWHM $\sigma$ of a Gaussian input wave packet with spectral envelope
\begin{equation}\label{eq:Gaussian}
    f(\omega) = \left[ \frac{4\ln(2)}{\pi \sigma^2} \right]^{1/4} e^{-2\ln(2)(\omega - \omega_0)^2/\sigma^2},
\end{equation}
where $\omega_0$ is the central frequency. We calculate $E_r$ and $F_r$ using Eqs.~(\ref{eq:E_r}) and (\ref{eq:F_r}), respectively, with the reflection amplitude from Eq.~(\ref{eq:t_and_r}), corresponding to a two-level emitter symmetrically coupled to a waveguide [Fig.~\ref{fig:two_level_emitter_waveguide}(a)]. We assume that the wave packet is centered at the transition frequency of the emitter (${\omega_0 = \omega_e}$), where perfect reflection occurs in the ideal lossless case (see Fig.~\ref{fig:R_T}). Fig.~\ref{fig:E_F_vs_sigma} shows that, in the absence of photon loss (solid blue curves), the reflection efficiency and fidelity are high when the pulse bandwidth is smaller than the linewidth of the emitter (${\sigma/\Gamma < 1}$), corresponding to pulse durations longer than the emission lifetime. When the pulse bandwidth exceeds the emitter linewidth (${\sigma/\Gamma > 1}$), the efficiency and fidelity are reduced significantly below unity because a large fraction of the wave packet does not interact with the emitter and hence does not get reflected.

\begin{figure}
    \centering
    \includegraphics[width=\linewidth]{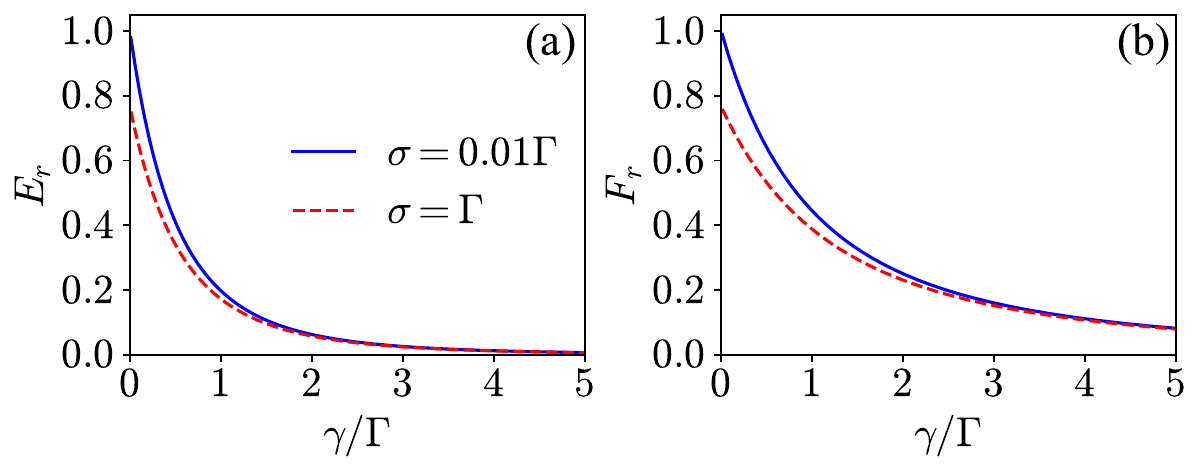}
    \caption{(a) Reflection efficiency $E_r$ and (b) reflection fidelity $F_r$ for a two-level emitter symmetrically coupled to a waveguide, calculated as a function of the emitter loss rate $\gamma$. The solid blue curves correspond to a narrow input wave packet with ${\sigma = 0.01\Gamma}$, and the dashed red curves correspond to a broader wave packet with ${\sigma = \Gamma}$.}
    \label{fig:E_F_vs_gamma}
\end{figure}

In Fig.~\ref{fig:E_F_vs_gamma}, we show how the reflection efficiency and fidelity scale with the loss rate $\gamma$ for a narrow wave packet with ${\sigma = 0.01\Gamma}$ (solid blue curves), and for a broader wave packet with ${\sigma = \Gamma}$ (dashed red curves). In both cases, the wave packet is centered at the transition frequency of the emitter as in Fig.~\ref{fig:E_F_vs_sigma}. As discussed in Section~\ref{subsubsec:methods}, photon loss is introduced into the reflection amplitude in Eq.~(\ref{eq:t_and_r}) via the substitution ${\Delta \rightarrow \Delta + i\gamma/2}$. Fig.~\ref{fig:E_F_vs_gamma} shows that the emitter loss rate $\gamma$ into non-guided modes must be very small compared to the emitter decay rate $\Gamma$ into the waveguide in order to obtain a high reflection efficiency and fidelity (${\gamma/\Gamma \ll 1}$), in addition to having a sufficiently narrow wave packet with ${\sigma/\Gamma < 1}$. When photon loss becomes the dominant mechanism (${\gamma/\Gamma > 1}$), the efficiency and fidelity become equally low for both of the considered wave packets, as the wave packet width is less significant in this regime.

Note that the transmission efficiency $E_t$ and the transmission fidelity $F_t$ can be made arbitrarily close to one by sufficiently detuning the input wave packet from the emitter, as discussed at the beginning of Section~\ref{sec:theory}. In the large detuning regime, effectively no scattering occurs so emitter imperfections do not degrade the transmission efficiency and fidelity.


\subsection{Noise in realistic quantum emitters}\label{subsec:methods_6}

Spontaneous emission loss into non-guided modes (i.e., imperfect emitter-waveguide coupling) is one possible imperfection that reduces the efficiency and fidelity of realistic single-photon switches based on quantum emitters. There are also other sources of noise that are present for quantum emitters in the solid state (e.g., QDs), which can degrade the switch performance. For example, coupling to phonons leads to frequency shifts in photon emission through phonon sideband transitions~\cite{Iles-Smith2017} and pure dephasing~\cite{Muljarov2004, Reigue2017, Denning2020} at elevated temperatures. This will naturally lead to changes in the photon wave packet, resulting in decoherence of the scattered photon. Other noise sources that reduce the photon coherence include charge noise and spin noise~\cite{Kuhlmann2013}. Charge noise occurs when charge fluctuations in the host lattice environment lead to variations in the local electric field at the position of the emitter, which shifts the transition frequency via the Stark effect (spectral wandering). Spin noise arises from fluctuating nuclear spins that result in variations in the magnetic field experienced by an electron spin in the emitter via the hyperfine interaction. The effects of these noise processes can be reduced by, for example, working at cryogenic temperatures to minimize phonon coupling, stabilizing the charge environment with a diode structure that provides electrical control of the quantum emitter~\cite{Zhai2020}, and employing dynamical decoupling schemes that suppress the effects of interactions with nuclear spins~\cite{Lang2010, Dyte2025}.

When outlining the standard methods used to model photon scattering in the Schr\"{o}dinger picture (real-space approach, Section~\ref{subsec:methods_1}) and the Heisenberg picture (input-output formalism, Section~\ref{subsec:methods_2}), we mentioned that photon loss can be straightforwardly included with a simple substitution in the transmission and reflection amplitudes. However, the wave function method of Section~\ref{subsec:methods_1} does not provide a simple way to include other noise sources that are present in realistic quantum emitters, such as those mentioned above. This is evident in the literature that is reviewed in the next section, where a treatment of these noise processes in the context of single-photon routing is lacking. As mentioned in Section~\ref{subsec:methods_4}, taking a more sophisticated master equation approach would allow for the inclusion of Lindblad terms that account for different sources of noise, for example, pure dephasing via the Lindblad operator ${L_{\text{ph}} = \sqrt{\gamma_{\text{ph}}}\sigma_z}$, where $\gamma_{\text{ph}}$ is the dephasing rate and $\sigma_z$ is the atomic Pauli-Z operator. We also note that such Lindblad terms can be included in the Heisenberg equations within the input-output formalism of Section~\ref{subsec:methods_2}~\cite{Duda2024}.


\section{Theoretical studies of single-photon routing}\label{sec:theory}

\begin{figure}
    \centering
    \includegraphics[width=\linewidth]{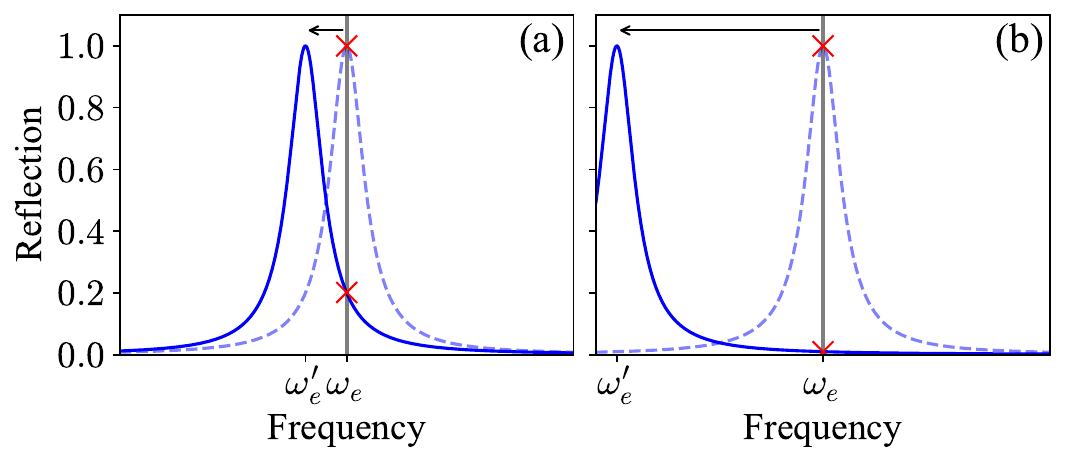}
    \caption{Single-photon switching with a two-level emitter in a waveguide [see Fig.~\ref{fig:two_level_emitter_waveguide}(a)]. The blue dashed and solid curves are the single-photon reflection probabilities before and after tuning the transition frequency of the emitter from $\omega_e$ to $\omega_e^{\prime}$, respectively [calculated using Eq.~(\ref{eq:R})]. The vertical black lines show the frequency of an input photon that is initially resonant with the emitter, and the red crosses mark the reflection probability at this frequency, before and after tuning. The frequency shift is (a) one linewidth and (b) five linewidths.}
    \label{fig:two_level_emitter_switch}
\end{figure}

As we have seen in Section~\ref{sec:methods}, the direction of propagation of a single photon in a waveguide can be controlled via scattering from a quantum emitter. When the photon is on resonance with the emitter, and the emitter couples to the waveguide equally in both directions, perfect reflection occurs in the absence of loss (Fig.~\ref{fig:R_T}). Conversely, when the photon is sufficiently detuned from the emitter (by more than the linewidth of the emitter), near-perfect transmission can be achieved. Hence, by adjusting the photon-emitter detuning, it becomes possible to switch between two output paths (the transmission path and the reflection path)~\cite{Zhou2008}. This is a basic example of a single-photon switch based on scattering from a quantum emitter, the operation of which is illustrated in Fig.~\ref{fig:two_level_emitter_switch}. In practice, the photon-emitter detuning can be controlled by tuning the transition frequency of the emitter~\cite{Houck2008, Nowak2014, Hallett2018}. In Fig.~\ref{fig:two_level_emitter_switch}(a), we show that a frequency shift of one linewidth (FWHM) reduces the reflection probability of an initially resonant photon from $1$ to approximately $0.2$, while in (b) the reflection is reduced from $1$ to below $0.01$ under a frequency shift of five linewidths. Therefore, a ${>99\%}$ change in reflection can be achieved with emitter frequency tuning for an input photon that is initially resonant with the emitter.

\begin{figure}
    \centering
    \includegraphics[width=\linewidth]{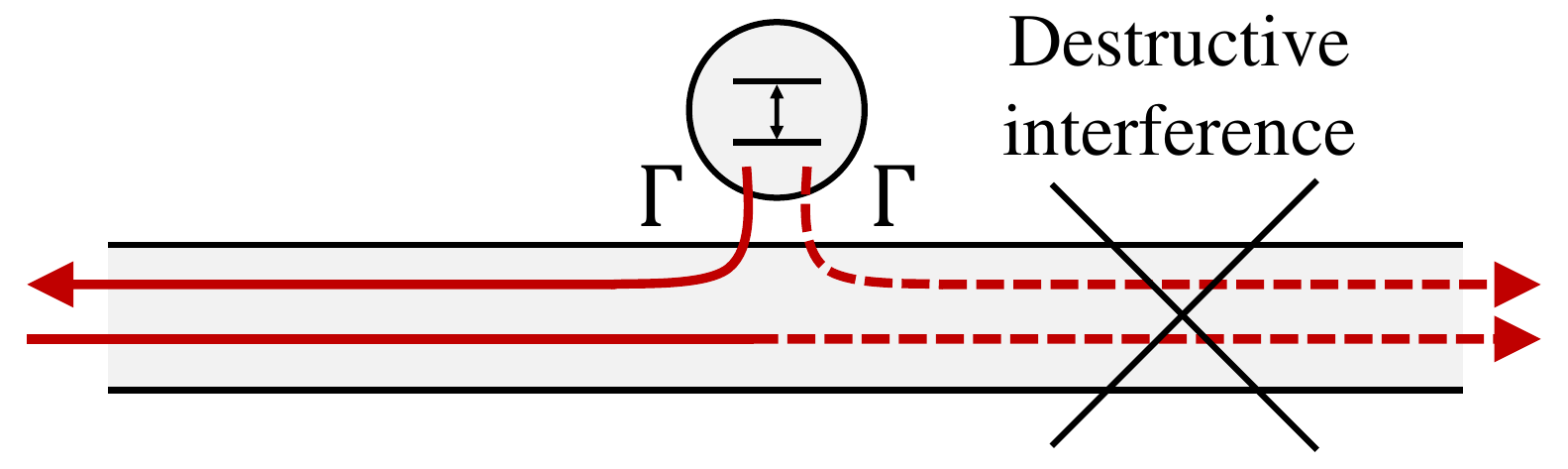}
    \caption{When an input photon in the waveguide is resonant with a two-level emitter that is symmetrically coupled to the waveguide, the two possible transmission paths (direct transmission and rightward emission following absorption) interfere destructively, resulting in perfect reflection.}
    \label{fig:interference}
\end{figure}

The observed behavior of perfect reflection on resonance can be explained in terms of interference between different photon pathways~\cite{Gonzalez-Ballestero2016}. The situation where the emitter couples to the waveguide symmetrically (with decay rate $\Gamma$ in both directions) is shown in Fig.~\ref{fig:interference}, where we consider a right-moving input photon. There are two possible paths where the photon leaves the waveguide to the right---direct transmission (lower path in Fig.~\ref{fig:interference}), and absorption followed by rightward emission (upper path in Fig.~\ref{fig:interference}). In the case of symmetric coupling and a resonant input photon, these two paths interfere destructively, resulting in zero transmission and perfect reflection [this can be viewed as the cancellation of the two terms on the right-hand side of the input-output relation for ${\mu = R}$ in Eq.~(\ref{eq:input_output})]. When the photon is detuned from the emitter, the path involving the interaction with the emitter has a reduced amplitude, so the interference is not perfect and transmission becomes possible. For detunings much larger than the emitter linewidth, there is essentially no interference and the only possible path is the direct transmission path. In addition, when the coupling becomes asymmetric, the destructive interference in transmission again becomes imperfect (even on resonance), as the relative amplitudes of the rightward and leftward emission paths are modified from the case of symmetric coupling.

A two-level emitter coupled to a waveguide is a simple single-photon switch, but there are multiple reasons why we may want to consider more complex setups. Firstly, as is evident from Fig.~\ref{fig:R_T}, reflection is only perfect at a single frequency (on resonance with the emitter). This limits the switching efficiency for realistic photon wave packets that have a distribution of frequencies (e.g., Gaussian wave packets), as only one frequency component can be reflected perfectly. It is therefore desirable to design switches with a larger bandwidth than the single-emitter case. For this purpose, routers with multiple emitters have been considered~\cite{Chang2011, Yan2020, Kim2010, Song2025, Yan2018_2}. An array of ${N>1}$ quantum emitters coupled to a waveguide acts as a distributed Bragg reflector, with a wider bandwidth of near-perfect reflection compared to having a single emitter~\cite{Duda2024, Chang2011}.

Furthermore, using a single waveguide with two propagating directions only allows for two output ports (a transmission port and a reflection port). However, a quantum network of practical use is likely to have more than two possible routing directions, so it is necessary to consider systems that allow for more input and output channels. One way of achieving this is to couple quantum emitters to two~\cite{Li2015, Chen2016, Zhang2022, Poudyal2020, Gonzalez-Ballestero2016, Yan2022, Yan2018, Cheng2017_2, Li2018, Yang2022, Wang2022_3, Amgain2024, Cheng2016, Liu2025, Zhou2023} or more~\cite{Yan2014} waveguides. Each additional waveguide provides two extra output ports for routing photons (or one port if one end of the waveguide is terminated by a mirror, which can improve routing to other ports due to interference induced by reflections from the mirror~\cite{Liu2017, Huang2020, Li2021, Wang2021_2}). Alternatively, in the case of coupled-resonator waveguides, quantum emitters embedded at the crossing point of a T-shaped~\cite{Lu2015, Liu2019, Huang2018, Liu2017, Kim2025, Huang2019_2} or X-shaped~\cite{Huang2019, Li2020, Lu2014, Ahumada2019, Zhou2013, Du2021, Huang2018_2, Yan2017, Yan2016} waveguide have been considered.

It is also beneficial to consider alternative routing mechanisms, not necessarily based on emitter tuning. This could be due to having a limited tuning range in a physical implementation, or because other routing mechanisms may lead to improved figures of merit for switching. Using additional degrees of freedom, such as more atomic energy levels, cavities, and external driving lasers, enables a wider range of routing mechanisms to be explored.

In the following subsections, we review theoretical proposals of single-photon routing schemes. We organize the theoretical studies by routing mechanism, and discuss the underlying physics that enables the control of single-photon propagation in each case. Ultimately, the mechanisms rely on controlling the interference between different photon pathways to route photons in the desired direction by changing a relevant parameter. This corresponds to attaining constructive interference in the desired channel, and destructive interference in the undesired output channels, as we explained in the context of the simple example above. Towards the end of this section, we mention examples of related single-photon routing schemes that do not use quantum emitters.


\subsection{Chiral/directional emitter-waveguide coupling}\label{subsec:theory_chiral}

Considering again the two-level emitter coupled to a waveguide, the single-photon transmission and reflection amplitudes in Eqs.~(\ref{eq:t_real_space}) and (\ref{eq:r_real_space}) depend on three parameters: the photon-emitter detuning $\Delta$, and the decay rates $\Gamma_R$ and $\Gamma_L$ for emission to the right and left in the waveguide, respectively. Hence, instead of controlling the detuning, we can control the emitter-waveguide coupling to change the scattering behavior. In particular, when ${\Gamma_R \neq \Gamma_L}$, the interference in Fig.~\ref{fig:interference} is modified, which changes the transmission and reflection probabilities~\cite{Gonzalez-Ballestero2016}.

\begin{figure}
    \centering
    \includegraphics[width=\linewidth]{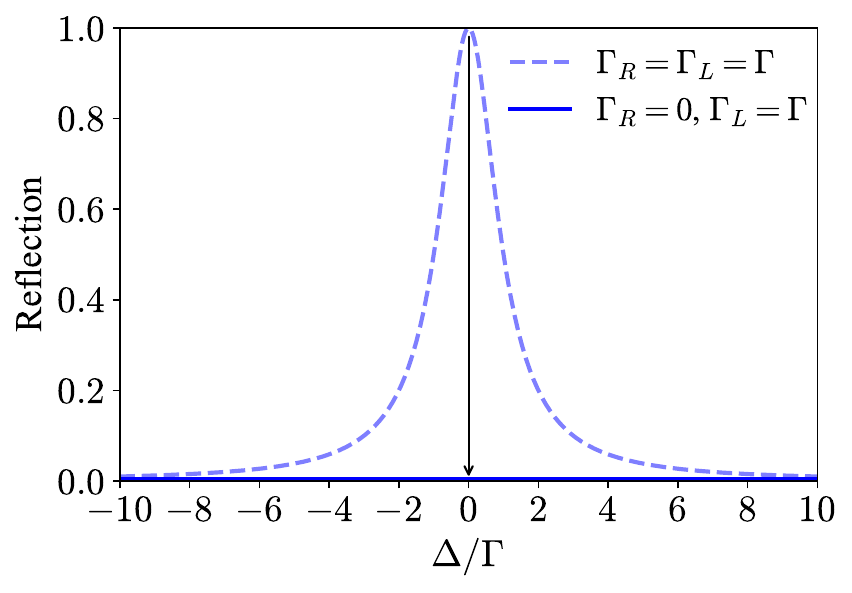}
    \caption{Single-photon switching with a two-level emitter chirally coupled to a waveguide. The blue dashed and solid curves are the single-photon reflection probabilities for symmetric (${\Gamma_R = \Gamma_L = \Gamma}$) and chiral (${\Gamma_R = 0}$, ${\Gamma_L = \Gamma}$) coupling, respectively [calculated using Eq.~(\ref{eq:r_real_space})]. The reflection switches from one to zero on resonance with the emitter (${\Delta = 0}$) when the coupling is switched from symmetric to chiral, as indicated by the arrow.}
    \label{fig:chiral_switch}
\end{figure}

The extreme case of asymmetric emitter-waveguide coupling occurs when ${\Gamma_R = 0}$ or ${\Gamma_L = 0}$, and is called chiral (or unidirectional) coupling~\cite{Lodahl2017}. Here an incoming photon can only be scattered in one direction. For a right-moving input photon, ${\Gamma_R = 0}$ implies that no interaction takes place and the photon is transmitted, and ${\Gamma_L = 0}$ implies that the photon cannot be reflected and must be scattered to the right, again being transmitted. This is markedly different from the symmetric coupling case in Fig.~\ref{fig:interference}, where perfect reflection occurs on resonance. By switching from symmetric coupling (${\Gamma_R = \Gamma_L = \Gamma}$) to chiral coupling (${\Gamma_R = 0}$ and ${\Gamma_L = \Gamma}$ or vice-versa), we switch from perfect reflection to perfect transmission at the transition frequency of the emitter, as shown in Fig.~\ref{fig:chiral_switch}.

Single-photon routing based on a two-level emitter chirally coupled to a waveguide was proposed by Yan et al. in 2011~\cite{Yan2011}. The more asymmetric the coupling, the closer the transmission on resonance becomes to one. This was later extended to a two-level emitter chirally coupled to two waveguides, where there are four possible output ports~\cite{Cheng2016, Yang2022}. If the coupling between the emitter and the waveguides is symmetric, the input photon can be transferred to the second waveguide with a maximum probability of 0.5, and the photon is routed into this waveguide in both directions with equal probability~\cite{Cheng2016}. In order to route an input photon in one waveguide towards one output port in the other waveguide deterministically, the coupling to both waveguides must be chiral to break the symmetry of the system. For example, if the emitter couples only to the right-moving waveguide mode in both waveguides, a right-moving input photon in one waveguide can be routed to the right in the second waveguide with a probability of one~\cite{Cheng2016}. In the case of coupled-resonator waveguides, the symmetry can be broken by making the cavity-cavity coupling on either side of the emitter asymmetric~\cite{Huang2019, Li2020}.

More recent work has focused on chirally coupling more than one two-level emitter to waveguides, which can enhance the routing performance~\cite{Li2018, Poudyal2020, Zhou2023, Amgain2024, Liu2025}. In such schemes, the emitters can interact indirectly via the waveguide, or directly via the inclusion of additional interaction terms in the Hamiltonian, proportional to $\sigma_i^+\sigma_j^-$ (where the subscripts $i$ and $j$ label different emitters). The dipole-dipole interaction between a pair of two-level emitters can improve the routing probabilities in the presence of imperfect directionality and emitter loss~\cite{Li2018, Liu2025}. Physically, the dipole-dipole interaction provides additional photon pathways in the system (corresponding to an exchange of excitations between emitters), the interference of which can be controlled with the interaction strength to achieve routing in the desired direction~\cite{Yan2018_3, Jiang2018, Song2025}. More photon pathways can be created by coupling the emitters to the waveguides at multiple points to form “giant atoms”, and the phase shift between the coupling points can be used to control the routing~\cite{Zhou2023} (see also Refs.~\cite{Zhao2020, Wang2021_2, Zhang2022, Chen2022, Zhang2023, Zhu2025, Wang2025_2, Chai2025, Wang2025_3, Legon2025} for routing with giant atoms). Increasing the number of emitters can further improve the routing probabilities in the presence of loss~\cite{Poudyal2020} and disorder in the emitter separations~\cite{Amgain2024}. Note that the strength of the direct dipole-dipole interaction decreases quickly with increasing emitter separation~\cite{Poudyal2020, Amgain2024}. When the emitter separation is significantly larger than the wavelength of photons in the waveguides, this dipole-dipole interaction is negligible, and it is in this regime that the routing bandwidth exhibits a clear broadening~\cite{Chang2011, Yan2020, Kim2010, Song2025, Yan2018_2} (the emitters essentially form a distributed Bragg reflector in this regime, and only interact via the waveguide).

In practice, chiral emitter-waveguide coupling can be realized by coupling a quantum emitter with a circularly polarized transition to the local circularly polarized electric field of a nanophotonic waveguide~\cite{Lodahl2017}. The directional coupling gives rise to nonreciprocal photon scattering, where the light-matter interaction depends on the propagation direction of the input photon. For example, if a quantum emitter couples only to the right-moving waveguide mode, a right-moving photon would be scattered from the emitter while a left-moving photon would propagate past it with no interaction. This nonreciprocity could be used to implement a single-photon diode, where photons propagating in one direction are routed to a different channel and photons propagating in the opposite direction are not routed. For more details on the physical origins of directional light-matter interactions and the resulting nonreciprocity, see the reviews in Refs.~\cite{Lodahl2017, Suarez-Forero2025}.

As we will see in Section~\ref{sec:experiment}, there have been no experimental demonstrations of single-photon switching based on chiral coupling. This is likely because it is challenging to control emitter-waveguide coupling when the quantum emitter is embedded within the waveguide and cannot be moved to a different position in the waveguide where the coupling would be modified (e.g., where the local electric field polarization would be different). It is generally easier to use the transition frequency as a tunable parameter rather than the emitter-waveguide coupling rates. Nevertheless, a wavelength-tunable chirality has been demonstrated with a quantum emitter embedded in a cavity~\cite{Hallett2022, Martin2025}, which may be applied to photon switching. In order to tune the directionality while keeping the properties of the emitter fixed it may be beneficial to use a nonlinear medium, the optical properties of which can be tuned to adjust the coupling (e.g., refractive index modulation).


\subsection{Strong emitter-cavity coupling}\label{subsec:theory_cavity}

\begin{figure}
    \centering
    \includegraphics[width=\linewidth]{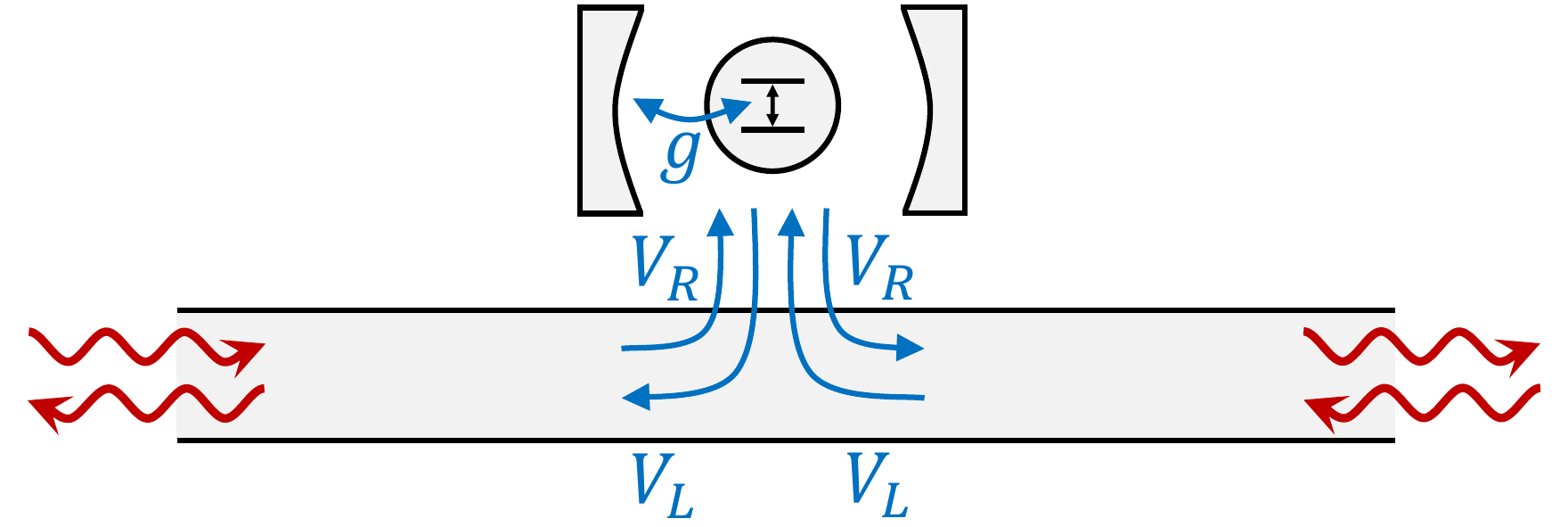}
    \caption{An extension of the system in Fig.~\ref{fig:two_level_emitter_waveguide}(a), where the two-level emitter couples to a single-mode cavity with coupling rate $g$. The cavity also couples to the right- and left-moving waveguide modes with coupling rates $V_R$ and $V_L$, respectively.}
    \label{fig:two_level_emitter_cavity}
\end{figure}

Alternative routing mechanisms beyond controlling the photon-emitter detuning or emitter-waveguide coupling require additional control parameters. One way to achieve this is to couple a quantum emitter to a cavity, which is further coupled to the input and output waveguide channels. A simple example is shown in Fig.~\ref{fig:two_level_emitter_cavity}, where a two-level emitter couples to a single-mode cavity with coupling rate $g$. The emitter-cavity interaction can be described by the widely used Jaynes-Cummings Hamiltonian~\cite{Jaynes1963, Rephaeli2012},
\begin{equation}\label{eq:H_JC}
    H_{\text{JC}} = \omega_e \sigma^+ \sigma^- + \omega_c c^{\dagger} c +  g \sigma^+c + g^* \sigma^- c^{\dagger},
\end{equation}
where $\omega_c$ is the cavity resonance frequency, $c^{\dagger}$ and $c$ are the creation and annihilation operators for the cavity mode, respectively, and, as before, $\omega_e$ is the transition frequency and $\sigma^\pm$ are the raising and lowering operators of the emitter. When we include the right- and left-moving waveguide modes, we obtain a Hamiltonian similar to the emitter-waveguide Hamiltonian in Eq.~(\ref{eq:H_k}):
\begin{figure}
    \centering
    \includegraphics[width=0.6\linewidth]{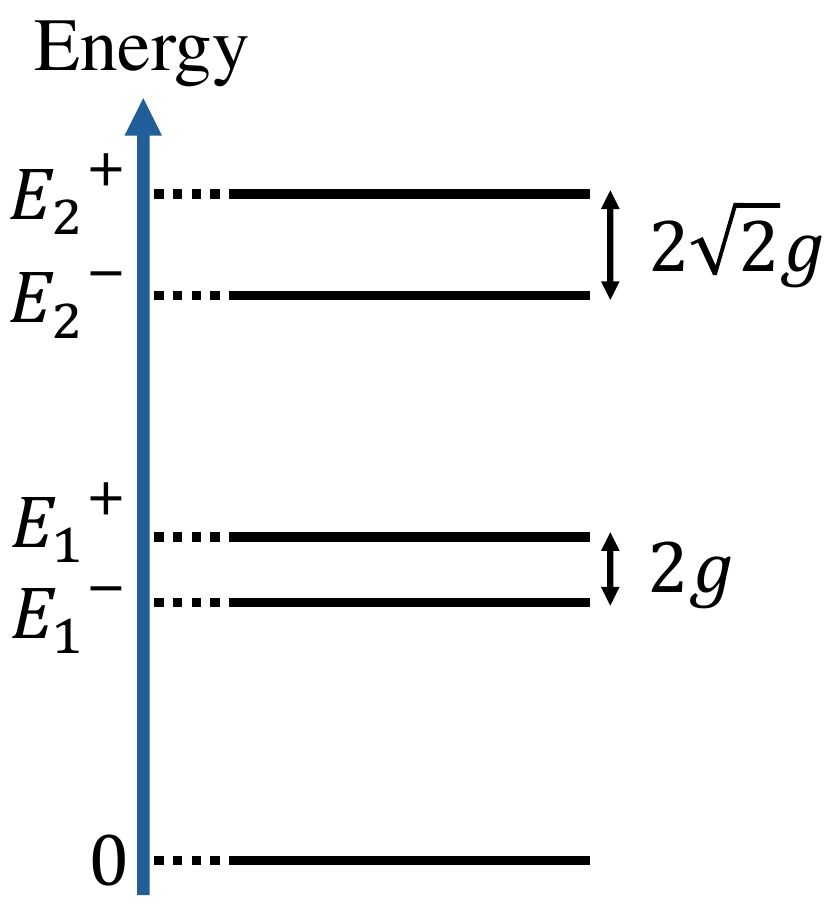}
    \caption{Energy spectrum of the Jaynes-Cummings Hamiltonian [Eq.~(\ref{eq:H_JC})] up to the two-excitation manifold. The energy eigenvalues are given by ${E_n^{\pm} = n\omega_c \pm \sqrt{n}g}$ where $n$ is the photon number, assuming that the emitter and cavity are in resonance (${\omega_e = \omega_c}$) and the coupling rate $g$ is real. The energy splitting between $E_n^+$ and $E_n^-$ is $2\sqrt{n}g$ for a fixed $n$.}
    \label{fig:JC_ladder}
\end{figure}
\begin{align}
\begin{split}
    H_{\text{cav}} =&\; H_{\text{JC}} + \int_0^{\infty} \omega(k) a_R^{\dagger}(k) a_R(k) dk\\
    &+ \int_{-\infty}^0 \omega(k) a_L^{\dagger}(k) a_L(k) dk\\
    &+ \int_0^{\infty} \left[ \frac{V_R}{\sqrt{2\pi}} a_R^{\dagger}(k) c + \frac{V^*_R}{\sqrt{2\pi}} a_R(k) c^{\dagger} \right]dk\\
    &+ \int_{-\infty}^0 \left[ \frac{V_L}{\sqrt{2\pi}} a_L^{\dagger}(k) c + \frac{V^*_L}{\sqrt{2\pi}} a_L(k) c^{\dagger} \right]dk,
\end{split}
\end{align}
where $V_R$ and $V_L$ are now the coupling rates between the cavity mode and the right- and left-moving waveguide modes, respectively (within the Markov approximation). We note that, for symmetric cavity-waveguide coupling, these can be calculated as ${V_\mu = \omega_c/2Q}$ for ${\mu \in \{L,R\}}$, where $Q$ is the quality factor of the coupled cavity. With this Hamiltonian, the transmission and reflection amplitudes for a right-moving input photon are modified from Eqs.~(\ref{eq:t_real_space}) and (\ref{eq:r_real_space}) to~\cite{Rephaeli2012, Duda2024}
\begin{equation}
    t_{\text{cav}} = \frac{\Delta_c - \frac{|g|^2}{\Delta_e} - \frac{i}{2}(|\Gamma_R| - |\Gamma_L|)}{\Delta_c - \frac{|g|^2}{\Delta_e} + \frac{i}{2}(|\Gamma_R| + |\Gamma_L|)}
\end{equation}
and
\begin{equation}\label{eq:r_cavity}
    r_{\text{cav}} = \frac{-i\sqrt{\Gamma_L \Gamma_R^*}}{\Delta_c - \frac{|g|^2}{\Delta_e} + \frac{i}{2}(|\Gamma_R| + |\Gamma_L|)},
\end{equation}
where $\Delta_e$ is the photon-emitter detuning, $\Delta_c$ is the photon-cavity detuning, and ${\Gamma_{\mu} = V_{\mu}^2/v_g}$ is the cavity decay rate in the $\mu$ direction in the waveguide. The emitter loss rate $\gamma$ and the cavity loss rate $\kappa$ can be included in these results using the substitutions ${\Delta_e \rightarrow \Delta_e + i\gamma/2}$ and ${\Delta_c \rightarrow \Delta_c + i\kappa/2}$~\cite{Rephaeli2013}. By adding the cavity, we have introduced two additional parameters that can be used to control photon routing---the cavity detuning $\Delta_c$, and the emitter-cavity coupling rate $g$.

Theoretical proposals of single-photon switches based on quantum emitters in cavities often exploit the strong coupling regime of cavity QED. In this regime, the coupling rate $g$ is larger than the emitter loss rate $\gamma$ and the cavity loss rate $\kappa$. Such coupling results in the formation of dressed states, called the Jaynes-Cummings ladder. The Jaynes-Cummings ladder consists of the energy levels of the Hamiltonian $H_{\text{JC}}$ in Eq.~(\ref{eq:H_JC}), which has eigenvalues ${E_n^{\pm} = n\omega_c \pm \sqrt{n}g}$ (assuming that $g$ is real and the emitter is on resonance with the cavity, i.e., ${\omega_e = \omega_c}$), where $n$ is a positive integer. The first few energy levels are shown in Fig.~\ref{fig:JC_ladder}, up to the two-excitation manifold (${n=2}$). The energy spectrum of the Jaynes-Cummings Hamiltonian is anharmonic, as $E_n^{\pm}$ is not a linear function of the photon number $n$. For a fixed $n$, there are two states with energy splitting ${E_n^+ - E_n^- = 2\sqrt{n}g}$. In particular, in the single-photon regime (${n=1}$), there is a doublet with an energy separation of $2g$, which is called vacuum Rabi splitting~\cite{Khitrova2006}.

\begin{figure}
    \centering
    \includegraphics[width=\linewidth]{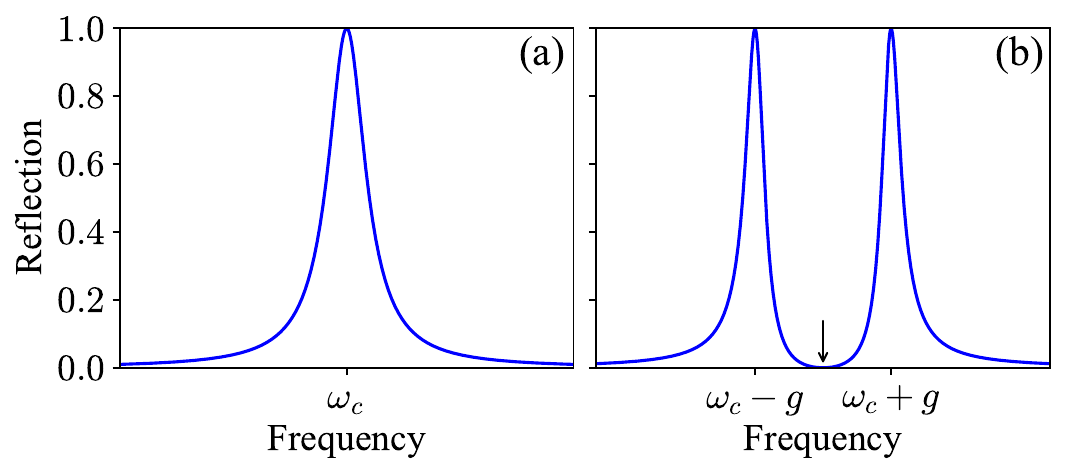}
    \caption{Single-photon switching with a two-level emitter in a waveguide-coupled cavity (Fig.~\ref{fig:two_level_emitter_cavity}). (a) and (b) show the single-photon reflection spectrum [calculated using Eq.~(\ref{eq:r_cavity})] in the weak (${g \ll \kappa, \gamma}$) and strong (${g \gg \kappa, \gamma}$) emitter-cavity coupling regimes, respectively. The transition from (a) to (b) results in Rabi splitting, which causes photons with frequency $\omega_c$ that were resonant with the cavity to be transmitted through the waveguide (the drop in reflection at this frequency is indicated by the arrow).}
    \label{fig:Rabi_splitting_switch}
\end{figure}

The rich energy level structure of the Jaynes-Cummings system arising from the emitter-cavity interaction enables a wider range of routing mechanisms. In particular, both the anharmonicity of the Jaynes-Cummings ladder and the Rabi splitting in the single-excitation manifold have been used for switching single photons. First consider the case of Rabi splitting, shown in Fig.~\ref{fig:Rabi_splitting_switch}. In the weak emitter-cavity coupling regime (${g \ll \kappa, \gamma}$), shown in Fig.~\ref{fig:Rabi_splitting_switch}(a), a reflection peak occurs on resonance with the cavity, similar to the case of a two-level emitter coupled to a waveguide (see Fig.~\ref{fig:R_T}). In the strong coupling regime (${g \gg \kappa, \gamma}$), shown in Fig.~\ref{fig:Rabi_splitting_switch}(b), the reflection peak splits into a doublet with resonance frequencies ${E_1^{\pm} = \omega_c \pm g}$, as the Rabi splitting becomes resolvable. As a result, the emitter-cavity system becomes transparent to photons at the bare cavity resonance frequency $\omega_c$ [as indicated by the arrow in Fig.~\ref{fig:Rabi_splitting_switch}(b)], due to the formation of the dressed states (which are also called polariton modes). Hence, photons at the frequency $\omega_c$ can be switched between the transmission and reflection paths by tuning the emitter-cavity coupling rate $g$ between the weak and strong coupling regimes. When the loss rates $\kappa$ and $\gamma$ become comparable to the cavity decay rates $\Gamma_\mu$ into the waveguide, the maximum reflection can drop significantly below one due to photon loss to other (non-guided) channels, similar to Fig.~\ref{fig:R_T_loss}.

Using Rabi splitting to route single photons has been proposed by various authors~\cite{Shen2005_2, Shen2009, Cheng2012, Zhou2013_2, Parkins2014, Xiong2015, Qin2016, Cao2017, Li2017, Dong2019, Li2023, Li2023_2, Duda2024, Berndsen2024, Lan2024, Wu2025}. One approach is to tune the frequency of the emitter to shift the two resonance peaks on and off resonance with incoming photons~\cite{Shen2005_2}. Another method involves controlling the splitting of the peaks either by adjusting the emitter-cavity detuning~\cite{Shen2009} or the emitter-cavity coupling rate $g$ as outlined above~\cite{Cheng2012, Zhou2013_2, Parkins2014, Xiong2015, Qin2016, Cao2017, Li2017, Duda2024}. Additional control over the Rabi splitting can be obtained by introducing time dependence to parameters such as the emitter-cavity coupling rate~\cite{Li2023, Li2023_2} or the transition frequency of the emitter~\cite{Lan2024}, as this provides more control parameters like the amplitude and frequency of the modulation. Furthermore, coupling the cavity at multiple points in the waveguide~\cite{Cheng2012} or coupling together multiple emitter-cavity systems~\cite{Li2017} introduces more photon pathways, the interference of which can be controlled with the extra coupling parameters. Using multiple separated emitters in cavities leads to more interference that determines the positions of the reflection peaks~\cite{Zhou2013_2} and increases the reflection bandwidth in the weak emitter-cavity coupling regime~\cite{Duda2024, Berndsen2024}, allowing photon wave packets to be routed with a higher efficiency. Multiple waveguide-coupled cavities can also increase the probability of routing to a different waveguide due to the interference~\cite{Dong2019}. In systems with multiple waveguides, using chiral/directional cavity-waveguide coupling allows for deterministic routing to a chosen output port by breaking the symmetry of the system~\cite{Lan2024, Li2023_2}. The coupling rates to different waveguides~\cite{Wu2025} and the phases of emitter-cavity coupling rates~\cite{Yang2018, Wang2022_2, Li2025} can also be tuned to control routing in multi-waveguide systems (these are alternative ways of controlling interference of different photon pathways). Note that cavities can either be side coupled or directly coupled to a waveguide, and the difference between the two cases is that the transmission and reflection spectra are swapped (e.g., a side-coupled cavity reflects resonant photons, whereas a directly-coupled cavity transmits on resonance)~\cite{Shen2009}.

\begin{figure}
    \centering
    \includegraphics[width=0.6\linewidth]{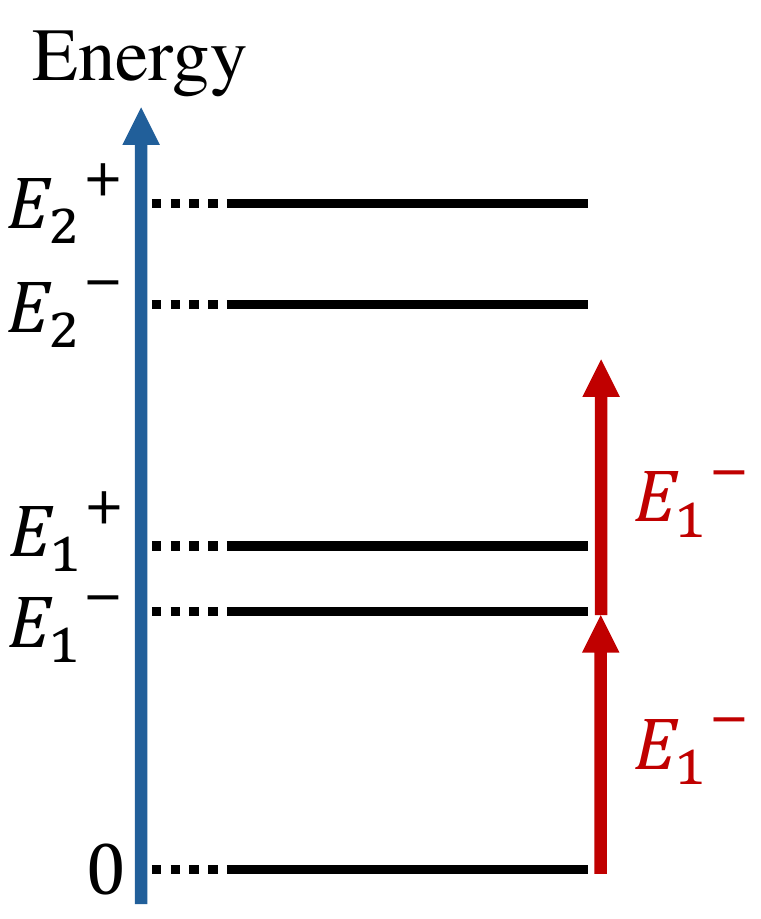}
    \caption{Photon blockade based on the Jaynes-Cummings system. Due to the unequal energy level spacing, the absorption of one photon that is resonant with the emitter-cavity system prohibits the absorption of a second photon with the same energy ($E_1^-$ in this example; see the red arrows). This means that one photon is reflected by the cavity, but a second photon is transmitted through the waveguide as it is detuned from the two-excitation states (${E_2^\pm \neq 2E_1^\pm}$).}
    \label{fig:JC_photon_blockade}
\end{figure}

In addition to Rabi splitting, the anharmonicity of the Jaynes-Cummings ladder enables routing via photon blockade, where the propagation direction of one photon is switched with a second control/gate photon~\cite{Shi2011, Rephaeli2012, Liu2014, Xu2016, Rinaldi2025}. The working principle of photon blockade based on the emitter-cavity system is illustrated in Fig.~\ref{fig:JC_photon_blockade}. A single photon with energy ${E_1^\pm = \omega_c \pm g}$ is reflected by the cavity because it is resonant with one of the states in the doublet, as shown in Fig.~\ref{fig:Rabi_splitting_switch}(b). However, a second photon with the same energy would not be resonant with another transition due to the unequal energy level spacing (${E_2^\pm  = 2\omega_c \pm \sqrt{2}g \neq 2E_1^\pm = 2\omega_c \pm 2g}$). Since the second photon is detuned from the emitter-cavity system, it would be transmitted through the waveguide without scattering from the cavity. The absorption of the first photon therefore blocks the reflection of the second photon, provided that the second photon reaches the cavity before the first photon leaves the cavity (otherwise the two photons would just be reflected sequentially). Alternatively, if each photon has energy $E_2^\pm/2$, then one photon would be detuned and two photons would be resonant with the emitter-cavity system, so one photon would serve to block the transmission of the initially-detuned second photon in this case.

In the large emitter-cavity detuning (dispersive coupling) regime, an effective Hamiltonian can be derived where the transition frequency of the emitter depends on the photon number in the cavity~\cite{Xu2016}. Therefore, it is also possible to realize photon blockade in this regime, where the presence or absence of a gate photon in the cavity determines whether the emitter is on or off resonance with incoming photons. Dipole induced transparency also enables switching in the weak emitter-cavity coupling regime, provided that a sufficiently large Purcell factor can be attained~\cite{Waks2006}. This transparency effect is a result of destructive interference of the cavity field, and does not require the emitter-cavity coupling rate to be larger than the loss rates in the system.

To make switches based on photon blockade more resource efficient, it may be desirable to retrieve the control photon for later switching. This can be achieved by using a second filter cavity that blocks the control photon~\cite{Rinaldi2025}, or using a second waveguide to separate the two photons~\cite{Xu2016}.


\subsection{Driving fields and electromagnetically induced transparency}\label{subsec:theory_EIT}

\begin{figure}
    \centering
    \includegraphics[width=\linewidth]{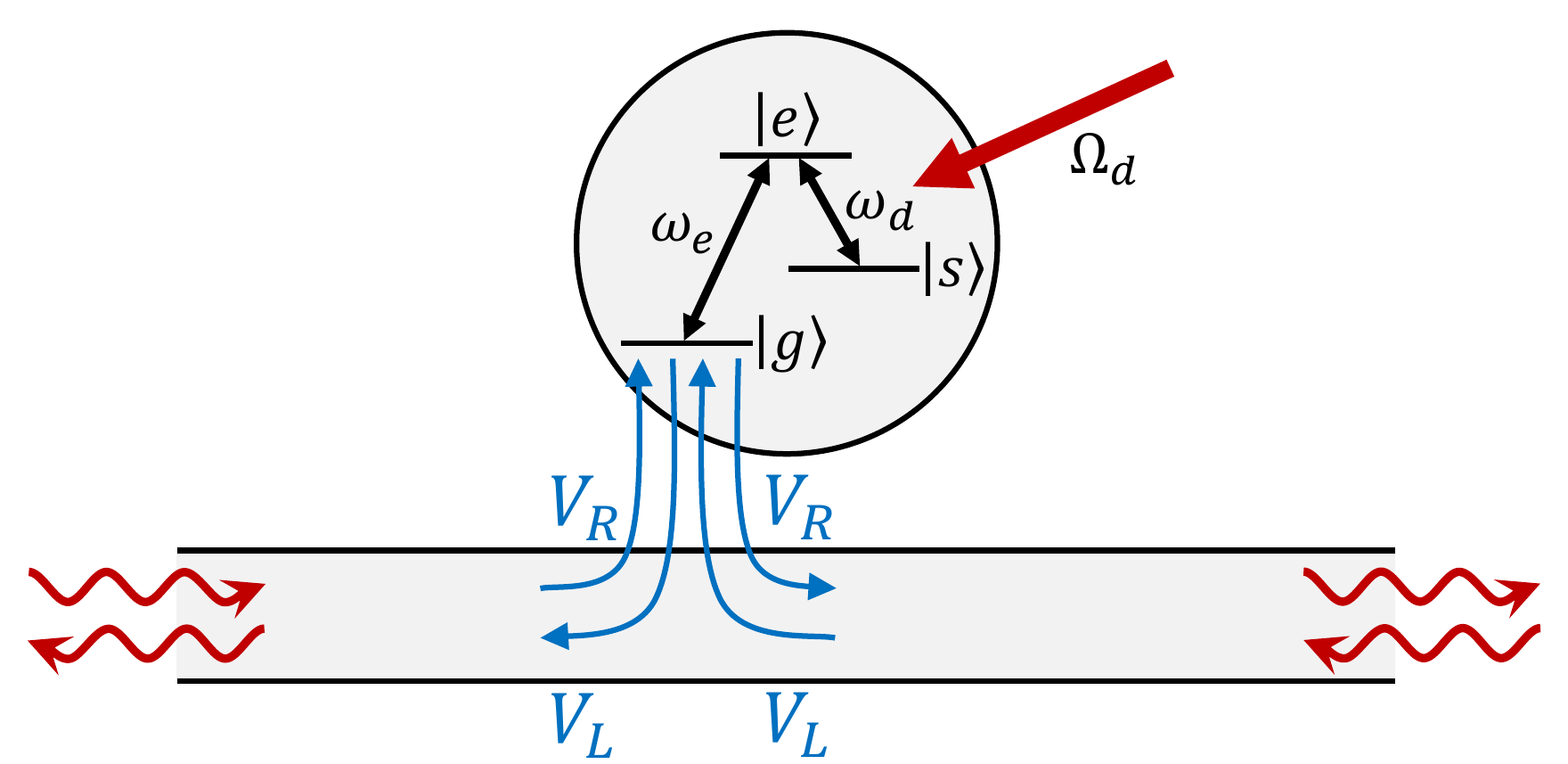}
    \caption{A three-level $\Lambda$-type emitter coupled to a waveguide. The ${\ket{g} \leftrightarrow \ket{e}}$ transition with frequency $\omega_e$ is coupled to the right- and left-moving waveguide modes with coupling rates $V_R$ and $V_L$, respectively. The ${\ket{s} \leftrightarrow \ket{e}}$ transition with frequency $\omega_d$ is driven by a laser field with Rabi frequency $\Omega_d$, and does not couple to the waveguide ($\ket{s}$ is a metastable state).}
    \label{fig:three_level_emitter_waveguide}
\end{figure}

There are other physical mechanisms that can produce a transparency window in a system that is otherwise opaque to incoming photons. A common approach that does not require the use of a cavity is to use a three-level atomic system driven by an external field (e.g., a laser). This can give rise to electromagnetically induced transparency (EIT), where different excitation pathways interfere destructively to produce a narrow transparency window within an absorption line of the atomic medium~\cite{Fleischhauer2005}. A simple example of a system that can undergo EIT, consisting of a three-level $\Lambda$-type emitter coupled to a waveguide, is shown in Fig.~\ref{fig:three_level_emitter_waveguide}. Compared to the two-level emitter in Figs.~\ref{fig:two_level_emitter_waveguide} and \ref{fig:two_level_emitter_cavity}, we have included a third, metastable atomic state $\ket{s}$. The ${\ket{s} \leftrightarrow \ket{e}}$ transition is driven by a field with Rabi frequency $\Omega_d$, and only the ${\ket{g} \leftrightarrow \ket{e}}$ transition couples to the waveguide. In the absence of the driving field (${\Omega_d = 0}$), this system is identical to the waveguide-coupled two-level emitter considered previously. However, in the driven three-level scheme (${\Omega_d\neq0}$), there are multiple pathways that can transfer the emitter to the excited state $\ket{e}$, including the direct ${\ket{g} \rightarrow \ket{e}}$ path and the indirect ${\ket{g} \rightarrow \ket{e} \rightarrow \ket{s} \rightarrow \ket{e}}$ path. These paths interfere destructively under resonant driving of the ${\ket{s} \leftrightarrow \ket{e}}$ transition, resulting in a transparency window for waveguide photons that are resonant with the ${\ket{g} \leftrightarrow \ket{e}}$ transition.

In the strong driving regime, where the Rabi frequency $\Omega_d$ is larger than the transition linewidth, the absorption spectrum splits into a doublet. This effect is called Autler-Townes splitting (or the AC Stark effect)~\cite{Autler1955, Fleischhauer2005}, and results from the formation of dressed emitter-field states due to the interaction between the emitter and the driving field. This is analogous to the formation of dressed states in the emitter-cavity system considered previously. However, in this case the splitting is induced by coupling to a driving field rather than a cavity mode. The reflection spectrum is similar to that in Fig.~\ref{fig:Rabi_splitting_switch}(b), except that here the separation between the two reflection peaks is given by $2\Omega_d$~\cite{Zhou2013, Lu2014}. Like EIT, Autler-Townes splitting produces a transparency window for photons in the waveguide that would otherwise be reflected in the absence of the driving field. In both cases, switching the driving field on/off enables switching between transmission and reflection.

There have been numerous theoretical proposals where single-photon switching is realized by controlling the strength of the driving field in a multi-level emitter scheme~\cite{Manzacca2007, Gong2008, Zhou2013, Hai2013, Lu2014, Yan2016, Sabet2017, Cheng2017_2, Yan2018, Yan2018_2, Huang2018_2, Huang2019_2, Ahumada2019, Aghamalyan2019, Wang2022_3, Yadav2025, Liu2026}. In systems with a single waveguide, changing the Rabi frequency of the driving laser from ${\Omega_d = 0}$ to ${\Omega_d \neq 0}$ opens a transparency window that switches photons from the reflection path to the transmission path, as mentioned above~\cite{Yan2018_2, Yadav2025}. In systems with multiple waveguides, the presence or absence of the driving field can be used to determine whether photons always propagate in the same waveguide or are transferred to another waveguide. For example, this can be achieved using a cyclic three-level scheme, where the driving field links two transitions that couple to separate waveguides~\cite{Zhou2013, Yan2018, Huang2019_2, Cai2024}. The routing probability to a chosen output port in another waveguide is limited when the emitter-waveguide coupling is symmetric, but it can be increased to one by breaking the symmetry with chiral/directional coupling~\cite{Cheng2017_2, Yan2018, Yan2022, Wang2022_3}. Alternatively, additional emitters acting as “atomic mirrors” can be used to direct photons to the desired output port~\cite{Huang2018_2} (see also Refs.~\cite{Zhou2008_2, Li2015, Huang2018, Li2024} for routing with atomic mirrors). As with the other routing mechanisms, using multiple separated emitters increases the switching bandwidth, allowing wider photon wave packets to be routed with a higher efficiency~\cite{Yan2018_2, Ahumada2019}.

Some of the early work on single-photon switching with driving fields involves the use of a gate photon to control the interference in EIT. In 1998, Harris and Yamamoto proposed that using a fourth atomic level coupled via another field, which can have the energy of a single photon, can completely destroy the quantum interference and make the atom opaque to probe photons~\cite{Harris1998}. In a later proposal, Bermel et al. showed that a detuned gate photon can induce a Stark shift of the EIT spectrum, which can be used for switching if the frequency shift is greater than the EIT linewidth~\cite{Bermel2006}. Here the emitter was coupled to a cavity, and the frequency shift could be increased by increasing the emitter-cavity coupling strength. In another cavity-based proposal, the Rabi splitting in the strong emitter-cavity coupling regime was switched off by applying a control field to detune the emitter from the cavity~\cite{Manzacca2007}, enabling switching between transmission and reflection as in Fig.~\ref{fig:Rabi_splitting_switch} (except that the driving field is tuned, not the coupling rate parameter $g$). Alternatively, switching in the weak emitter-cavity coupling regime can be realized by using a driven three-level emitter in a ring resonator, where the interference between counter-propagating modes in the resonator can control the routing~\cite{Aghamalyan2019}.

Other single-photon routers based on driven multi-level emitters introduce additional control parameters that can be tuned to control photon propagation. For example, the topology of a coupled-resonator waveguide can be used to induce a frequency shift in the transmission spectrum~\cite{Yadav2025}. In other work, an effective multi-level atom was created from two two-level emitters with a Rydberg interaction, and the Rydberg interaction strength serves as an extra parameter that can be tuned to control the interference between different photon pathways (e.g., by changing the emitter separation)~\cite{Du2021, Wang2025}.


\subsection{Emitter state switching}\label{subsec:theory_state_switching}

\begin{figure}
    \centering
    \includegraphics[width=\linewidth]{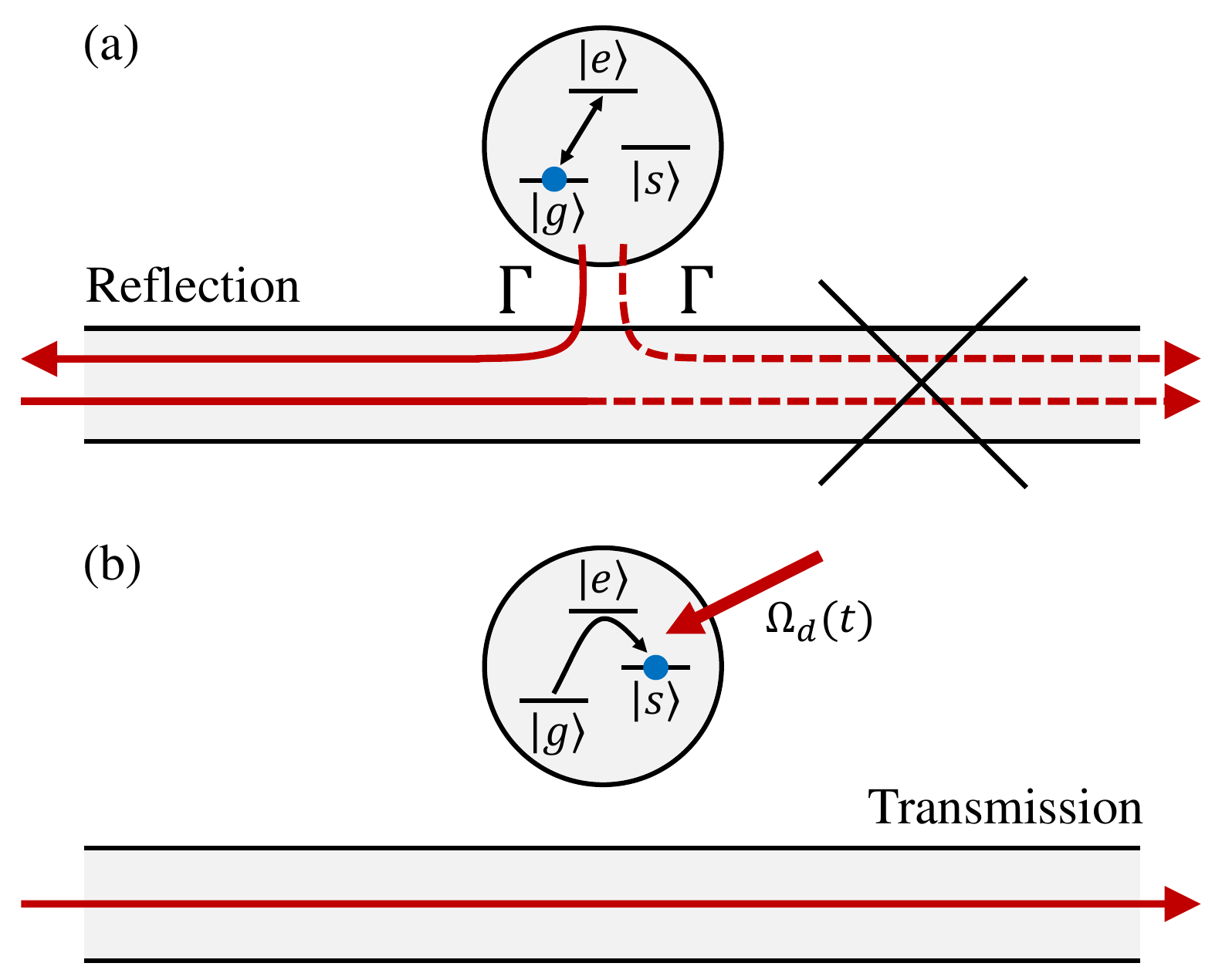}
    \caption{Single-photon switching by controlling the state of a three-level $\Lambda$-type emitter. In (a), the emitter is initially in state $\ket{g}$ and perfect reflection occurs for photons resonant with the ${\ket{g} \leftrightarrow \ket{e}}$ transition due to destructive interference in transmission (as in Fig.~\ref{fig:interference}). In (b), the emitter is prepared in the metastable state $\ket{s}$ via the absorption of a gate photon and a control pulse. The ${\ket{s} \leftrightarrow \ket{e}}$ transition is decoupled from the waveguide, so subsequent photons are not scattered and hence are transmitted perfectly.}
    \label{fig:emitter_state_switch}
\end{figure}

The final routing mechanism that we discuss in detail in this section is emitter state switching. In this case, the direction in which a single photon is scattered from an emitter depends on the quantum state of the emitter. A simple example of this mechanism is shown in Fig.~\ref{fig:emitter_state_switch}, which is based on the same system as in Fig.~\ref{fig:three_level_emitter_waveguide}. The three-level $\Lambda$-type emitter couples to the waveguide via the ${\ket{g} \leftrightarrow \ket{e}}$ transition, while the ${\ket{s} \leftrightarrow \ket{e}}$ transition is decoupled. If the emitter is prepared in the ground state $\ket{g}$ [Fig.~\ref{fig:emitter_state_switch}(a)], a single photon that is resonant with the ${\ket{g} \leftrightarrow \ket{e}}$ transition is perfectly reflected, due to destructive interference between the direct transmission path and the absorption-emission path (assuming an equal decay rate $\Gamma$ in both directions---this is the same interference mechanism as in Fig.~\ref{fig:interference}). Now consider the case where the emitter is prepared in the metastable state $\ket{s}$ by applying a control pulse with (time-dependent) Rabi frequency $\Omega_d(t)$, shown in Fig.~\ref{fig:emitter_state_switch}(b). The control pulse drives the emitter into the state $\ket{s}$ after it is excited into state $\ket{e}$ via the absorption of a photon from the waveguide, storing the atomic excitation in the metastable state. Since the ${\ket{s} \leftrightarrow \ket{e}}$ transition is decoupled from the waveguide, subsequent photons are transmitted without scattering from the emitter. Hence, photons are reflected if the emitter is prepared in the state $\ket{g}$, or transmitted if the emitter is prepared in the state $\ket{s}$.

The switching scheme outlined above was first proposed by Chang et al. in 2007~\cite{Chang2007}. It involves using a gate photon to excite the emitter, followed by a control pulse to store the gate photon by transferring the emitter to a metastable state. The arrival times of the gate photon and the control pulse must be precisely timed for the ${\ket{g} \rightarrow \ket{e} \rightarrow \ket{s}}$ process to be successful. Similar routing schemes based on switching the state of a quantum emitter with a gate photon were later developed~\cite{Witthaut2010, Neumeier2013, Yan2014_2, Manzoni2014, Kyriienko2016, Sala2016, Hu2016, Hu2017, Luo2023}. In 2010, Witthaut and S{\o}rensen proposed a scheme that does not require precise timing between the gate photon and the control pulse, but instead uses an adiabatic passage where the Rabi frequency $\Omega_d(t)$ is varied slowly~\cite{Witthaut2010}. In 2013, Neumeier et al. used two coupled two-level emitters to implement the switching, where the absorption of a gate photon by one emitter prevented the absorption of a target photon by the second emitter (causing it to be transmitted), due to the detuning between the target photon and the doubly-excited state~\cite{Neumeier2013}. In later work, a three-level ladder-type scheme was used where one transition couples to a waveguide and a second transition couples to a cavity mode~\cite{Yan2014_2, Luo2023}. In this case, a control pulse is not required, as the photon number in the cavity acts as the control parameter that determines the atomic state. In particular, the absence of a photon in the cavity means that the emitter remains in its two lowest energy levels and behaves like a two-level emitter that can perfectly reflect resonant photons. When the cavity contains a photon, the emitter can be excited to its third level through coupling to the cavity mode, after absorbing a gate photon. This energy level is decoupled from the waveguide, so subsequent waveguide photons would be transmitted without scattering from the emitter as long as it remains in this state.

Other emitter state switching schemes use some of the previously mentioned routing mechanisms and condition them on the state of the emitter. This includes state-dependent Rabi splitting, which is present or absent depending on whether the emitter is prepared in a state that couples strongly to the cavity~\cite{Stolyarov2020}, and state-dependent EIT, where the position of the EIT window is tuned using the state of a qubit~\cite{Xia2018}. There are also proposals where coupling between different cavities is mediated by a quantum emitter~\cite{Xia2013, Li2019}. This cavity-cavity interaction can be switched on or off depending on the state of the emitter, allowing photons to be routed into different waveguides that couple to the cavities. In the dispersive emitter-cavity coupling regime, a state-dependent shift of the cavity resonance frequency can be used to control photon routing~\cite{Cumbrado2025}.

In the context of switching with a gate photon, the term “single-photon transistor” is used~\cite{Chang2007, Neumeier2013, Manzoni2014, Kyriienko2016}, in analogy to an electronic transistor where electrical signals are controlled with a gate voltage. In addition to switching between different output directions, a transistor can amplify a signal passing through it. The gain of a single-photon transistor can be defined as the number of photons that can be switched with a single gate photon~\cite{Neumeier2013, Kyriienko2016}.


\subsection{Routing schemes without quantum emitters}

All the routing mechanisms above rely on controlling the properties of quantum emitters coupled to nanophotonic structures, such as the energy level structure and coupling parameters. We now briefly mention other routing mechanisms that do not involve quantum emitters but are still based on nanostructures. In particular, there is a number of photon routing proposals based on cavities without coupled emitters~\cite{Liao2009, Liao2010, Agarwal2012, Jiang2012, Yan2015, Li2016, Yan2016_2, Gao2018, Huang2018_3, Zhu2019, Zhu2019_2, Yan2021, Liu2021, Yang2021, Ren2022}. Some are based on optomechanically induced transparency, where the interaction between an optical cavity mode and a mechanical mode is controlled with a driving field that creates radiation pressure~\cite{Agarwal2012, Li2016, Gao2018, Yang2021}. This optomechanical coupling results in splitting of the cavity mode spectrum, creating a transparency window similar to EIT (here the mechanical mode replaces the emitter in EIT schemes)~\cite{Agarwal2012}. An optomechanical system consisting of a Bose-Einstein condensate in a cavity has also been considered~\cite{Chen2012}, where the collective oscillations of the condensate are used for switching instead of the motion of a mirror. Other cavity-based approaches, beyond optomechanical systems, require tuning of cavity frequencies~\cite{Liao2010, Yan2016_2} or coupling rates~\cite{Liao2009, Zhu2019, Yan2021, Liu2021, Ren2022}, or interference with additional control fields~\cite{Yan2015, Zhu2019_2}, to route photons to different output ports. In the case of ring resonators, backscattering between the counter-propagating resonator modes can be used to control routing~\cite{Huang2018_3}. In general, the absence of emitters means that the parameters of the nanostructure must be controlled to enable photon switching.

There are also other types of non-emitter-based switches that have been proposed in the literature. One example is cross-phase modulation in a ring resonator structure via the use of a pump beam that shifts the resonance frequency~\cite{Ruf2022}. Another example is linear-optical programmable interferometers, where a signal photon is routed to a chosen output port of an interferometer conditioned on the state of a control photon~\cite{Lemr2013, Lemr2013_2}. As mentioned in the Introduction, such classical switching architectures can route photons with high fidelity at the cost of a larger resource overhead compared to emitter-based switches, due to the probabilistic nature of routing with linear optics.

One advantage of single-photon switching without using emitters is that we avoid undesirable emitter properties that limit the switching performance, such as spontaneous emission loss and dephasing. These properties inevitably make emitter-based schemes probabilistic in practice. For example, photon loss due to spontaneous emission into non-guided modes leads to reduced transmission/reflection probabilities (see Fig.~\ref{fig:R_T_loss}), which limits the routing efficiency to below unity. However, the fast and versatile tunability of quantum emitters makes them very appealing switching elements, and producing high-quality quantum emitters with better optical properties is an active area of research in different physical platforms. Without emitters, the nanostructures in which photons propagate must be controlled instead (e.g., through cavity mode tuning~\cite{Fushman2007, Brunswick2025, Luxmoore2012}), which may lead to slower switching and can be more difficult to implement in a practical setting, depending on the material platform and control mechanism.


\subsection{Summary of theoretical proposals}

A variety of routing mechanisms have been proposed to control the photon propagation direction via scattering from quantum emitters coupled to nanophotonic structures. For emitters coupled to waveguides, the photon-emitter detuning or the emitter-waveguide coupling can be tuned to control the routing. Introducing interactions with cavities or driving fields can lead to dressing of the atomic states, which produces a transparency window that can be tuned with the cavity/field parameters. Using a gate photon that prepares the atomic state can enable state-dependent photon switching, where photons in a waveguide are reflected or transmitted depending on whether the emitter is prepared in a state that is coupled or decoupled from the waveguide. Common to all the routing mechanisms is the requirement to control the interference between different photon pathways using the available parameters.

In the vast majority of theoretical proposals, authors consider a single-photon input with a single frequency. In this case, the performance of the proposed switches is quantified by the efficiency of routing a photon with that frequency, which is simply given by the transmission and reflection probabilities as in Eqs.~(\ref{eq:E_t_one_frequency}) and (\ref{eq:E_r_one_frequency}). The photon frequency is typically chosen to coincide with a resonance in the system, for example, the transition frequency of an emitter or the resonance frequency of a cavity mode. The transmission/reflection at this frequency can usually be tuned from zero to one or vice-versa (or arbitrarily close to these values) by varying the relevant parameter(s), implying that the switching efficiency can be perfect for photons at this frequency (in the absence of photon loss, which clearly reduces the efficiency). A small number of proposals consider input photon wave packets with a finite spectral width, and evaluate the routing performance using equations similar to Eqs.~(\ref{eq:E_t})-(\ref{eq:F_r})~\cite{Duda2024, Witthaut2010, Gao2018, Zhu2019}. Some authors define other, related figures of merit that quantify the switching efficiency, such as an extinction ratio~\cite{Rinaldi2025}, a contrast between transmission/reflection probabilities for the two switch states~\cite{Neumeier2013, Xia2018, Stolyarov2020}, or an error probability associated with preparing the emitter state~\cite{Manzoni2014}. Generally, near-unity efficiencies and fidelities for switching photon wave packets are achievable within some parameter range, where the wave packet width fits within the bandwidth of the switch and losses are small compared to the relevant processes.

In the next section, we turn to the experimental demonstrations of single-photon switches. First, in Section~\ref{sec:FOM_experiment}, we discuss the figures of merit that are relevant in practical implementations, including the efficiency and fidelity, but also other factors like speed, operation time, scalability, and compatibility. In Section~\ref{sec:structures}, we provide an overview of the types of nanostructures that have been coupled with quantum emitters for photon switching. Then, in Section~\ref{sec:experiment}, we review experimental photon switching results across different physical implementations, including semiconductor QDs, neutral atoms, superconducting qubits, and solid-state defects. We also mention other platforms that have been considered for emitter-based single-photon switching, such as atomic ensembles and single organic molecules. For each platform, we discuss the routing mechanisms that have been used to control photon propagation.



\section{The ideal switch: Experimental figures of merit}\label{sec:FOM_experiment}

In this section we collect the experimental figures of merit that define an ideal single-photon switch. Building on Section~\ref{subsec:methods_5}, where efficiency and fidelity were defined in terms of scattering amplitudes, we translate the figures of merit into measurable quantities and place them alongside platform-dependent considerations: switching speed, operation time, scalability, and compatibility with existing photonic and electronic technologies. Together, these metrics provide a common basis for comparing implementations across platforms.


\subsection{Efficiency and fidelity}\label{subsec:FOM_experiment_1}

The efficiency of a single-photon switch is the probability that an input photon is routed to the intended output port after a switching operation~\cite{Duda2024}. Efficiency is a primary performance metric, directly reflecting how reliably the device performs routing in a quantum photonic circuit. Deviations from an efficiency of unity typically arise from inelastic loss into non-guided modes~\cite{Zhou2008}, imperfect emitter-waveguide coupling and mode mismatch~\cite{Daveau2017, Sun2018, Hoi2011, Giesz2016, Lenzini2016}, and rapid decoherence of the control degrees of freedom~\cite{Baur2014, Hu2017, Loo2012, Murray2016}. In practice, the transmission (reflection) efficiency is commonly inferred from the change in transmission (reflection) of a probe pulse between the “on” and “off” states~\cite{Bose2012}. However, many papers instead use an “extinction ratio” or “on/off ratio”~\cite{Sipahigil2016, Baur2014}, referring specifically to the reduction in transmission when the switch is operated. Extinction alone is only sufficient when transmission is the sole figure of interest; in general, transmission and reflection can change independently and loss may increase~\cite{Bose2012}. To demonstrate that the extinction of a probe pulse corresponds to a reflected signal rather than loss, the reflected and transmitted light can be measured simultaneously~\cite{Hoi2011}.

The fidelity of a single-photon switch quantifies how faithfully the switching process preserves the quantum state of the input photon~\cite{Northup2014, Duda2024}. The ideal fidelity of a switch is ${F=1}$, corresponding to perfect state preservation (see Section~\ref{subsec:methods_5}). High fidelity is essential for technological applications, since state changes translate directly into information loss. Fidelity can be obtained via quantum process tomography by comparing the measured process to the theoretical ideal. Because analyzing the output quantum state can be experimentally demanding, fidelity has been estimated from a calibrated error model~\cite{Wang2021} and may be bounded using interference measurements. Many experimental works do not explicitly quote fidelity, reflecting a focus on other switch characteristics or on demonstrating basic functionality.


\subsection{Speed and operation time}\label{subsec:FOM_experiment_2}

The switch speed quantifies how quickly the device responds to a change in control, e.g., how fast it can be toggled between transmitting and reflecting states. This directly sets the usable clock rate of a quantum photonic circuit and the tolerance to timing jitter and slow reconfiguration. The maximum speed depends on the physical switching mechanism. In cavity-based schemes, intrinsic timescales are set by the emitter-cavity coupling rate $g$ and the cavity linewidth $\kappa$~\cite{Volz2012, Bose2012}. In the strong coupling regime, switching events can occur on timescales comparable to $1/g$ in principle, typically in the range of tens of picoseconds for state-of-the-art nanophotonic cavities~\cite{Volz2012, Englund2012}. The switching speed also depends on the external control mechanism. For example, the speed of an optically-controlled switch would be limited by the duration of the pulse used for switching, e.g., sub-picosecond pulses can enable picosecond-scale switching. All-optical switching has the advantage of ultrafast operation and the potential for low energy consumption, which makes it especially appealing for scalable quantum technologies~\cite{Miller2010}.

In electrically-driven devices, the quantum confined Stark effect and electro-absorption provide efficient modulation at GHz frequencies~\cite{Liu2008, Hallett2018}, and lithographically-defined electrodes can support multi-GHz bandwidths. However, parasitic heating and crosstalk between densely packed tuning channels can reduce usable speed in practice. Strain-based actuators, while versatile for coarse tuning and stabilization, are limited to MHz operation due to mechanical inertia, making them less suitable for high-speed modulation.

In atomic and ensemble-based systems, EIT provides another route to photon switching, as mentioned in Section~\ref{subsec:theory_EIT}. EIT relies on coherent interference between excitation pathways, controlled by an optical or microwave field~\cite{Harris1990, Fleischhauer2005}. Here, the achievable speed is limited by the transparency window, which is set by the Rabi frequency and dephasing of the control field. Reported EIT switches typically operate on microsecond or longer timescales, though they offer high efficiencies and compatibility with long-lived memories~\cite{Baur2014, Tiarks2014, Gorniaczyk2014, Yu2020}.  

Finally, in superconducting microwave resonators, switching speeds are tied to qubit anharmonicity, resonator linewidth, and the strength of the applied control pulse. Fast operations on the order of a few nanoseconds are possible~\cite{Hoi2011}, although the millikelvin cryogenic environment imposes practical engineering constraints.

Beyond the switch speed set by the intrinsic switching mechanism and the external control method, an experimentally crucial, and sometimes implicit, speed limit is set by the single-photon pulse that is being switched. For a transform-limited wave packet, shortening the pulse in time necessarily broadens its spectrum. In switches based on scattering from quantum emitters, the relevant interaction bandwidth is set by the emitter radiative linewidth in the guided mode, or in cavity-assisted schemes, by the dressed resonance linewidth. Consequently, when the pulse bandwidth greatly exceeds this linewidth, a large fraction of spectral components are effectively off-resonant and do not scatter, reducing extinction and routing efficiency~\cite{Roy2017, Duda2024}. In the opposite limit of very long pulses, the response approaches the continuous-wave monochromatic spectrum that is often used as a proxy for performance~\cite{Roy2017}. In the intermediate regime where the pulse duration is comparable to the emitter lifetime, the frequency-dependent complex scattering amplitude across the pulse spectrum can lead to significant distortion of the input wave packet~\cite{Ralph2015, Pettersson2026}. The time-bandwidth trade-off is the physical origin of the speed versus efficiency/fidelity trade-off discussed in Section~\ref{subsec:methods_5} and quantified in Fig.~\ref{fig:E_F_vs_sigma}.

Another relevant parameter is the operation time (also called switch memory or switch lifetime): the time the switch can remain in a given state before it decoheres or must be reset. This is especially relevant to emitter state switching (Section~\ref{subsec:theory_state_switching}), where the emitter coherence time sets how long the switch can operate before repump/reset. The operating time can be limited by a range of system-dependent factors. For embedded solid-state emitters like QDs, decoherence can be caused by coupling to surrounding nuclear spins~\cite{Millington-Hotze2024} or phonon processes~\cite{Sipahigil2016}. For photon blockade effects, the operation time is limited to the time for which the blockade persists~\cite{Baur2014,Yu2020}. The optimal switch memory depends on the system in question---routing methods focusing on single switching events (such as all-optical switching with detuned pulses) or blockade effects may benefit from a shorter switch memory, so that the switch can be deterministically operated from its relaxed state for each event. For schemes that rely on continuous routing of a photon stream, longer switch memories are preferable to ensure consistent operation.

In summary, while physical switching mechanisms that have been studied suggest picosecond- to nanosecond-scale operation is possible in emitter-based switches, the achievable speed is also determined by the control mechanism---ultrafast (picosecond-scale) for optical schemes, hundreds of picoseconds to nanosecond-scale for electrical tuning, and much slower for ensemble-based EIT or strain control. Both the choice of the physical system and the external mechanism used to control it determine the maximum achievable switching speed, in addition to the duration of the photon wave packets being switched.


\subsection{Scalability and compatibility}\label{subsec:FOM_experiment_3}

Scaling emitter-based single-photon switches from one-off demonstrations to densely-integrated photonic circuits hinges on (i) device yield and emitter placement, (ii) spectral uniformity and tuning density, (iii) physical footprint and wiring budget, and (iv) packaging and thermal stability. Deterministic spatial alignment of quantum emitters in nanophotonic structures is a primary bottleneck. For self-assembled QDs, in situ optical or electron-beam registration enables the lithographic construction of cavities or waveguides around pre-measured QDs, increasing the usable device yield from a few percent towards reliable wafer-scale arrays~\cite{Pregnolato2019}. Site-controlled QD growth enhances placement determinism, while pick-and-place/transfer-printing strategies move pre-characterized emitters or membranes onto pre-fabricated photonic devices to relax growth constraints while preserving precise alignment~\cite{Schnauber2019, Kim2017TransferPrinting, Park2021Heterogeneous}. In diamond and silicon carbide, implanted color centers with wafer-scale nanofabrication offer foundry-like routes to large numbers of devices, though maintaining lifetime-limited linewidths uniformly across arrays remains challenging~\cite{Sipahigil2016, PRXQuantum.1.020102, Radulaski2019}. Neutral atom platforms face a different yield problem---probabilistic loading and subwavelength positioning relative to guided modes; progress includes trapping and interaction engineering near photonic crystal waveguides~\cite{Goban2014}, but deterministic per-site loading at scale is still under development. In superconducting circuits, artificial atoms are lithographically defined so spatial yield is high; scaling constraints shift to resonator/coupler crowding and readout fan-out.

Even with perfect emitter positioning, spectral inhomogeneity makes tuning of individual devices necessary, and this intersects directly with compatibility requirements on detectors and fiber infrastructure. Electrical Stark tuning~\cite{Hallett2018, Pedersen2020, Nowak2014}, local deformation~\cite{Martn-Snchez2017}, and microheaters~\cite{Yu2016} are standard; practical scaling requires hundreds to thousands of independent channels with low crosstalk and stable calibrations. Visible/near-infrared operation (${\sim 600}$-$900$~nm) aligns with high-efficiency silicon single-photon avalanche detectors and convenient free-space/short-link optics, while telecom-wavelength operation in the O-band ($1260$-$1360$~nm), C-band ($1530$-$1565$~nm), or L-band ($1565$-$1625$~nm) leverages low-loss single-mode fibers (SMF-28), dense wavelength division multiplexing (DWDM) components, and mature filtering. Regardless of the operation wavelength, switches should ideally preserve temporal, spectral, and polarization properties: low added timing jitter and path-length drift support the implementation of time-bin protocols, low back-reflection and background fluorescence keeps the signal-to-noise ratio small, and minimal polarization-dependent loss or birefringence protects polarization encoding. Practically, telecom photons often require indium gallium arsenide (InGaAs) avalanche photodiodes (APDs) or superconducting nanowire single-photon detectors (SNSPDs), so output power budgets, spectral filtering interfaces, and stable mode profiles should be engineered for fiber-coupled detectors, while dispersion and differential group delay remain controlled for interference-based protocols over intended link lengths and DWDM spacings.

As device arrays grow in size, operational uptime becomes increasingly limited by packaging rather than device physics. Three concerns dominate: reproducible fiber input-output interfaces, cryogenic-compatible electrical access (wire bonding or flip-chip bonding), and mechanical and thermal robustness. To sustain high rates of information transfer, the targets are low insertion loss (sub-decibel per switching event), low back-reflection, and low inter-channel crosstalk. For multi-port devices, robust fiber attachment using V-groove fiber arrays and strain relief simplifies alignment and qualification under temperature and humidity cycling for field deployment. Integrated diamond nanophotonics has already demonstrated single-emitter switches and memories in on-chip cavities, underscoring both the promise and the practical challenges, especially fiber coupling and thermal drift~\cite{Sipahigil2016}. Neutral atom platforms must additionally combine ultra-high-vacuum hardware and optical access with their nanophotonic structures, whereas superconducting routers and switches operate in dilution refrigerators with mature microwave packaging and have achieved multi-port, nanosecond-scale routing in the microwave domain, indicating clear near-term routes to larger on-chip switch networks~\cite{Hoi2011, Abdumalikov2010, Wang2021}. On the electronics side, using fabrication processes compatible with standard complementary metal-oxide-semiconductor (CMOS) manufacturing and integrating compound-semiconductor or atomically thin materials where needed allows the photonic chip to be built alongside its control electronics, improving reproducibility and lowering manufacturing costs~\cite{Hybrid}.

No single platform dominates across all performance metrics. Both QD and color center nanophotonics offer compact, low-loss on-chip routing with strong light-matter coupling, but typically require deterministic placement, post-growth spectral tuning, and careful thermal management to achieve uniform performance across arrays~\cite{Hou2025, Hallett2018, PRXQuantum.1.020102}. Neutral atom platforms provide excellent coherence and reconfigurability~\cite{Wintersperger2023, Menon2024}, yet face challenges in deterministic loading, subwavelength positioning relative to nanostructures, and integration with ultra-high-vacuum hardware~\cite{Wintersperger2023}. Superconducting circuits scale easily in the number of elements and control lines and can realize multiport routing with nanosecond control, but they operate in the microwave domain and require cryogenic environments at the millikelvin level~\cite{Levenson-Falk_2025}.


\section{Switch architectures}\label{sec:structures}

\begin{figure*}
    \centering
    \includegraphics[width=0.98\linewidth]{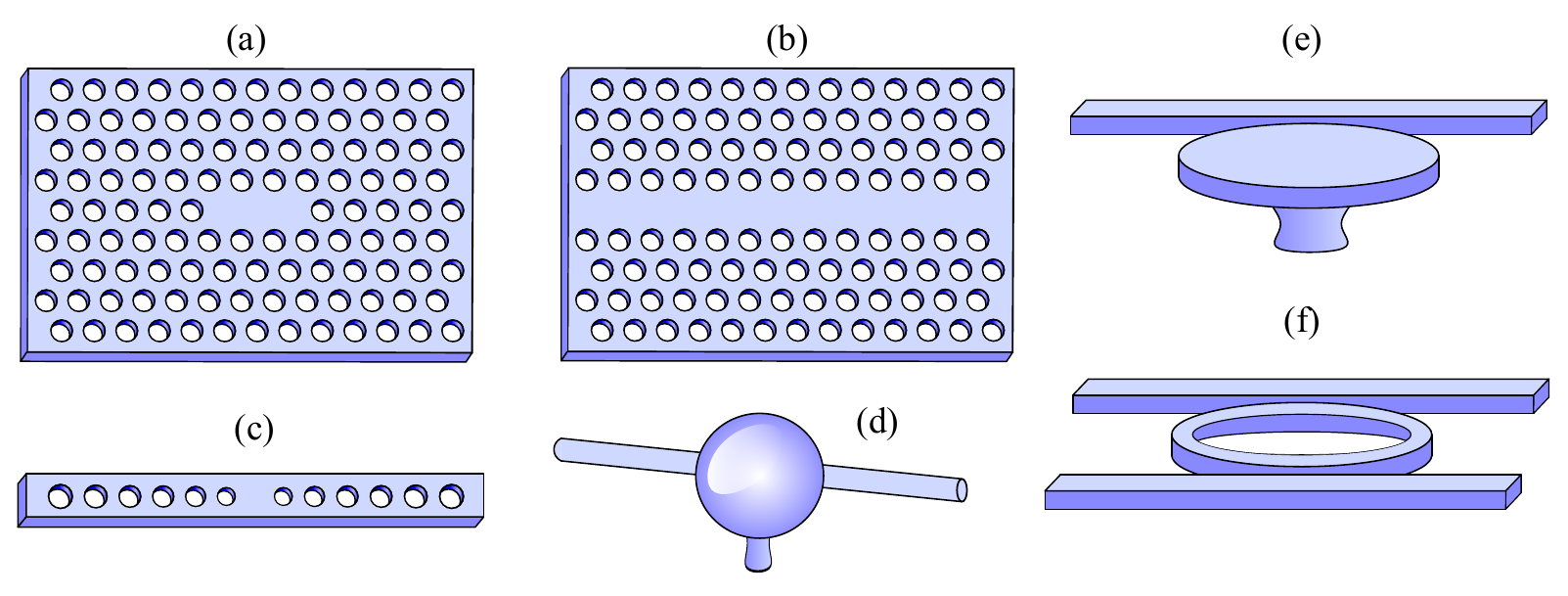}
    \caption{Examples of nanophotonic structures that can be coupled with quantum emitters for single-photon switching. (a) L3 photonic crystal cavity: three missing holes create an ultra-small-mode-volume defect. A single emitter in the cavity acts as a controllable scatterer, the properties of which can be tuned via, e.g., thermal, electrical, or strain control. (b) Photonic crystal waveguide: a line defect that can lead to slow light and a high $\beta$ factor (coupling efficiency), enabling near-deterministic emitter coupling and directional routing. (c) Nanobeam cavity: tapered Bragg mirrors lead to a high Q/V (quality factor to mode volume) ratio. An emitter positioned in the cavity enables few-photon switching. This cavity can be directly interfaced with a ridge waveguide. (d) Fiber-coupled microsphere: light propagating through a tapered fiber can evanescently excite ultra-high-Q whispering-gallery modes in the sphere. Coupling a controllable quantum emitter to the sphere can be used for single-emitter switching/routing with very low loss. (e) Microdisk: high-Q whispering-gallery modes evanescently couple to a nearby waveguide. An emitter near the rim of the disk can modulate transmission/reflection through the waveguide. (f) Planar ring resonator: the add-drop geometry defines multiple routing ports in the evanescently coupled waveguides. An emitter in the evanescent field or in the ring enables switching between the through and drop ports via thermo-optic or carrier tuning.}
    \label{fig:devices}
\end{figure*}

In Section~\ref{sec:FOM_experiment}, we outlined figures of merit for single-photon switches that are relevant in a practical setting. Achieving these in emitter-based switches requires platforms that confine light at the nanoscale and couple it strongly to a quantum emitter. Nanophotonic architectures provide this link from theory to hardware. In the following, we survey the main architectures for single-photon switching and relate their operating principles to these metrics, outlining how device-level design choices set system-level performance. The nanophotonic structures we discuss are shown schematically in Fig.~\ref{fig:devices}, and include photonic crystal structures (Section~\ref{subsec:photonic_crystals}), whispering-gallery-mode resonators (Section~\ref{subsec:rings}), and superconducting microwave resonators (Section~\ref{subsec:superconducting_resonators}).

  
\subsection{Photonic crystals}\label{subsec:photonic_crystals}

Photonic crystals are nanostructures with a periodically changing refractive index. This periodicity creates a photonic band gap, corresponding to a frequency range in which photon propagation is forbidden, directly analogous to the electronic band gap in semiconductors~\cite{Yablonovitch1987, Joannopoulos2008}. In practice, two-dimensional devices are formed by etching a lattice of holes into a thin membrane [e.g., silicon (Si), silicon nitride (Si$_3$N$_4$), gallium arsenide (GaAs), indium phosphide (InP), diamond, or silicon carbide (SiC)] and introducing local defects to engineer either tightly confined cavity modes (point defects) or waveguides (line defects). Crucially, these architectures are emitter-agnostic: they have been interfaced not only with QDs, but also with color centers in diamond and SiC, and rare-earth ions in crystalline hosts (including the telecom-band-compatible erbium ions Er$^{3+}$). In all cases, the photonic crystal provides the field confinement, mode control, and input-output engineering needed to realize strong light-matter interactions and few-photon nonlinearities.

Cavities arise by locally perturbing the lattice (e.g., removing or shifting a small number of holes) to move a defect mode into the band gap, preventing light from propagating into the periodic lattice and confining it to the defect. Standard designs include L3~\cite{Phillips2024,Faraon2010} and H1~\cite{Liu2018} defects in two-dimensional crystals and one-dimensional nanobeam cavities defined by tapered Bragg mirrors~\cite{Akahane2003, Deotare2009, Deotare2009_2,Brunswick2025,Zhong2024}. The L3 and nanobeam cavities are shown schematically in Fig.~\ref{fig:devices}(a) and Fig.~\ref{fig:devices}(c), respectively. Their hallmark is a combination of high quality (Q) factor and ultra-small mode volume (V), enabling large Purcell enhancement and access to the strong coupling regime of cavity QED with diverse emitters. Quantum emitters such as semiconductor QDs and diamond color centers have been coupled to photonic crystal cavities, acting as controllable reflectors/transmitters at the single-photon level~\cite{Sun2018, Sipahigil2016}. Spectral tuning to achieve the emitter-cavity resonance condition can be realized electrically~\cite{Faraon2010} (Stark shift), mechanically~\cite{Brunswick2025} (strain/piezo, cantilever), magnetically (Zeeman shift), or thermally~\cite{Phillips2024}, and in some material stacks via integrated electro-optic control, e.g., in lithium niobate (LiNbO$_3$), all of which are compatible with the photonic crystal cavity geometry.

Photonic crystal waveguides, on the other hand, are formed by removing one or more rows of holes to create a line defect that supports guided modes within the band gap. The simplest design, the W1 waveguide, provides efficient emitter-waveguide coupling, corresponding to a high $\beta$ factor~\cite{Lund-Hansen2008, Arcari2014} [see Fig.~\ref{fig:devices}(b)]. Glide-plane waveguides, formed by shifting one side of a W1 waveguide by half a lattice constant, suppress unwanted modes and provide a flatter, broadband dispersion. As a result, they are among the most promising designs for realizing directional coupling of single photons~\cite{Söllner2015, Siampour2023}. More recently, topological waveguides, such as valley-Hall waveguides, have been proposed as a route to switches that can be embedded in complex circuits, since their edge modes can, in principle, propagate robustly around sharp bends~\cite{Suarez-Forero2025, Shalaev2019, Add_drop, Yamaguchi_2019, Barik_2020, taper}. However, comparative measurements show that, for embedded QDs, topological (valley-Hall) interfaces currently yield lower directional contrast and $\beta$ factors than optimized glide-plane waveguides~\cite{PhysRevResearch.6.L022065, Hallacy2025}. Spin-Hall waveguides, which exploit the spin-orbit coupling of light to route orthogonal circular polarizations in opposite directions, have also been investigated~\cite{Barik2018}, although their edge modes often lie above the light line, restricting efficient emitter coupling and low-loss guiding~\cite{PhysRevResearch.6.L022065}.


\subsection{Ring, disk, and toroidal resonators}\label{subsec:rings}

Whispering-gallery resonators such as ring, disk, and toroidal resonators possess whispering-gallery modes (WGMs), which are optical modes that propagate around concave surfaces via continuous total internal reflection. They are characterized by their extremely high Q factors and small mode volumes, making them desirable for coupling to single-photon emitters. The circulating fields have a wavelength $\lambda$ that satisfies ${m\lambda = 2\pi R n_{\rm eff}}$, which corresponds to discrete resonances separated by ${{\rm FSR}\approx c/(2\pi R n_g)}$ (the free spectral range), where $R$ is the radius of the ring/disk, $n_{\rm eff}$ is the effective refractive index within the resonator, $n_g$ is the group index, $c$ is the vacuum speed of light, and $m$ is an integer. WGM microresonators commonly consist of raised microdisks [Fig.~\ref{fig:devices}(e)] or etched microrings [Fig.~\ref{fig:devices}(f)] in close proximity to nanofibers, to which their modes are evanescently coupled. The external coupling and loaded Q factor are determined by the resonator-waveguide gap and the coupling length. Many studies realize WGM resonators via microtoroid or microsphere cavities [Fig.~\ref{fig:devices}(d)], which have the highest Q factors~\cite{Yoshiki2014} and fiber-coupling efficiencies~\cite{Bai2018}, respectively. These resonators can couple cavity modes to high-Q-factor, low-mode-volume WGMs~\cite{Vernooy1998, Armani2003}. For emitter-based switching in such resonators, the relevant factors are Q/V (Purcell enhancement) and the cooperativity ${C=4g^2/(\kappa\gamma)}$ (defined in terms of the emitter-resonator coupling rate $g$, the intrinsic resonator loss rate $\kappa$, and the emitter loss rate $\gamma$), as well as interference effects such as dipole induced transparency/absorption that can strongly reshape the through and drop port spectra~\cite{Purcell1946, Waks2006}.

Tapered nanofibers are a useful tool for coupling light into microresonators, due to the high coupling efficiencies they provide~\cite{Knight1997, Cai2000, Spillane2003}. These tapers are typically cylindrical silica fibers with diameters of $1$-$4$~$\upmu$m, fabricated by heating and stretching a length of a single-mode fiber to create a narrow, tapered section~\cite{Knight1997}. The tapered fiber becomes a multi-mode fiber, allowing for improved coupling to the resonator via creation of a specific fundamental mode. By adjusting the taper size such that the propagation constant of the resonator mode is matched to the propagation constant of an appropriate fiber mode, in-coupling from the fiber to the resonator mode is significantly enhanced~\cite{Knight1997}. This has made tapered fibers a popular method for coupling to microresonator devices. However, such taper-resonator systems can suffer from losses as a result of the resonator transmitting light into radiation modes (due to the fiber-resonator interface) and unused higher-order taper modes~\cite{Spillane2003}.

In an overcoupled microtoroidal resonator, a single atom can act as a single-photon router, preferentially reflecting single-photon components while transmitting multiphoton components~\cite{Aoki2009}. Using a two-fiber add-drop bottle resonator, O’Shea et al. realized a fiber-optic switch controlled by one atom in the strong coupling regime with clear port-to-port routing~\cite{OShea2013}. A microsphere implementation achieved an all-optical, single-photon-activated switch, where a single reflected control photon toggled the device between high-reflection (${\sim65\%}$) and high-transmission (${\sim90\%}$) states~\cite{Shomroni2014}. Previously, a microtoroid “photon turnstile” established photon-number-dependent transport regulated by one atom~\cite{Dayan2008}. Efficient fiber-taper excitation of WGMs underpins these platforms~\cite{Knight1997, Cai2000, Spillane2003}. These ultra-high-Q WGM results establish the core, loss-averse, emitter-mediated input-output physics that integrated devices aim to reproduce.

While planar rings/disks have matured in terms of fabrication and on-chip coupling~\cite{Armani2003, Barclay2006, Manipatruni2007}, an unambiguous, low-loss, on-chip single-photon switch using a planar microring/microdisk and a single embedded emitter has not yet been reported to our knowledge. Instead, the enabling pieces exist separately: high-Q microring/disk structures with input/output ports, solid-state emitters with high Purcell and $\beta$ factors in other cavity types, and fast cavity or emitter tuning. Practical hurdles include sidewall-roughness backscattering between the clockwise- and counter-clockwise-propagating modes, modal crowding, and thermal crosstalk in dense layouts. Nevertheless, microrings offer attractive controls for a future integrated switch: thermo-optic heaters~\cite{Yu2016}, carrier injection/depletion in semiconductors~\cite{Manipatruni2007}, and electro-optic platforms (e.g., LiNbO$_3$) enable cavity tuning on nanosecond-to-microsecond scales. Bridging the remaining gap---achieving a large cooperativity with low parasitic loss and stable mode purity in a planar WGM resonator---is a key target for integrated quantum photonics.


\subsection{Superconducting microwave resonators}\label{subsec:superconducting_resonators}

Superconducting microwave resonators are the workhorse of circuit QED, where superconducting qubits interact coherently with confined microwave photons~\cite{Blais2020}. They are implemented as coplanar waveguides on a chip or as three-dimensional cavities, typically fabricated from aluminum or niobium and operated well below the superconducting transition temperature~\cite{Yu2021, Zikiy2023, Krasnok2024}. Their very low intrinsic loss supports high Q factors, which translate into long photon lifetimes and strong light-matter coupling~\cite{Reagor2013, Ye2025}.

For single-photon switching, these resonators play the role that optical cavities play in nanophotonics~\cite{Blais2004}. An embedded superconducting qubit provides a controllable two-level system that mediates nonlinear interactions between microwave photons, allowing switching and routing at the level of individual quanta~\cite{Wang2022, Ye2025}. The microsecond coherence times of state-of-the-art qubits support programmable control sequences and memory functionalities entirely in the microwave domain~\cite{Liu2023, Ganjam2024}. Combining these resonators with qubits allows for multi-component integration, which is a major strength of this platform. Lithographic processes allow many resonators and qubits to be fabricated on the same chip, with coupling strengths set by geometry, either capacitive or inductive~\cite{Blais2004, Poorgholam-Khanjari2025}. Resonators can be frequency multiplexed so that multiple modes share common wiring~\cite{Bakr2025}, and tunable elements such as superconducting quantum interference devices (SQUIDs) make it possible to adjust resonance frequencies and couplings in situ~\cite{Hoi2011}. These features make superconducting resonators a natural architecture for scaling networks of single-photon switches and routers.

There are practical constraints that differ from those in optical implementations. Operation requires dilution refrigerators at millikelvin temperatures, along with cryogenic amplification and careful microwave filtering to suppress thermal photons~\cite{Reed2010, Yeh2017, Yan2018_super}. Microwave photons are not suitable for long-distance transmission in standard media, which limits direct networking beyond the cryostat~\cite{Arnold2025}. As a result, hybrid approaches that couple superconducting circuits to optical transducers are being developed to combine strong microwave nonlinearities with low-loss transmission and the infrastructure developed around optical photons~\cite{vanThiel2025, Arnold2025}.


\section{Experimental demonstrations of single-photon switches by emitter platform}\label{sec:experiment}

\begin{figure*}
    \centering
    \includegraphics[width=0.85\linewidth]{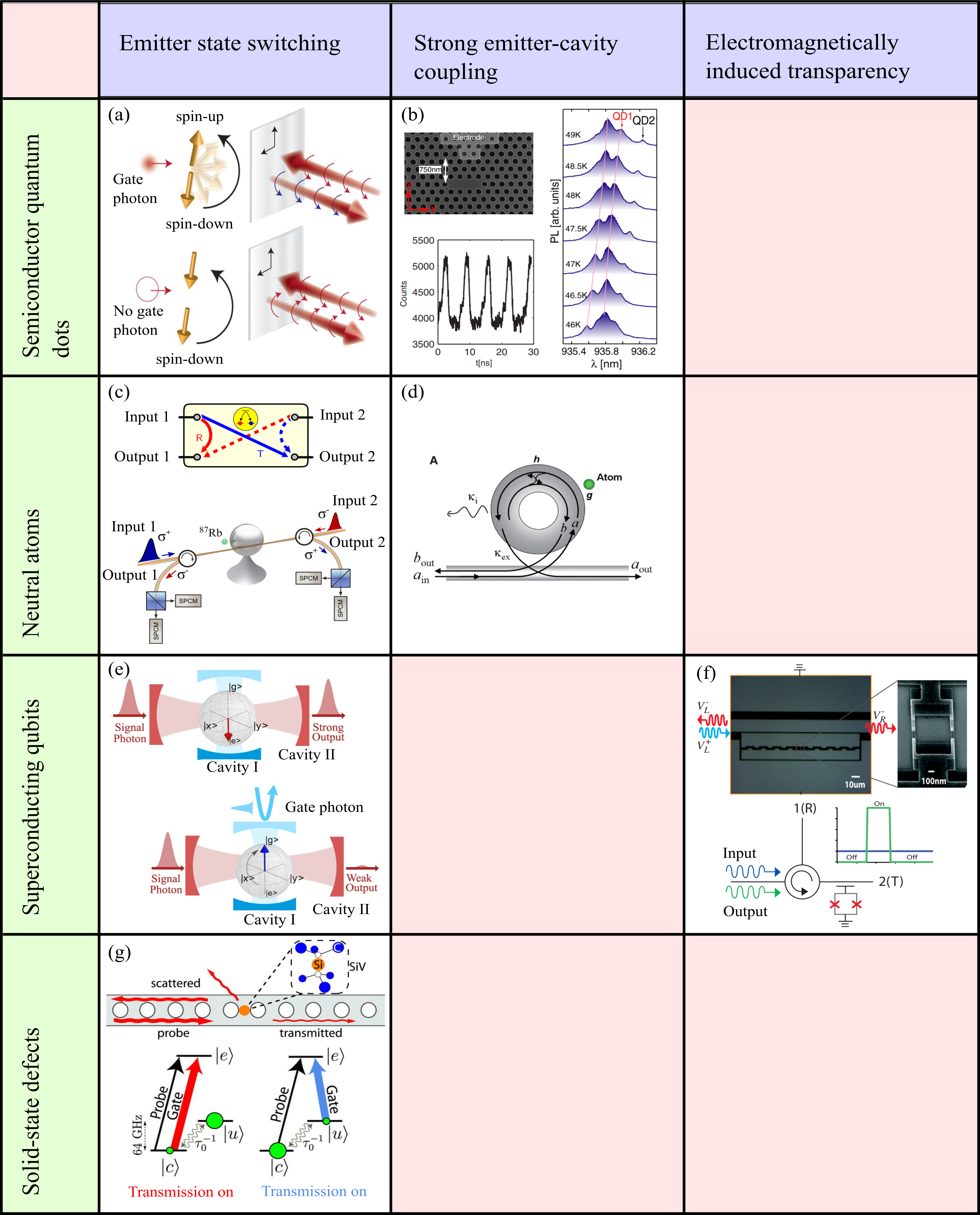}
    \caption{Experimental demonstrations of single-photon switches across physical platforms and operating mechanisms. The rows correspond to different emitter platforms: semiconductor QDs, neutral atoms, superconducting qubits, and solid-state defects. The columns group together different switching mechanisms: emitter state switching, strong emitter-cavity coupling, and EIT. Each panel illustrates a representative experiment. (a) Reproduced with permission from Science 361, 57 (2018). Copyright 2018 American Association for the Advancement of Science.~\cite{Sun2018} (b) Reproduced with permission from Phys. Rev. Lett. 104, 047402 (2010). Copyright 2010 American Physical Society.~\cite{Faraon2010}. (c) Reproduced with permission from Science 345, 903 (2014). Copyright 2014 American Association for the Advancement of Science.~\cite{Shomroni2014} (d) Reproduced with permission from Science 319, 1062 (2008). Copyright 2008 American Association for the Advancement of Science.~\cite{Dayan2008} (e) Wang et al., Nat. Commun., 13, 6104, 2022; licensed under a Creative Commons Attribution (CC BY) license.~\cite{Wang2022} (f) Reproduced with permission from Phys. Rev. Lett. 107, 073601 (2011). Copyright 2011 American Physical Society.~\cite{Hoi2011} (g) Reproduced with permission from Science 354, 847 (2016). Copyright 2016 American Association for the Advancement of Science.~\cite{Sipahigil2016}}
    \label{fig:switch_table}
\end{figure*}

Having discussed various solid-state nanostructures that can be coupled to quantum emitters in Section~\ref{sec:structures}, we now review the experimental demonstrations of single-photon switches. We organize the experimental results by the emitter platform, starting with semiconductor QDs (Section~\ref{subsec:qdots}), then moving to neutral atoms (Section~\ref{subsec:neutral_atoms}), superconducting qubits (Section~\ref{subsec:superconducting_qubits}), and solid-state defects (Section~\ref{subsec:solidstatedefects}). Representative experiments for each platform are shown in Fig.~\ref{fig:switch_table}, alongside the switching mechanism that was used in each case. We also provide a timeline of key experimental demonstrations in Fig.~\ref{fig:timeline}. At the end of this section, we briefly discuss single-photon switching in other platforms, including atomic ensembles and single organic molecules (Section~\ref{subsec:atomic_ensembles}).

\subsection{Semiconductor quantum dots}\label{subsec:qdots}

In recent decades, III-V semiconductor QDs have served as a versatile platform for demonstrating single-photon switches and other nanophotonic devices. Among these, GaAs/InAs systems in particular have underpinned many of the most advanced demonstrations of all-optical switching, and the broader field of QD photonics has reached a high level of technological maturity. QDs are often described as “artificial atoms”, and are nanometer-scale regions of low-band-gap semiconductor embedded within a higher-band-gap host. This band gap offset creates a three-dimensional potential well, confining electrons and holes into discrete excitonic states. As a result, a QD behaves like a quantum box that can be optically driven, emits single photons at discrete wavelengths, and can be integrated with nanophotonic structures such as cavities, waveguides, or doped heterostructures within the host wafer~\cite{Somaschi2016, Mller2018}. These atom-like properties make QDs one of the most promising solid-state platforms for integrated quantum photonic circuits.

QDs can be produced via Stranski–Krastanov (strain-driven) self-assembly~\cite{Ledentsov1996}, droplet epitaxy~\cite{Sala2020}, site-controlled/templated nucleation~\cite{Hou2025} (e.g., nanohole or prepatterned substrates), nanowire heterostructures grown by vapor–liquid–solid methods~\cite{Reimer2012} (axial or radial QDs), and interface-fluctuation~\cite{Zrenner1994, Gammon1996} or lithographically defined quantum-well islands~\cite{Bryant1996, Patel2010}, typically using molecular beam epitaxy (MBE) or metalorganic vapor phase epitaxy (MOVPE) depending on the material platform. In each case, epitaxial growth proceeds layer by layer, with careful control over atomic composition and strain, to engineer nanostructures that confine carriers within the desired potential wells~\cite{Arakawa2020}. Most research to date has centered on GaAs or InAs QDs, which combine several desirable properties for quantum light sources: high single-photon purity~\cite{Liu2018, Tomm2021}, long spin coherence times~\cite{Zhai2020}, and narrow optical linewidths that enable highly indistinguishable photon emission~\cite{Zhai2022}. Although GaAs/InAs systems remain the most widely studied, other material platforms are increasingly being explored to extend the spectral range and functional versatility of QDs. Strain-free QDs, for example, can be fabricated by embedding lattice-matched materials such as GaAs/AlGaAs~\cite{Zhai2020}, avoiding the inhomogeneous strain fields present in self-assembled systems~\cite{Ledentsov1996}. These dots benefit from improved control over position and size~\cite{Heyn2009}, making them attractive for scalable integration with photonic devices. In parallel, InP-based QDs, particularly InAs/InP, have emerged as a leading candidate for emission at telecom wavelengths~\cite{Cao2019}, where fiber transmission losses are minimal. InAs/InP dots combine the desirable single-photon properties of GaAs-based systems with spectral compatibility for quantum communication, and recent advances in growth and device engineering have demonstrated high-purity~\cite{Nawrath2019, Ha2020}, indistinguishable photon emission~\cite{Joos2024, Holewa2024} in this material platform. Together, these alternative systems expand the QD toolbox beyond GaAs/InAs, opening opportunities for new device architectures and applications ranging from quantum repeaters to integrated quantum networks.

In QD-cavity devices, strong emitter-cavity coupling can be used to realize efficient switching between different propagating directions (see Section~\ref{subsec:theory_cavity}). The quantum confined Stark effect is an established method for tuning QD properties, such as fine structure splitting~\cite{Bennett2010} and charge states~\cite{Martin2025_chargecontrol}, making it an attractive method for tuning QDs for single-photon switching applications. In 2010, Faraon et al. reported a QD switch based on the quantum confined Stark effect~\cite{Faraon2010} [Fig.~\ref{fig:switch_table}(b)]. They applied an electric field to an InAs QD in a photonic crystal cavity to shift the transition frequency of the dot such that it became detuned from the incident light, and thus the transmission could be controlled. They achieved a speed of ${\sim7}$~ns, limited by the RC constant of the transmission line, with an on/off ratio of $1.3$, limited by QD blinking and decoherence.  

All-optical cavity QED switching was performed in 2012 by Volz et al. and Englund et al. by the use of detuned laser pulses interacting with a strongly coupled QD-cavity system~\cite{Volz2012, Englund2012}. By utilizing a control (signal) pulse resonant with a lower (upper) transition of the Jaynes-Cummings ladder, the authors were able to control the transmission/scattering of the signal pulse with the presence or absence of the control pulse. The groups found switching speeds of approximately $50$~ps and $80$~ps, respectively. A more recent study by Mu\~{n}oz-Matutano et al. in 2020 demonstrated all-optical switching without the use of a cavity structure~\cite{Munoz-Matutano2020}. This was achieved by occupying the higher excitonic energy levels of the QD with a laser pulse, which quenches the recombination of the lower energy levels. This creates a single-photon stream that can be switched on/off when saturated by a control pulse. The authors reported a single-color exciton switching contrast of ${\sim 70\%}$ and a switching speed of ${\sim 1}$~ns. Faster switching on the order of $100$~ps is possible with lower excitation powers, at the cost of a reduced switching contrast.

An example of waveguide-cavity QED switching involving detuned laser pulses is the experiment of Bose et al. from 2012~\cite{Bose2012}. The system in this study was an emitter-cavity system coupled to a waveguide. The switch was controlled by the detuning between a probe pulse (which resonantly excites the QD) and a control pulse, which modifies the QD-cavity coupling. The input pulses and the output signal were delivered/extracted through the waveguide, allowing for on-chip switching. This method achieved a switching speed of ${\sim120}$~ps, limited by the pulse durations (and ultimately the QD-cavity coupling rate). The switching contrast was found to be $44\%$ for the scattered light and $13\%$ for the transmitted light, with the latter being lower due to the imperfect transmission contrast of the bare cavity. This method enables switching by as low as $140$ photons per pulse, although this number is limited by imperfect cavity-waveguide coupling. In a more recent study, Brooks et al. demonstrated photon routing using a QD embedded in a WGM microdisk cavity~\cite{Brooks2021}. The microdisk was coupled to two nanobeam waveguides that acted as bus and drop channels. By measuring the intensity at the bus and drop ports as a function of the laser-cavity detuning, a clear increase in scattering into the drop port could be seen when the laser was tuned onto the cavity mode, showing strong light-matter interactions between the QD and the incident photons. The authors describe two methods of routing control---varying the intensity of the incident light, and controlling the QD-cavity detuning by changing the sample temperature. For this system, the authors reported a routing efficiency of ${43 \pm 4\%}$, limited by suboptimal QD positioning. The authors predict that deterministic dot placement and electrical contacting of the sample would significantly increase this value.

So far, a high number of QD photon switching studies have used optical control pulses that modify the QD-cavity coupling. These studies demonstrate ultrafast switching speeds that are beneficial for quantum information processing applications, a key advantage of all-optical switching. These results underscore the viability of QD-cavity systems for high-speed photon switching, although the switches demonstrated in these studies required relatively high photon numbers to operate the switch.

In addition to strong QD-cavity coupling, spin-dependent switching with QDs has also been demonstrated (see Section~\ref{subsec:theory_state_switching} for a discussion of emitter state switching). Interfacing photons and QD spins is vital for the development of quantum repeaters and other quantum photonic devices~\cite{Carter2013}, and as such, spin-based photon switches will be easy to integrate with other quantum photonic nanotechnologies and can enable spin-photon entanglement~\cite{Laccotripes2024}. In 2018, Sun et al. demonstrated a spin-dependent switch based on a QD in a nanocavity, using the QD spin as a quantum memory~\cite{Sun2018} [Fig.~\ref{fig:switch_table}(a)]. A key result of this paper is that the switching of the spin state was achieved with a single-photon-level pulse, an important goal in developing a single-photon switch. The gate pulse duration was ${\sim 63}$~ps, and the number of photons per pulse that was coupled to the cavity was estimated to be $0.21$. However, the reported efficiency is $24\%$, limited by the spin preparation fidelity and finite cooperativity of the QD-cavity system. The authors state that the spin preparation could be improved by electrical gating and actively charged structures being integrated into their device. Higher cooperativities can be reached with higher-Q cavities. Also in 2018, Javadi et al. performed spin-dependent switching with a QD embedded in a nanophotonic waveguide~\cite{Javadi2018}. Although they report a spin preparation fidelity of up to $96\%$, the QD only blocks about $12\%$ of the probe light in the “off” state. The inhomogeneous broadening of the QD transition and the interaction between the electron spin and the nuclear spin ensemble are limiting factors in this case~\cite{Millington-Hotze2024}.

Spin-based photon switching has achieved control with very low photon numbers, in some cases requiring only a single gate photon to operate the switch. This provides a gateway to ultrafast all-optical switching, which leverages light-matter interactions between spin and photon states, and therefore a pathway towards scalable quantum optical circuits, as well as the operation of a switch in superposition and spin-photon entanglement. However, the studies of spin-based photon switches have reported very low efficiencies in contrast to other methods. This can be caused by a range of factors: electron spin dephasing due to interactions with the QD nuclear spin bath, spectral wandering from environmental charge noise~\cite{Liu2018}, and poor spin preparation. To further develop spin-photon systems for switching, these issues must be mitigated. Electrically gating devices provides the ability to tune the QD emission and thus reduce spectral wandering and the impact of charge noise~\cite{Kuhlmann2013}. To improve electron spin coherence, spin echo-based pulse sequences~\cite{Aebischer2024} and strain-based solutions~\cite{Millington-Hotze2024} can be used to reduce fluctuations of the magnetic field generated by the nuclear spins in the QD environment. These methods provide potential pathways for the improvement of spin-photon switches in future studies.


\subsection{Neutral atoms}\label{subsec:neutral_atoms}

Neutral atoms coupled to nanophotonic structures are another platform that has been used for single-photon switching and routing, with implementations based on whispering-gallery microresonators, photonic crystal cavities and waveguides, and tapered nanofibers. In the implementations discussed here, a single alkali atom (cesium or rubidium) is coupled to a high-quality microsphere, microtoroid, or bottle resonator that is evanescently interfaced to tapered optical fibers. The atom-resonator interaction produces a strong photon-number-dependent nonlinearity that can direct single photons to different output ports and imprint strong correlations on the transmitted and reflected fields~\cite{Aoki2009, OShea2013, Dayan2008, Shomroni2014}.

Early demonstrations established the basic routing mechanisms. In 2008, Dayan et al. used a cesium atom coupled to a microtoroidal resonator, showing regulated, photon-number-dependent transport, verified by a transition from Poissonian input statistics to a sub-Poissonian output~\cite{Dayan2008} [Fig.~\ref{fig:switch_table}(d)]. Without the atom, perfect reflection occurs on resonance due to destructive interference between the cavity field and the input field. The atom-resonator coupling modifies this interference, and is predicted to increase the single-photon transmission probability close to unity, but this reduces to a change of approximately $25\%$ when averaging over different atomic positions. A related experiment was performed by Aoki et al. in 2009, where they used a microtoroid resonator operated in the overcoupled regime, and showed that single photons are preferentially reflected while multi-photon components are transmitted. This manifests as antibunching in reflection and bunching in transmission, and persists for intracavity photon numbers from roughly $0.03$ to $0.7$~\cite{Aoki2009}. In 2013, O'Shea et al. used a bottle microresonator configured as an add-drop filter with two tapered fibers~\cite{OShea2013}. The presence of a single rubidium atom shifted the coupled resonator spectrum and redirected light between the bus and drop ports. This led to an increase in bus-fiber transmission from $3\%$ to $46\%$ and a decrease in bus-to-drop fiber transmission from $58\%$ to $12\%$ when an atom was coupled. The characteristic operation time was set by the atomic transit time near the resonator (about $5$$~\upmu$s).

Based on these ingredients, in 2014 Shomroni et al. realized a single-photon-activated switch in-fiber by coupling a single rubidium atom to a chip-based microsphere resonator~\cite{Shomroni2014} [Fig.~\ref{fig:switch_table}(c)]. A single reflected control photon toggled the device from a high-reflection state to a high-transmission state, routing subsequent target photons with probabilities ${R \approx 65\%}$ and ${T \approx 90\%}$, respectively. The average control energy was about $1.5$ photons per switching event (or ${\sim3}$ when linear losses are included), and no auxiliary control fields were required; control and target photons remain guided in-fiber throughout~\cite{Shomroni2014}. These results underline the appeal of neutral-atom cavity QED for single-photon routing: strong nonlinearities at the single-photon level, fiber compatibility, and clear signatures of photon-number-dependent transport.

Practical limitations are tied to atom delivery and coupling stability. Many experiments rely on releasing laser-cooled atoms past the resonator, so the operation time and repetition rate are set by atomic transit and loading, which restricts continuous operation and switching speed to microsecond time scales~\cite{Aoki2009, OShea2013, Dayan2008}. Variations in atom-field coupling during transit reduce the routing efficiency compared to position-optimized predictions~\cite{Dayan2008}. These challenges motivate deterministic trapping or conveyor loading near the resonator, along with stabilization of coupling and resonance conditions, to translate these proof-of-principle routers into continuously operating nodes. Within these constraints, neutral atoms coupled to nanostructures provide a clear pathway to single-photon switches and routers that are directly compatible with guided-wave architectures~\cite{Aoki2009, OShea2013, Dayan2008, Shomroni2014}.


\subsection{Superconducting qubits}\label{subsec:superconducting_qubits}

Many switching mechanisms involving superconducting qubits operate in the strong coupling regime of circuit QED. In 2021, Wang et al. developed a superconducting quantum router based on circuit QED~\cite{Wang2021}. This was achieved by coupling a transmon qubit to two coplanar waveguide resonators, where off-resonant photons passing through the transmission line were transmitted, whereas photons resonant with the cavity frequency were reflected. The state of the control qubit determined the resonance frequencies of the two cavities. Hence, by encoding a photonic qubit in the frequency of a signal photon, the authors ensured that the signal was always off-resonant when the control qubit was in the “off” state, and at least one frequency component was resonant when the control qubit was in the “on” state. This study highlights the ability to preserve the quantum state, and demonstrates switching of frequency-encoded superposition states for the control and signal photons. The authors reported a fidelity of ${\sim 96\%}$, showing high-quality quantum state preservation during the switching process. Another study by Wang et al. from 2022 demonstrated a single-photon transistor based on circuit QED~\cite{Wang2022} [Fig.~\ref{fig:switch_table}(e)]. This study made use of a similar dual-cavity system, where one cavity was single-sided to maximize reacquisition of gate photons, while the other was a dual-sided cavity designed to maximize signal transmission. The authors reported an extinction ratio of about $15$~dB at the few-photon level, as well as high transistor gain (${> 50}$~dB) for switching larger numbers of signal photons with a single gate photon.

In 2010, Abdumalikov et al. demonstrated EIT in a superconducting quantum circuit, leading to a $96\%$ change in transmission when changing the control field amplitude~\cite{Abdumalikov2010} (see Section~\ref{subsec:theory_EIT} for a discussion on EIT-based switching). The authors used a macroscopic three-level artificial flux atom coupled to a transmission line to demonstrate the EIT mechanism. In this experiment, the system was probed with continuous waves rather than single-photon pulses. In 2011, Hoi et al. performed a pulsed experiment~\cite{Hoi2011} [Fig.~\ref{fig:switch_table}(f)], achieving an extinction of around $99\%$ and a maximum switching speed of ${\sim 10}$~ns, given by the duration of the control pulses. The switching was implemented using a transmon qubit coupled to a superconducting transmission line. The authors used control pulses resonant with the transition frequency of the qubit to induce Autler-Townes splitting, making the transmon transparent to an incoming probe signal. Although the switching operation was demonstrated at the single-photon level, the weak probe consisted of attenuated laser pulses rather than photons produced by a true single-photon source.


\subsection{Solid-state defects}\label{subsec:solidstatedefects}

Solid-state defect platforms span several material families that offer stable, optically addressable emitters with pathways to nanophotonic integration. In wide-band-gap hosts such as diamond and SiC, color centers (e.g., NV, SiV and SnV in diamond; divacancy and V$_\text{Si}$ in SiC) can be created by implantation and annealing or by in situ growth, then embedded in photonic crystal cavities, nanobeam waveguides, or microrings to enhance the light-matter coupling and enable spin-photon interfaces. Crystalline hosts doped with rare-earth ions (e.g., Er$^{3+}$, Yb$^{3+}$ in YSO, YAG, or similar) combine ultra-narrow homogeneous linewidths with exceptionally long spin coherence times, making them promising candidates for quantum memories and telecom-band photon-spin links when coupled to microcavities or integrated waveguides. In silicon photonics, implanted or site-controlled defects such as the G- and T-centers provide single-photon emission compatible with CMOS processes, allowing direct integration with low-loss silicon/silicon nitride waveguides, gratings, and ring resonators. Complementary platforms include 2D materials like hexagonal boron nitride (hBN), where point defects act as bright room-temperature single-photon sources and can be transferred onto pre-fabricated photonic circuits. Together, these defect-based systems offer a spectrum of operating wavelengths, spin properties, and fabrication routes, with a common emphasis on scalable, CMOS-compatible nanophotonic integration.

In 2016, Sipahigil et al. demonstrated cavity and waveguide QED switching using SiV centers in diamond~\cite{Sipahigil2016} [Fig.~\ref{fig:switch_table}(g)]. Their system consisted of 1D nanophotonic devices (cavities and waveguides) etched into diamond, with implanted Si$^+$ ions subsequently annealed to create SiV centers strongly coupled to the photonic modes. By pumping the SiV into a state coupled to (or decoupled from) the probe, they achieved reduced (or enhanced) transmission, demonstrating single-photon-level switching with a switching time of ${\sim 30}$~ns (given by the duration of the gate pulse), a switch memory time of ${\sim 10}$~ns (limited by a thermal phonon relaxation process betweeen the two SiV states), and an extinction of ${\sim 38\%}$.

The main challenges associated with using solid-state defects in nanophotonic devices like single-photon switches include achieving transform-limited linewidths in solid-state environments, reducing spectral diffusion from charge noise, and deterministic placement of emitters in photonic nanostructures. Progress in heteroepitaxy, implantation control, and charge stabilization techniques continues to improve device reproducibility and scalability. These advances pave the way for defect-based solid-state photonic quantum networks with single-photon switching, routing, and memory functionality.


\begin{figure*}[t]
    \centering
    \includegraphics[width=0.98\linewidth]{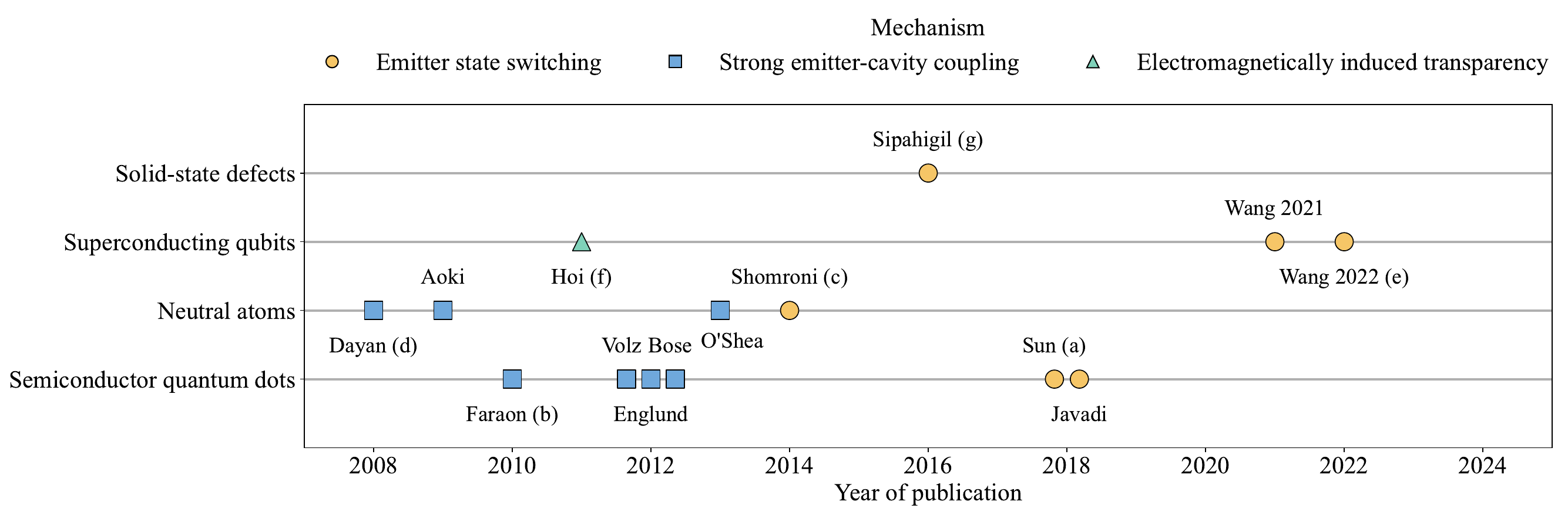}
    \caption{Timeline of experimental single-photon switching demonstrations discussed in the review. Points are arranged by emitter platform and publication year. Marker color and shape denote the switching mechanism: emitter state switching (gold circles), strong emitter-cavity coupling (blue squares), and EIT (green triangles). The timeline includes the representative experiments highlighted in Fig.~\ref{fig:switch_table} together with additional experimental papers discussed in Section~\ref{sec:experiment}: Dayan (d) Ref.~\cite{Dayan2008}, Aoki Ref.~\cite{Aoki2009}, Faraon (b) Ref.~\cite{Faraon2010}, Hoi (f) Ref.~\cite{Hoi2011}, Volz Ref.~\cite{Volz2012}, Englund Ref.~\cite{Englund2012}, Bose Ref.~\cite{Bose2012}, O'Shea Ref.~\cite{OShea2013}, Shomroni (c) Ref.~\cite{Shomroni2014}, Sipahigil (g) Ref.~\cite{Sipahigil2016}, Sun (a) Ref.~\cite{Sun2018}, Javadi Ref.~\cite{Javadi2018}, Wang 2021 Ref.~\cite{Wang2021}, and Wang 2022 (e) Ref.~\cite{Wang2022}.}
    \label{fig:timeline}
\end{figure*}

\subsection{Other platforms}\label{subsec:atomic_ensembles}

We now give a brief overview of photon switching experiments performed in other platforms, first considering atomic ensembles. Atomic ensemble experiments are usually performed in free space as opposed to nanophotonic structures, but there has been significant progress in photon switching with these systems in the last few decades, and therefore a short survey is presented here. Atomic ensemble devices typically consist of a vapor or gas-phase sample of identical atoms (usually a single isotope of rubidium), often at low temperatures, however warm gas experiments have also been performed~\cite{Novikova2004, Dawes2005}. EIT has been used in a number of switching experiments with atomic ensembles, in the form of Rydberg-EIT. This involves applying a control pulse to promote an atom in the ensemble to a Rydberg state (an excited state with high principal quantum number $n$). Due to the van der Waals interaction between the atoms, which scales with a high power of $n$, the energy levels of surrounding atoms within the “blockade radius” are shifted out of resonance with subsequent pulses, preventing their excitation~\cite{Urban2009}. In a switching experiment, a control pulse applied to the ensemble will prevent the absorption of a target pulse within the blockade radius, enabling EIT. This technique has been used to develop a number of optical switching experiments.

In 2009, Bajcsy et al. presented an EIT-induced switching scheme in the slow-light regime by using a laser-cooled atomic cloud in a photonic crystal fiber~\cite{Bajcsy2009}. They obtained a ${\sim 50\%}$ reduction in transmission with a control pulse of around $700$ photons, and a transmission change ${>90\%}$ can be obtained by increasing the number of switch photons to ${\sim 2000}$. In 2014, Gorniaczyk et al. used a free-space atomic ensemble to demonstrate a single-photon transistor, obtaining a switching contrast of ${\sim 40\%}$ and a speed (control pulse duration) of $500$~ns~\cite{Gorniaczyk2014}, which was chosen to maximize the storage efficiency of gate photons. The operation time was limited to ${\sim1}$~$\upmu$s by the coherence time of the Rydberg excitation. A similar experiment was performed in the same year by Tiarks et al.~\cite{Tiarks2014}, who obtained an extinction ratio of $0.89$ at a switch speed of ${\sim 200}$~$\upmu$s, determined by the duration of the pulse sequence (note that, according to their definition of the extinction ratio, the ideal value is $0$ when signal photon transmission is completely blocked by the gate pulse). When the pulse duration exceeds the lifetime of the excited Rydberg states (${\sim 100}$~$\upmu$s), the extinction reduces significantly as the Rydberg blockade effect disappears. Similarly, in 2014 Baur et al. developed a switch leveraging EIT-induced Rydberg blockade, reporting an extinction ratio of $0.812$ without postselection, which improves to $0.051$ when postselecting on a retrieved gate photon~\cite{Baur2014}. In a more recent study, Yu et al. developed a similar Rydberg ensemble-based switch and demonstrated its use on entangled photon pairs~\cite{Yu2020}. They reported a switching contrast of ${77.6\%}$, a switch memory time of ${\sim 800}$~ns (determined by the excited state lifetime), and stated that the switch does not dramatically degrade the entanglement fidelity (${F\approx80}$-$85\%$), showing robust coherence across the switching process.

Although they can show impressive figures of merit, atomic ensemble systems take up significant physical space and can lose atom number over many cycles, requiring a new sample~\cite{Baur2014}. In addition, the switching speeds for atomic ensemble systems are significantly lower than other emitter-based switching platforms, due to the pulse sequences used to create Rydberg-EIT.

Single organic molecules are another relevant class of emitters that have been used to perform photon switching. In 2009, Hwang et al. demonstrated a single-molecule optical transistor, showing that a single dye molecule could coherently attenuate or amplify a tightly focused laser beam under the control of a second gating beam~\cite{Hwang2009}. Later, Maser et al. reported coherent nonlinear optics with a single molecule, including switching of a laser beam with tens of pump photons~\cite{Maser2016}. In a more integrated geometry, T\"urschmann et al. demonstrated chip-based all-optical control of single molecules coherently coupled to dielectric nanoguides, showing that an external optical beam can switch light propagation in an on-chip platform~\cite{Tuerschmann2017}.


\section{Conclusion and outlook}\label{sec:conclusion}

The tunability of quantum emitter properties, including transition frequencies and coupling parameters, makes them controllable single-photon scatterers that can act as effective single-photon switches. Such single-photon switching devices are a key component in various quantum technology proposals based on photonic qubits, including quantum communication protocols~\cite{Karavias2024} and quantum computing architectures~\cite{Bartolucci2021, Chan2025, PsiQuantum2025}. Single-photon switches can also enable effective multiplexing and demultiplexing of single photons in reconfigurable photonic integrated circuits~\cite{Meyer-Scott2020}, leading to controlled photon routing in an on-chip architecture. As the size and complexity of photonic quantum systems grows, there is a greater need for actively routing quantum information to one of possibly many outputs.

In the theory sections of this review, we presented several theoretical methods that are commonly used to model single-photon scattering in waveguide-based nanophotonic structures with quantum emitters (Section~\ref{sec:methods}). We showed how such methods can be used to derive single-photon transmission and reflection amplitudes by applying them to an elementary system consisting of a two-level emitter coupled to a waveguide. We then demonstrated how a quantum emitter coupled to a waveguide can act as a single-photon switch by controlling the photon-emitter detuning. Finally, we reviewed proposals of controllable single-photon routing from the theoretical literature, which often go beyond the simple two-level-emitter-waveguide setup and consider additional degrees of freedom that provide control over single-photon transport (Section~\ref{sec:theory}). This leads to other routing mechanisms including chiral emitter-waveguide coupling~\cite{Yan2011}, strong emitter-cavity coupling~\cite{Shen2009}, EIT and other schemes using external driving fields~\cite{Gong2008}, as well as emitter state switching~\cite{Chang2007}. The basis for all of these routing methods is the ability to control the interference between different photon pathways to direct photons to the desired output port. Introducing additional control parameters is possible via, e.g., multiple coupling points (giant atoms~\cite{Wang2021_2}), dipole-dipole interactions~\cite{Song2025}, and atomic mirrors~\cite{Li2015}, which provide more photon pathways that can be used to control the interference. Moving from point-like interactions to distributed systems consisting of multiple spatially-separated emitters increases the switching bandwidth, allowing wider photon wave packets to be routed with a higher efficiency and fidelity~\cite{Duda2024, Chang2011}.

On the experimental front, single-photon switching with quantum emitters coupled to nanostructures has seen significant progress across multiple platforms over the past two decades. In Section~\ref{sec:FOM_experiment}, we outlined figures of merit associated with single-photon switches, including efficiency and fidelity, speed and operation time, and scalability and compatibility. In Sections~\ref{sec:structures} and \ref{sec:experiment}, we reviewed the different types of nanophotonic structures that have been coupled with quantum emitters and the experimental results of single-photon switch demonstrations across different physical platforms. This includes semiconductor QDs, neutral atoms, superconducting qubits, and solid-state defects. Superconducting circuits currently have the highest switching efficiencies, with on-off ratios of around $99\%$ being reported~\cite{Hoi2011}. However, their operation in the microwave regime limits their compatibility with optical communication channels, which requires the development of efficient optical-microwave interfaces. The highest switching speeds have been demonstrated with all-optical control of QD-cavity systems, and are on the order of tens of picoseconds~\cite{Volz2012, Englund2012}. This is not unexpected, as these switches are ultimately limited by the emitter-cavity coupling rate (which can be very high in the strong coupling regime), as well as the bandwidth of the control pulse. However, it will prove a significant challenge to create single-photon nonlinearities (e.g., a single control photon operating the switch) using these schemes. This is because the minimum energy required to operate such a switch is limited by the imperfect coupling of the control pulse to the cavity mode~\cite{Nozaki2010}. Progress in this area could be made by shaping the control pulse to optimize coupling to the cavity~\cite{Bose2012}. The control pulse-cavity mode overlap must be improved if these high-speed all-optical switches are to be implemented in photonic quantum technologies with single-photon-level operation.

Neutral atoms have been efficiently coupled to high-quality whispering-gallery resonators for switching applications, but their operation time is limited by the atomic transit time past the resonator, and the way the atoms are delivered to the resonator leads to coupling instability and makes it challenging to construct stable planar photonic devices. Solid-state defects such as color centers in silicon or silicon carbide are a promising platform for CMOS-compatible nanophotonic integration and room-temperature operation. This contrasts with superconducting qubit and QD systems, which currently require cryogenic environments for effective operation.

As is evident from Figs.~\ref{fig:switch_table} and \ref{fig:timeline}, the switching mechanism that has received the most experimental attention, especially in the last decade, is emitter state switching. All four of the emitter platforms shown in these figures have been able to demonstrate switching via the state of quantum emitters, showing that this switching method is relatively well investigated across the platforms. This may be because this method provides a clear pathway to switching with a single control photon, in contrast to schemes that require driving fields to induce blockade or EIT effects. The next most commonly reported technique is based on strong emitter-cavity coupling. This mechanism has received the most attention in semiconductor QD systems possibly due to the ease of coupling QDs to photonic crystal cavities, and the ability to couple these cavities to other nanostructures such as waveguides. Switching based on cavity QED has also received significant attention in neutral-atom systems for similar reasons, as the ease of strongly coupling atoms to high-quality whispering-gallery resonators makes these systems an appealing target for emitter-cavity switches.

On the other hand, despite receiving significant numbers of theoretical proposals (see Section~\ref{subsec:theory_EIT}), EIT-based single-photon switching has so far only been reported in superconducting systems, out of the four platforms in Fig.~\ref{fig:switch_table}. This gap in the literature could be a result of EIT requiring long coherence times, which are not often observed in solid-state systems~\cite{Fleischhauer2005}. This is due to a range of effects, for example, in QD systems EIT is significantly affected by inhomogeneous broadening~\cite{Marcinkeviius2008}. By contrast, EIT-based switching has proven to be very effective in atomic ensembles, as these systems have long coherence times and a long blockade range provided by Rydberg interactions~\cite{Urban2009}.

\begin{figure*}[t]
    \centering
    \includegraphics[width=0.9\linewidth]{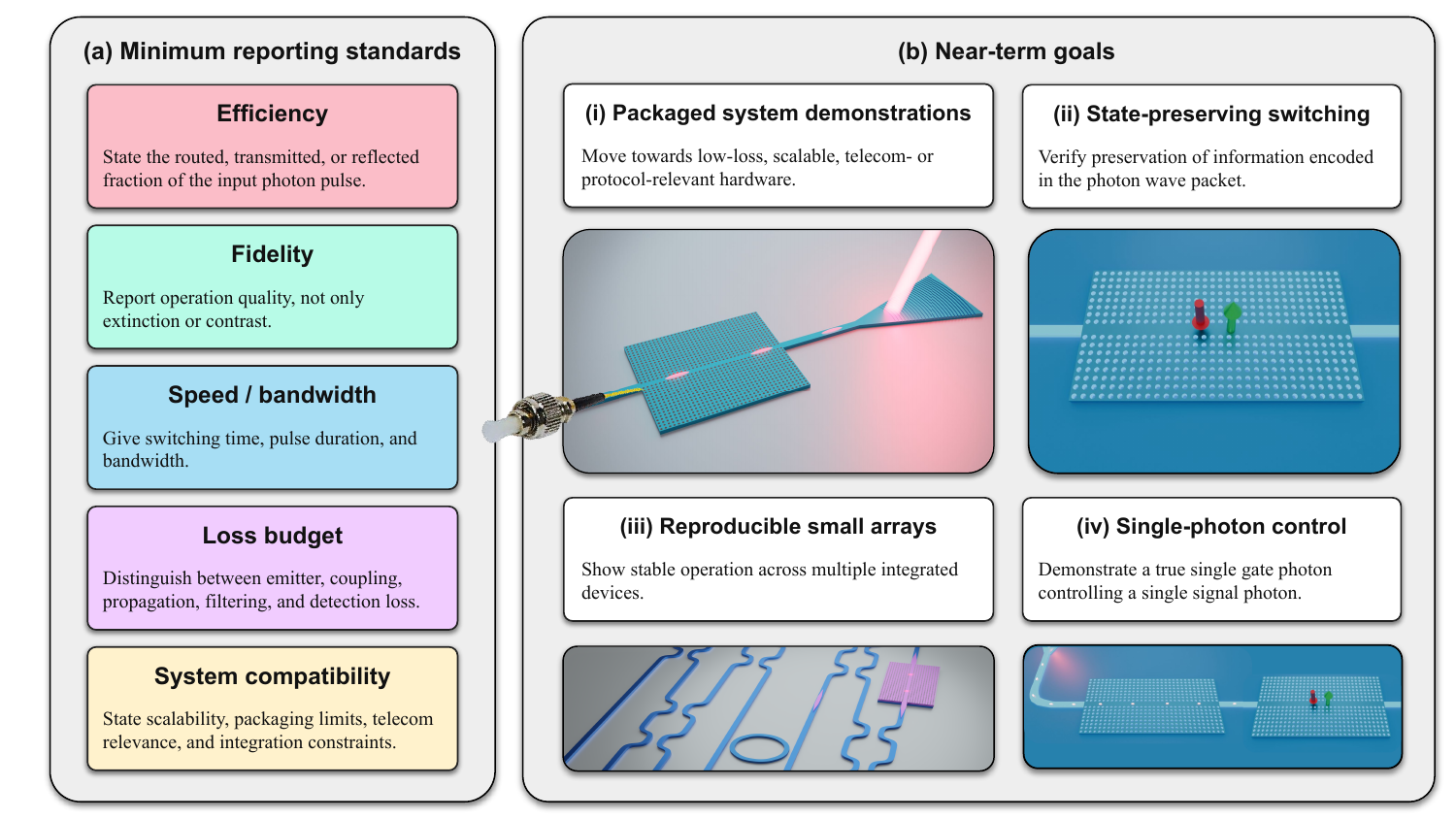}
    \caption{Summary of recommended reporting standards and near-term targets for quantum-emitter-based single-photon switches. (a) Minimum reporting standards for future demonstrations: report the routed, transmitted, or reflected fraction of the input photon pulse (efficiency), the operation quality beyond extinction or contrast alone (fidelity), the switching time together with pulse duration and bandwidth, a complete loss budget separating emitter, coupling, propagation, filtering, and detection losses, and system-level compatibility including scalability, packaging constraints, and telecom relevance. (b) Near-term goals for the research field, illustrated schematically: (i) packaged, low-loss, scalable, telecom- or protocol-relevant hardware; (ii) state-preserving switching that maintains the information encoded in the photon wave packet; (iii) reproducible operation across small integrated arrays; and (iv) true single-photon control, where a single gate photon controls a single signal photon.}
    \label{fig:report_goals}
\end{figure*}

Another switching method that has received significant attention in the theoretical literature is chiral/directional emitter-waveguide coupling (Section~\ref{subsec:theory_chiral}). Chiral emitter-waveguide coupling is now firmly established at the level of spin-dependent, highly directional emission and nonreciprocal transport~\cite{Lodahl2017}. Experiments on QDs in photonic crystal and nanobeam waveguides, as well as atoms coupled to optical nanofibers, have demonstrated near-unity directionality and robust spin-to-path conversion, the essential ingredients for chiral routing. However, a fully reconfigurable, low-loss multiport switch based on chiral coupling has not been realized to date. The central challenge is achieving fast, in situ control of the coupling asymmetry without having to physically relocate the quantum emitter along the mode profile of the nanostructure. Practical routes include tuning a different degree of freedom: (i) electro-optic or thermo-optic refractive-index control to shift local chiral ``hot spots''; (ii) spectral tuning of the quantum emitter relative to nearby cavity/slow-light resonances that provide a strongly wavelength-dependent directionality~\cite{Martin2025}; or (iii) spin control via Zeeman or Stark tuning to select which circularly polarized transition couples to the forward/backward mode, thereby toggling the routed port. Early demonstrations validate high-directionality spin-path interfaces and path-spin control, but they have not yet delivered calibrated, low-loss, reconfigurable operation at MHz-GHz rates with multiport readout. Looking ahead, multi-emitter waveguide QED provides a natural pathway to enhance the routing probability and realize deterministic port selection via interference and collectively engineered decay channels~\cite{hallett2025controll, Lodahl_super}; translating these concepts from theory and cold-atom or proof-of-principle settings to solid-state chiral platforms is a compelling near-term target.

Many demonstrations of emitter-based single-photon switches have so far relied on attenuated laser pulses rather than single photons from true single-photon sources. Moving the field forward will require experiments that realize genuine single-photon control, in which a single gate photon switches a single signal photon, together with continued progress towards scalable, packaged, and telecom-compatible devices for larger quantum networks. As summarized in Fig.~\ref{fig:report_goals}(b), we highlight four near-term directions: (i) demonstrations of packaged hardware compatible with telecom wavelengths or other protocol-relevant operating regimes; (ii) state-preserving switching, with preservation of information encoded in the photon wave packet; (iii) reproducible small arrays, with stable operation across multiple integrated devices; and (iv) single-photon control, in which a single gate photon controls a single signal photon. The studies reviewed here already provide a strong foundation of concepts and methods for pursuing these goals, and further refinement of these switching schemes should help drive the development of the high-quality devices required for next-generation photonic quantum technologies.

To enable meaningful comparison between platforms and to assess suitability for quantum technology applications, we suggest a small set of reporting conventions for single-photon switching experiments, summarized in Fig.~\ref{fig:report_goals}(a). Within these reporting standards, we would like to highlight key areas of improvement.

Firstly, we note that there is a significant absence of fidelity measurements in the photon switching experiments. Measurement of the fidelity of photonic states after switching is crucial to confirm that the switch can faithfully preserve the information encoded in the routed wave packet. There are multiple reasons for this under-reporting of the switching fidelity. For example, the fidelity of quantum information transfer may have been out of scope for first demonstrations of specific mechanisms. In addition, fidelity is more challenging to measure than efficiency, requiring consistent quantum state readout and quantum state tomography to extract a value (such as the method used in Ref.~\cite{Wang2021}). Future experimental work on single-photon switching for quantum technology applications should focus on how well the proposed switch can preserve the input photon wave packet. A high fidelity is essential to maintain the indistinguishability of photons routed through a quantum network, which is a necessary requirement in quantum technologies using photonic qubits, e.g., for two-photon interference in quantum computing schemes~\cite{Kok2007}.

Pulse duration (or equivalently spectral width) should be stated for control and signal fields, since this determines how well the input wave packet matches the switching bandwidth and whether temporal reshaping is expected. Second, where possible, transmitted, reflected, and lost (out-of-mode) fractions of the input signal should be quantified simultaneously, so that a quoted extinction ratio can be distinguished from genuine redirection into a desired output port. Third, fidelity should be reported whenever the intended application relies on preserving the photonic state, either via quantum process tomography or via bounds obtained from interference/indistinguishability measurements, since high extinction alone does not guarantee that the output wave packet remains suitable for high-visibility interference.

One near-term route to sharpening the connection between device physics and circuit-level utility is to treat the temporal profile of the input photon wave packet as a primary design parameter, rather than an experimental detail. As emphasized in Sections~\ref{subsec:methods_5} and \ref{subsec:FOM_experiment_2}, pulse duration fixes the spectral width of the incident photon, and therefore determines how much of the wave packet actually overlaps the relevant scattering bandwidth~\cite{Roy2017, Duda2024}. When the pulse is too short, most spectral components can be off-resonant, reducing the efficiency and fidelity of the routing process. On the other hand, when the input pulse is very long, experiments tend towards the steady-state response that is easier to measure but does not capture reshaping effects that become important when the pulse duration approaches the emitter lifetime~\cite{Liao2015, Manzoni2014}.

Many of the practical limitations of solid-state emitters ultimately appear as loss or decoherence channels in the switching process. Even if the quantum emitter couples efficiently to the guided mode of the nanostructure, phonon-assisted emission, phonon sidebands, and temperature-dependent pure dephasing reduce the coherent fraction of the scattered field, degrading both indistinguishability and switch fidelity~\cite{Iles-Smith2017, Muljarov2004, Reigue2017, Denning2020}. Charge and spin noise can likewise produce spectral wandering on experimentally relevant timescales, complicating calibration and reducing device stability~\cite{Kuhlmann2013, Zhai2020}. These effects motivate more realistic models that can predict performance under repeated operation, particularly when the control degree of freedom must remain coherent during switching. Nevertheless, the outlook remains positive: many of the improvements required for quantum-emitter-based single-photon switches, including reduced photon loss, longer coherence times, and more mature photonic integration, are already being driven by the broader development of quantum emitters as high-quality sources of single photons and more complex photonic resource states (e.g., cluster states). Continued progress in these areas may naturally support the emergence of practical single-photon switches for future photonic quantum technologies.


\begin{acknowledgments}
The authors thank Daniel Hodgson and Jake Iles-Smith for helpful discussions. This work was supported by the Engineering and Physical Sciences Research Council [grant numbers EP/W524360/1, EP/V026496/1, and EP/Z533208/1].
\end{acknowledgments}


\section*{Author declarations}

\subsection*{Conflict of interest}

The authors have no conflicts to disclose.


\section*{Data availability}
Data sharing is not applicable to this article as no new data were created or analyzed in this study.

\newpage
\onecolumngrid
\appendix


\section{Real-space approach---Hamiltonian and Schr\"{o}dinger equation}\label{app:real-space_model}

In this appendix, we first show how the $k$-space Hamiltonian in Eq.~(\ref{eq:H_k_linearised}) is transformed into the position-space Hamiltonian given in Eq.~(\ref{eq:H_x}), and then use the resulting Hamiltonian to solve the Schr\"{o}dinger equation [Eq.~(\ref{eq:Schrodinger})]. When we substitute the Fourier transform in Eq.~(\ref{eq:Fourier_transform}) into Eq.~(\ref{eq:H_k_linearised}), we obtain
\begin{align}
\begin{split}
    H =&\; \omega_e \sigma^+ \sigma^- + \int_{-\infty}^{\infty} dk\; v_gk\; \left( \frac{1}{\sqrt{2\pi}} \int_{-\infty}^{\infty} a_R^{\dagger}(x)e^{i(k+k_0)x}dx \right) \left( \frac{1}{\sqrt{2\pi}} \int_{-\infty}^{\infty} a_R(x')e^{-i(k+k_0)x'}dx' \right)\\[0.05in]
    &- \int_{-\infty}^{\infty} dk\; v_gk\; \left( \frac{1}{\sqrt{2\pi}} \int_{-\infty}^{\infty} a_L^{\dagger}(x)e^{i(k-k_0)x}dx \right) \left( \frac{1}{\sqrt{2\pi}} \int_{-\infty}^{\infty} a_L(x')e^{-i(k-k_0)x'}dx' \right)\\[0.05in]
    &+ \int_{-\infty}^{\infty} dk \left[ \frac{V_R}{\sqrt{2\pi}} \left( \frac{1}{\sqrt{2\pi}} \int_{-\infty}^{\infty} a_R^{\dagger}(x)e^{i(k+k_0)x}dx \right)\sigma^- + \frac{V^*_R}{\sqrt{2\pi}} \left( \frac{1}{\sqrt{2\pi}} \int_{-\infty}^{\infty} a_R(x)e^{-i(k+k_0)x}dx \right)\sigma^+ \right]\\[0.05in]
    &+ \int_{-\infty}^{\infty} dk \left[ \frac{V_L}{\sqrt{2\pi}} \left( \frac{1}{\sqrt{2\pi}} \int_{-\infty}^{\infty} a_L^{\dagger}(x)e^{i(k-k_0)x}dx \right)\sigma^- + \frac{V^*_L}{\sqrt{2\pi}} \left( \frac{1}{\sqrt{2\pi}} \int_{-\infty}^{\infty} a_L(x)e^{-i(k-k_0)x}dx \right)\sigma^+ \right]\\[0.1in]
    =&\; \omega_e \sigma^+ \sigma^- + v_g \int_{-\infty}^{\infty} dx \int_{-\infty}^{\infty} dx' \left[ a_R^{\dagger}(x)a_R(x')e^{ik_0(x-x')} - a_L^{\dagger}(x)a_L(x')e^{-ik_0(x-x')} \right] \left( \frac{1}{2\pi} \int_{-\infty}^{\infty} dk\; k\;e^{ik(x-x')} \right)\\[0.05in]
    &+ \int_{-\infty}^{\infty} dx \left[ V_R a_R^{\dagger}(x) \sigma^- \left( \frac{1}{2\pi} \int_{-\infty}^{\infty} dk\; e^{i(k+k_0)x} \right) + V^*_R a_R(x) \sigma^+ \left( \frac{1}{2\pi} \int_{-\infty}^{\infty} dk\; e^{-i(k+k_0)x} \right) \right]\\[0.05in]
    &+ \int_{-\infty}^{\infty} dx \left[ V_L a_L^{\dagger}(x) \sigma^- \left( \frac{1}{2\pi} \int_{-\infty}^{\infty} dk\; e^{i(k-k_0)x} \right) + V^*_L a_L(x) \sigma^+ \left( \frac{1}{2\pi} \int_{-\infty}^{\infty} dk\; e^{-i(k-k_0)x} \right) \right].
\end{split}
\end{align}
To evaluate the integrals over $k$, we use
\begin{subequations}
\begin{equation}
    \frac{1}{2\pi} \int_{-\infty}^{\infty} dk\; e^{\pm i(k+k_0)x} = \frac{1}{2\pi} \int_{-\infty}^{\infty} dk\; e^{\pm ikx} = \delta(x),
\end{equation}
\begin{equation}
    \frac{1}{2\pi} \int_{-\infty}^{\infty} dk\; e^{\pm i(k-k_0)x} = \frac{1}{2\pi} \int_{-\infty}^{\infty} dk\; e^{\pm ikx} = \delta(x),
\end{equation}
\end{subequations}
where we made the substitutions ${k \pm k_0 = k' \rightarrow k}$. In addition,
\begin{equation}
    \frac{1}{2\pi} \int_{-\infty}^{\infty} dk\; k\;e^{ik(x-x')} = \frac{1}{2\pi} \int_{-\infty}^{\infty} dk\; i\frac{\partial}{\partial x'} \left[e^{ik(x-x')}\right] = i\frac{\partial}{\partial x'} \left[ \frac{1}{2\pi} \int_{-\infty}^{\infty} dk\; e^{ik(x-x')} \right] = i\frac{\partial}{\partial x'} \delta(x-x').
\end{equation}
This leads to
\begin{align}
\begin{split}
    H =&\; \omega_e \sigma^+ \sigma^- + iv_g \int_{-\infty}^{\infty} dx \int_{-\infty}^{\infty} dx' \left[ a_R^{\dagger}(x)a_R(x')e^{ik_0(x-x')} - a_L^{\dagger}(x)a_L(x')e^{-ik_0(x-x')} \right] \frac{\partial}{\partial x'} \delta(x-x')\\[0.05in]
    &+ \int_{-\infty}^{\infty} dx \left[ V_R a_R^{\dagger}(x) \sigma^- + V^*_R a_R(x) \sigma^+ + V_L a_L^{\dagger}(x) \sigma^- + V^*_L a_L(x) \sigma^+ \right] \delta(x)\\[0.1in]
    =&\; \omega_e \sigma^+ \sigma^- +iv_g \int_{-\infty}^{\infty} dx\; a_R^{\dagger}(x) e^{ik_0x} \int_{-\infty}^{\infty} dx'\; a_R(x') e^{-ik_0x'} \frac{\partial}{\partial x'} \delta(x-x')\\[0.05in]
    &- iv_g \int_{-\infty}^{\infty} dx\; a_L^{\dagger}(x) e^{-ik_0x} \int_{-\infty}^{\infty} dx'\; a_L(x') e^{ik_0x'} \frac{\partial}{\partial x'} \delta(x-x')\\[0.05in]
    &+ V_R a_R^{\dagger}(0) \sigma^- + V_R^* a_R(0) \sigma^+ + V_L a_L^{\dagger}(0) \sigma^- + V_L^* a_L(0) \sigma^+.
\end{split}
\end{align}
We now evaluate the integrals over $x'$ via integration by parts, which moves the derivatives from the delta functions to the operators. Note that integration by parts produces a boundary term where $x'$ is set to $\pm \infty$, but these terms vanish assuming that all fields decay to zero in the limit ${x' \rightarrow \pm \infty}$ (there are no contributions from mode operators evaluated at infinity). Integrating over $x'$ in this way gives
\begin{align}
\begin{split}
    H =&\; \omega_e \sigma^+ \sigma^- -iv_g \int_{-\infty}^{\infty} dx\; a_R^{\dagger}(x) e^{ik_0x} \int_{-\infty}^{\infty} dx'\; \frac{\partial}{\partial x'} \left[a_R(x') e^{-ik_0x'}\right] \delta(x-x')\\[0.05in]
    &+ iv_g \int_{-\infty}^{\infty} dx\; a_L^{\dagger}(x) e^{-ik_0x} \int_{-\infty}^{\infty} dx'\; \frac{\partial}{\partial x'} \left[a_L(x') e^{ik_0x'}\right] \delta(x-x')\\[0.05in]
    &+ V_R a_R^{\dagger}(0) \sigma^- + V_R^* a_R(0) \sigma^+ + V_L a_L^{\dagger}(0) \sigma^- + V_L^* a_L(0) \sigma^+\\[0.1in]
    =&\; \omega_e \sigma^+ \sigma^- -iv_g \int_{-\infty}^{\infty} dx\; a_R^{\dagger}(x) e^{ik_0x} \frac{\partial}{\partial x} \left[a_R(x) e^{-ik_0x}\right] + iv_g \int_{-\infty}^{\infty} dx\; a_L^{\dagger}(x) e^{-ik_0x} \frac{\partial}{\partial x} \left[a_L(x) e^{ik_0x}\right]\\[0.05in]
    &+ V_R a_R^{\dagger}(0) \sigma^- + V_R^* a_R(0) \sigma^+ + V_L a_L^{\dagger}(0) \sigma^- + V_L^* a_L(0) \sigma^+\\[0.1in]
    =&\; \omega_e \sigma^+ \sigma^- + \int_{-\infty}^{\infty} dx \left[ -iv_g a_R^{\dagger}(x) \frac{\partial}{\partial x}a_R(x) + iv_g a_L^{\dagger}(x) \frac{\partial}{\partial x}a_L(x) \right]\\[0.05in]
    &+ V_R a_R^{\dagger}(0) \sigma^- + V_R^* a_R(0) \sigma^+ + V_L a_L^{\dagger}(0) \sigma^- + V_L^* a_L(0) \sigma^+ - v_gk_0 \int_{-\infty}^{\infty} dx \left[ a_R^{\dagger}(x) a_R(x) + a_L^{\dagger}(x) a_L(x) \right].
\end{split}
\end{align}
The last integral above corresponds to a constant phase that depends on the linearization point of the waveguide dispersion ($k_0$). This term can be dropped without any observable consequences, which gives us the desired real-space Hamiltonian from Eq.~(\ref{eq:H_x}):
\begin{align}
\begin{split}
H =&\; \omega_e \sigma^+ \sigma^- + \int_{-\infty}^{\infty} dx \left[ -iv_g a_R^{\dagger}(x) \frac{\partial}{\partial x}a_R(x) + iv_g a_L^{\dagger}(x) \frac{\partial}{\partial x}a_L(x) \right] + V_R a_R^{\dagger}(0) \sigma^- + V_R^* a_R(0) \sigma^+\\[0.05in]
&+ V_L a_L^{\dagger}(0) \sigma^- + V_L^* a_L(0) \sigma^+.
\end{split}
\end{align}

When we apply this Hamiltonian to the single-photon state $\ket{\psi_k}$ in Eq.~(\ref{eq:state}), we obtain:
\begin{align}
\begin{split}
    H\ket{\psi_k} =&\; \omega_e u_e \sigma^+ \sigma^- \ket{0,e} - iv_g \int_{-\infty}^{\infty} dx \int_{-\infty}^{\infty} dx'\; u_{R,k}(x') a_R^{\dagger}(x) \frac{\partial}{\partial x} [a_R(x)] a_R^{\dagger}(x') \ket{0,g}\\[0.05in]
    &+ iv_g \int_{-\infty}^{\infty} dx \int_{-\infty}^{\infty} dx'\; u_{L,k}(x') a_L^{\dagger}(x) \frac{\partial}{\partial x} [a_L(x)] a_L^{\dagger}(x') \ket{0,g} + V_R u_e a_R^{\dagger}(0)\sigma^- \ket{0,e}\\[0.05in]
    &+ V_R^* \int_{-\infty}^{\infty} dx\; u_{R,k}(x)a_R(0)a_R^{\dagger}(x)\sigma^+ \ket{0,g} + V_L u_e a_L^{\dagger}(0)\sigma^- \ket{0,e} + V_L^* \int_{-\infty}^{\infty} dx\; u_{L,k}(x)a_L(0)a_L^{\dagger}(x)\sigma^+ \ket{0,g},
\end{split}
\end{align}
where we used the bosonic commutation relations ${[a_{\mu}(x), a_{\nu}^{\dagger}(x')] = \delta_{\mu\nu}\delta(x-x')}$ (all other terms are zero due to commuting operators, resulting in annihilation operators acting on the vacuum state). To simplify the double integrals above, we also need to use the commutator $[\partial_x a_{\mu}(x), a_{\mu}^{\dagger}(x')] = -\partial_{x'}\delta(x-x')$ for ${\mu \in \{L,R\}}$, which follows from the Fourier transform in Eq.~(\ref{eq:Fourier_transform_2}). Using this result, and simplifying the other terms, gives
\begin{align}
\begin{split}
    H\ket{\psi_k} =&\; \omega_e u_e \ket{0,e} + iv_g \int_{-\infty}^{\infty} dx\; a_R^{\dagger}(x)\int_{-\infty}^{\infty} dx'\; u_{R,k}(x') \frac{\partial}{\partial x'} \delta(x-x') \ket{0,g}\\[0.05in]
    &- iv_g \int_{-\infty}^{\infty} dx\; a_L^{\dagger}(x) \int_{-\infty}^{\infty} dx'\; u_{L,k}(x') \frac{\partial}{\partial x'} \delta(x-x') \ket{0,g} + V_R u_e \int_{-\infty}^{\infty} dx\; \delta(x)a_R^{\dagger}(x) \ket{0,g}\\[0.05in]
    &+ V_R^* u_{R,k}(0) \ket{0,e} + V_L u_e \int_{-\infty}^{\infty} dx\; \delta(x) a_L^{\dagger}(x) \ket{0,g} + V_L^* u_{L,k}(0) \ket{0,e}.
\end{split}
\end{align}
Integrating over $x'$ by parts allows us to move the derivatives from the delta functions to the amplitudes. As before, we neglect boundary terms where the amplitudes would be evaluated at infinity, such that
\begin{align}
\begin{split}
    H\ket{\psi_k} =&\; \omega_e u_e \ket{0,e} - iv_g \int_{-\infty}^{\infty} dx\; a_R^{\dagger}(x)\int_{-\infty}^{\infty} dx'\; \delta(x-x') \frac{\partial}{\partial x'}u_{R,k}(x')  \ket{0,g}\\[0.05in]
    &+ iv_g \int_{-\infty}^{\infty} dx\; a_L^{\dagger}(x) \int_{-\infty}^{\infty} dx'\; \delta(x-x') \frac{\partial}{\partial x'} u_{L,k}(x') \ket{0,g} + V_R u_e \int_{-\infty}^{\infty} dx\; \delta(x)a_R^{\dagger}(x) \ket{0,g}\\[0.05in]
    &+ V_R^* u_{R,k}(0) \ket{0,e} + V_L u_e \int_{-\infty}^{\infty} dx\; \delta(x) a_L^{\dagger}(x) \ket{0,g} + V_L^* u_{L,k}(0) \ket{0,e}\\[0.1in]
    =&\; \omega_e u_e \ket{0,e} - iv_g \int_{-\infty}^{\infty} dx\; a_R^{\dagger}(x) \frac{\partial}{\partial x}u_{R,k}(x)  \ket{0,g} + iv_g \int_{-\infty}^{\infty} dx\; a_L^{\dagger}(x) \frac{\partial}{\partial x} u_{L,k}(x) \ket{0,g}\\[0.05in]
    &+ V_R u_e \int_{-\infty}^{\infty} dx\; \delta(x)a_R^{\dagger}(x) \ket{0,g} + V_R^* u_{R,k}(0) \ket{0,e} + V_L u_e \int_{-\infty}^{\infty} dx\; \delta(x) a_L^{\dagger}(x) \ket{0,g} + V_L^* u_{L,k}(0) \ket{0,e}\\[0.1in]
    =&\; \left[ \omega_e u_e + V_R^* u_{R,k}(0) + V_L^* u_{L,k}(0) \right]\ket{0,e} + \int_{-\infty}^{\infty} dx\; \left[ -iv_g \frac{\partial}{\partial x}u_{R,k}(x) + V_R u_e \delta(x) \right]a_R^{\dagger}(x)\ket{0,g}\\[0.05in]
    &+ \int_{-\infty}^{\infty} dx\; \left[ iv_g \frac{\partial}{\partial x}u_{L,k}(x) + V_L u_e \delta(x) \right]a_L^{\dagger}(x)\ket{0,g}.
\end{split}
\end{align}
When we equate this to the right-hand side of Eq.~(\ref{eq:Schrodinger}),
\begin{equation}
    v_gk\ket{\psi_k} = \int_{-\infty}^{\infty} dx\; v_gk\; u_{R,k}(x) a_R^{\dagger}(x)\ket{0,g} + \int_{-\infty}^{\infty} dx\; v_gk\;u_{L,k}(x) a_L^{\dagger}(x)\ket{0,g} + v_gk\;u_e\ket{0,e},
\end{equation}
and compare coefficients of the states $\ket{0,e}$, $a_R^{\dagger}(x)\ket{0,g}$, and $a_L^{\dagger}(x)\ket{0,g}$, we arrive at the three simultaneous equations in Eqs.~(\ref{eq:u_e})-(\ref{eq:u_L}).


\section{Discrete coordinate scattering approach---Hamiltonian and Schr\"{o}dinger equation}\label{app:discrete_model}

Here we derive the dispersion relation of the coupled-resonator waveguide in Fig.~\ref{fig:two_level_emitter_waveguide}(b) by diagonalizing the waveguide Hamiltonian in $k$-space. We then solve the Schr\"{o}dinger equation using the full Hamiltonian and the dispersion relation. Substituting the Fourier transform in Eq.~(\ref{eq:discrete_FT}) (and its Hermitian conjugate) into Eq.~(\ref{eq:H_cavities}) leads to
\begin{align}
\begin{split}
    \omega_c& \sum_j a_j^{\dagger} a_j - \xi\sum_j \left( a_{j+1}^{\dagger} a_j + a_j^{\dagger} a_{j+1} \right)\\
    &= \omega_c \sum_j \left( \frac{1}{\sqrt{N}} \sum_k b_k^{\dagger}e^{-ikj} \right) \left( \frac{1}{\sqrt{N}} \sum_{k'} b_{k'}e^{ik'j} \right) - \xi \sum_j \left[ \left( \frac{1}{\sqrt{N}} \sum_k b_k^{\dagger}e^{-ik(j+1)} \right) \left( \frac{1}{\sqrt{N}} \sum_{k'} b_{k'}e^{ik'j} \right) \right.\\
    &\left.\hspace{3.5in} + \left( \frac{1}{\sqrt{N}} \sum_k b_k^{\dagger}e^{-ikj} \right) \left( \frac{1}{\sqrt{N}} \sum_{k'} b_{k'}e^{ik'(j+1)} \right) \right]\\[0.1in]
    &= \omega_c \sum_k \sum_{k'} b_k^{\dagger} b_{k'} \left( \frac{1}{N} \sum_j e^{-i(k-k')j} \right) - \xi \sum_k \sum_{k'} b_k^{\dagger} b_{k'} \left( e^{-ik} + e^{ik'} \right) \left( \frac{1}{N} \sum_j e^{-i(k-k')j} \right).
\end{split}
\end{align}
In a one-dimensional chain with periodic boundary conditions, the allowed wave numbers are ${k = (2\pi n/L) = 2\pi n/(Na)}$, where ${L = Na}$ is the total length of the chain, $a$ is the lattice constant, and $n$ is an integer. We have set ${a=1}$ for simplicity, so the wave number is an integer multiple of $2\pi/N$. Hence,
\begin{equation}
    \frac{1}{N}  \sum_j e^{-i(k-k')j} = \delta_{kk'},
\end{equation}
which follows from the identity
\begin{equation}
    \delta_{nm} = \frac{1}{N} \sum_j e^{-2\pi i (n-m)j/N}
\end{equation}
for integers $n$ and $m$ (${N \rightarrow \infty}$). Substituting this into the above expression for the waveguide Hamiltonian gives
\begin{align}
\begin{split}
    \omega_c \sum_j a_j^{\dagger} a_j - \xi\sum_j \left( a_{j+1}^{\dagger} a_j + a_j^{\dagger} a_{j+1} \right) =&\; \omega_c \sum_k \sum_{k'} b_k^{\dagger} b_{k'} \delta_{kk'} - \xi \sum_k \sum_{k'} b_k^{\dagger} b_{k'} \left( e^{-ik} + e^{ik'} \right) \delta_{kk'}\\
    =& \sum_k \left[ \omega_c - \xi \left( e^{-ik} + e^{ik} \right) \right] b_k^{\dagger} b_k\\
    =& \sum_k \left[ \omega_c - 2\xi \text{cos}(k) \right] b_k^{\dagger} b_k\\
    =& \sum_k \omega_k b_k^{\dagger} b_k,
\end{split}
\end{align}
where ${\omega_k = \omega_c - 2\xi \text{cos}(k)}$ is the dispersion relation.

When we apply the full Hamiltonian in Eq.~(\ref{eq:H_discrete}) to the single-photon state $\ket{\psi_k}$ in Eq.~(\ref{eq:state_discrete}), we obtain:
\begin{align}
\begin{split}
    H_{\text{crw}}\ket{\psi_k} =&\; \omega_e u_e \sigma^+ \sigma^- \ket{0,e} + \omega_c \sum_j \sum_{j'} u_{j',k} a_j^{\dagger} a_j a_{j'}^{\dagger} \ket{0,g} - \xi \sum_j \sum_{j'} u_{j',k} a_{j+1}^{\dagger} a_j a_{j'}^{\dagger} \ket{0,g} - \xi \sum_j \sum_{j'} u_{j',k} a_j^{\dagger} a_{j+1} a_{j'}^{\dagger} \ket{0,g}\\
    &+ gu_e a_0^{\dagger} \sigma^- \ket{0,e} + g^* \sum_j u_{j,k} a_0 \sigma^+ a_j^{\dagger} \ket{0,g}.
\end{split}
\end{align}
This can be simplified using the definitions of $\sigma^+$ and $\sigma^-$, as well as the commutation relations ${[a_j,a_{j'}^{\dagger}] = \delta_{jj'}}$:
\begin{align}
\begin{split}
    H_{\text{crw}}\ket{\psi_k} =&\; \omega_e u_e \ket{0,e} + \omega_c \sum_j u_{j,k} a_j^{\dagger} \ket{0,g} - \xi \sum_j u_{j,k} a_{j+1}^{\dagger} \ket{0,g} - \xi \sum_j u_{j+1,k} a_j^{\dagger} \ket{0,g}\\
    &+ gu_e a_0^{\dagger} \ket{0,g} + g^* u_{0,k} \ket{0,e}\\[0.1in]
    =&\; \omega_e u_e \ket{0,e} + \omega_c \sum_j u_{j,k} a_j^{\dagger} \ket{0,g} - \xi \sum_j u_{j-1,k} a_{j}^{\dagger} \ket{0,g} - \xi \sum_j u_{j+1,k} a_j^{\dagger} \ket{0,g}\\
    &+ gu_e \sum_j \delta_{j0} a_j^{\dagger} \ket{0,g} + g^* u_{0,k} \ket{0,e}\\
    =&\; (\omega_e u_e + g^* u_{0,k})\ket{0,e} + \sum_j \left[ \omega_cu_{j,k} - \xi(u_{j-1,k} + u_{j+1,k}) + gu_e\delta_{j0} \right]  a_j^{\dagger} \ket{0,g},
\end{split}
\end{align}
where, to get from the first line to the second line, we changed the summation index from $j$ to ${n = j+1}$ in the third term, and then relabeled $n$ with $j$. When we equate this to the right-hand side of Eq.~(\ref{eq:Schrodinger_2}),
\begin{equation}
    \left[ \omega_c - 2\xi\text{cos}(k) \right] \ket{\psi_k} = \sum_j \left[ \omega_c - 2\xi\text{cos}(k) \right] u_{j,k} a_j^{\dagger} \ket{0,g} + \left[ \omega_c - 2\xi\text{cos}(k) \right] u_e \ket{0,e},
\end{equation}
and compare coefficients of the states $\ket{0,e}$ and $a_j^{\dagger}\ket{0,g}$, we arrive at the two simultaneous equations in Eqs.~(\ref{eq:u_e_discrete}) and (\ref{eq:u_j_discrete}).

\newpage
\twocolumngrid


%

\end{document}